\RequirePackage{rotating}
\documentclass[fleqn,usenatbib]{mnras}

\usepackage{newtxtext}
\usepackage[varg]{newtxmath}
\usepackage[utf8]{inputenc}

\usepackage[T1]{fontenc}
\usepackage{ae,aecompl}
\usepackage{xcolor}
\usepackage{multirow}

\usepackage{graphicx}	
\usepackage{amsmath}	

\usepackage{amssymb}	
\usepackage{hyperref}
\usepackage{caption}
\usepackage{adjustbox}
\usepackage{subcaption}
\usepackage{rotating}

\usepackage{siunitx}
\usepackage{booktabs}
\usepackage{etoolbox}

\newcounter{ionstage}
\renewcommand{\ion}[2]{\setcounter{ionstage}{#2}%
  \ensuremath{\mathrm{#1\,\scriptstyle\Roman{ionstage}}}}
\newcommand\hii{\ion{H}{2}}
\newcommand\nii{[\ion{N}{2}]}

\newcommand\Wav[1]{\ensuremath{\lambda #1}}
\newcommand\Mach{\ensuremath{\mathcal{M}}}
\newcommand\shock{\ensuremath{_{\mathrm{sh}}}}
\newcommand\sound{\ensuremath{c_{\mathrm{s}}}}
\newcommand\ws{\ensuremath{_{\mathrm{ws}}}}
\newcommand\jet{\ensuremath{_{\mathrm{jet}}}}

\title[HH~529~II and III in the Orion Nebula]{Photoionized Herbig-Haro objects in the Orion Nebula through deep high-spectral resolution spectroscopy I: HH~529~II and III}

\author[J. E. M\'endez-Delgado et al.]
{J. E. M\'endez-Delgado$^{1,2}$ \thanks{E-mail: jemd@iac.es},
C. Esteban$^{1,2}$, J. Garc{\'{\i}}a-Rojas$^{1,2}$, W. J. Henney$^{3}$  
\newauthor 
A. Mesa-Delgado$^{4}$ and K. Z. Arellano-C\'ordova$^{1}$\\
\\
$^{1}$Instituto de Astrof\'isica de Canarias (IAC), E-38205 La Laguna, Spain\\
$^{2}$Departamento de Astrof\'isica, Universidad de La Laguna, E-38206 La Laguna, Spain\\
$^{3}$Instituto de Radioastronom\'ia y Astrof\'isica, Universidad Nacional Aut\'onoma de M\'exico, Apartado Postal 3-72, 58090 Morelia, Michoac\'an, M\'exico\\
$^{4}$ Calle Camino Real 64, Icod el Alto, Los Realejos, 38414, Tenerife, Spain}

\date{Accepted XXX. Received YYY; in original form ZZZ}

\pubyear{2021}

\begin{document}
\label{firstpage}
\pagerange{\pageref{firstpage}--\pageref{lastpage}}
\maketitle

\begin{abstract}

We present the analysis of physical conditions, chemical composition and kinematic properties of two bow shocks ---HH~529~II and HH~529~III--- of the fully photoionized Herbig-Haro object HH~529 in the Orion Nebula. The data  were obtained with the Ultraviolet and Visual Echelle Spectrograph at the 8.2m Very Large Telescope and 20 years of \textit{Hubble Space Telescope} imaging. We separate the  emission of the high-velocity components of HH~529~II and III from the nebular one, determining $n_{\rm e}$ and $T_{\rm e}$ in all components through multiple diagnostics, including some based on recombination lines (RLs). We derive ionic abundances of several ions, based on collisionally excited lines (CELs) and RLs. We find a good agreement between the predictions of the temperature fluctuation paradigm ($t^2$) and the abundance discrepancy factor (ADF) in the main emission of the Orion Nebula. However, $t^2$ can not account for the higher ADF found in HH~529 II and III. We estimate a 6\%  of Fe in the gas-phase of the Orion Nebula, while this value increases to 14\% in HH~529~II and between 10\% and 25\% in HH~529~III. We find that such increase is probably due to the destruction of dust grains in the bow shocks. We find an overabundance of C, O, Ne, S, Cl and Ar of about 0.1 dex in HH~529~II-III that might be related to the inclusion of H-deficient material from the source of the HH~529 flow. We determine the proper motions of HH~529 finding multiple discrete features. We estimate a flow angle with respect to the sky plane of $58 \pm 4^{\circ}$ for HH~529.

\end{abstract}




\section{Introduction}
\label{sec:introduction}

\begin{figure*}
\includegraphics[width=\textwidth]{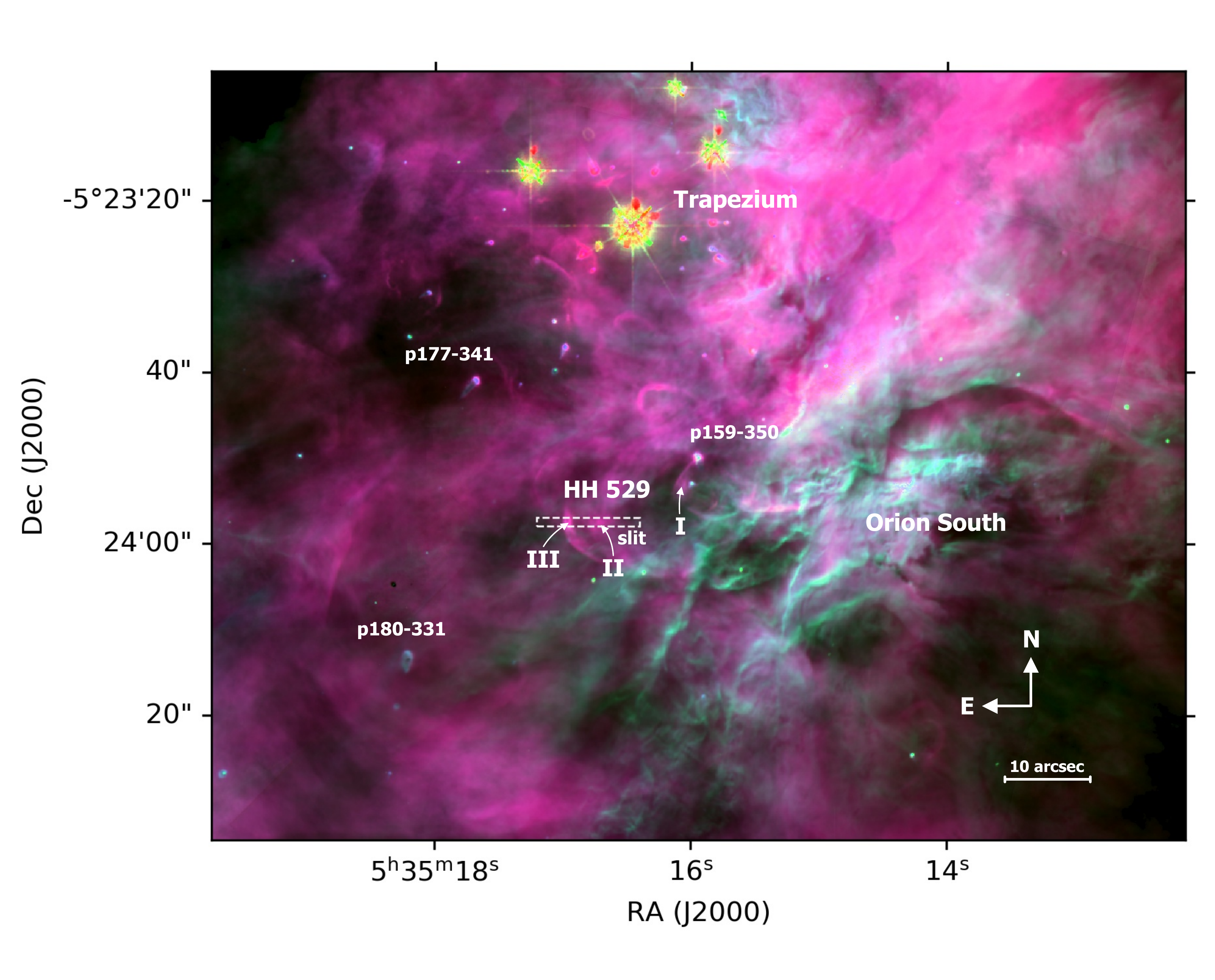}
\caption{Composite WFPC2 \textit{HST} image of the central Orion Nebula. Three narrow filters were used for the color scale: F502N, F658N and F656N for red, green and blue, respectively \citep[images obtained by][]{Bally:1998a}. The slit position of our observations is indicated. It covers HH~529~III and HH~529~II. HH~529 bowshocks I, II and III are indicated following the designation given by \citet{odellyhenney08}. The position of some protoplanetary discs (proplyds) are also indicated.}
\label{fig:hst}
\end{figure*} 

Herbig-Haro (HH) objects are small emission nebulae associated with  outflows from young stars interacting with the surrounding environment \citep{Schwartz83}. Since their discovery by George Herbig and Guillermo Haro \citep{Herbig50,Herbig51, Herbig52, Haro52, Haro53} a multitude of them have been discovered and studied. Through the \textit{Hubble Space Telescope} \textit{(HST)}, multiple velocity features associated with HH objects have been observed in the Orion Nebula with unprecedented detail. There are several works dedicated to determine the nature and physical properties of many outflows from stars in the Orion Nebula \citep[see][and references therein]{bally00,bally01,odellyhenney08,Odell15}. These have revealed that the Orion Nebula is a complex environment with multiple gas interactions. These high velocity systems cover a wide range of velocities with noticeable differences in the conditions of their emitting gas.

Through the radiation field of the massive stars of the Orion Nebula, HH objects can be photoionized under conditions where the shock between the ambient gas and the HH merely serves to create a dense blob where we can determine the physical conditions and chemical abundances using the standard methods developed to study ionized nebulae \citep{Reipurth01}. Moderate velocity ($v<100 \text{ km s}^{-1}$) shocks in H\thinspace II regions are predicted to be strongly radiative, showing only a thin high-$T_{\rm e}$ cooling zone immediately behind the shock, which contributes little to the total emission \citep{henney02}. The bulk of the shocked gas returns to thermal equilibrium at the same $T_{\rm e}$ as the ambient gas, hence the combined front (shock plus cooling zone) can be considered isothermal. However, there are few works in the literature dedicated to analyse the chemical composition of photoionized HH objects, isolating their emission from that of the nebula in which they are immersed. Using high-spectral resolution spectroscopy, \citet{Blagrave06} and \citet{mesadelgado09} were able to separate the emission of HH~529~III+II and HH~202~S, respectively, from the main emission of the Orion Nebula. This permitted the analysis of the chemical composition of the ionized gas under the peculiar physical conditions of the HHs and the effects  of their interaction with the surrounding nebular gas, such as the chemical effects of dust destruction. 

As \citet{mesadelgado08} showed through long slit spectra, there are important spatial variations in the physical conditions of the Orion Nebula due to the presence of HH objects. These variations also affect some chemical properties of the gas. For example, these authors found an increase in the discrepancy between the abundances obtained from recombination lines (RLs) and collisionally excited lines (CELs) for the same heavy element at the locations of HH objects. Therefore, it is important to investigate the physical and chemical influence that HH objects exert on the gas of ionized nebula and test our knowledge of photoionized regions by analysing objects with complex conditions.

This work aims to be the first in a series devoted to the analysis of photoionized HHs in the Orion Nebula using very high resolution spectroscopy from the Ultraviolet and Visual Echelle Spectrograph (UVES) \citep{Dodorico00} attached to the UT2 (Kueyen) of the Very Large Telescope (VLT). This paper is dedicated to two bow shocks associated with HH~529: HH~529~II and HH~529~III. HH~529 consists of a series of shocks flowing toward the east in the central region of the Orion Nebula. It is divided into three main shocks designated as HH~529~I, HH~529~II and HH~529~III, numbered from west to east \citep{odellyhenney08}. We spatially separate the emission from HH~529~II and HH~529~III and isolate the blueshifted high-velocity emission of the gas of the shock from the nebular one. We analyse our high-spectral resolution observations that cover a wide spectral range (3100-10400 \AA) through 4 spatial cuts, obtaining 7 1D spectra: 4 corresponding to the main emission of the Orion Nebula, one for HH~529~II, another one for HH~529~III and one additional 1D spectrum corresponding to the sum of all the 1D spectra. This last spectrum simulates a single low-spectral resolution longslit observation, including the mixing of the HH emission with that of the nebular gas, summing up the emission of all the velocity components for each emission line. In this paper we  analyse the physical conditions, chemical composition and kinematic properties of HH~529~II and HH~529~III as well as the Orion Nebula in several small and nearby areas. 

The paper is organized as follows: in Section~\ref{sec:data} we describe the observations and the reduction process for the spectroscopic data, as well as the {\it HST} imaging used to calculate the proper motions of HH~529 in the plane of the sky. In Section~\ref{sec:line_inten} we describe the emission line measurements, identifications and the reddening correction as well as a comparison between our observations and those from \citet{Blagrave06} over the common spectral range (3500-7500 \AA). In Section~\ref{sec:physical_conditions} we derive the physical conditions of the gas throughout different methods, using CELs, RLs and continuum emission. In Section~\ref{sec:chemical_abundances} we derive ionic abundances using both RLs and CELs. In Section~\ref{sec:temp_fluc} we describe the temperature fluctuations paradigm and estimate values of $t^2$, based on the different temperature diagnostics. In Section~\ref{sec:ADF} we discuss the abundance discrepancy (AD) between ionic abundances derived with CELs and RLs. In Section~\ref{sec:total_abun} we analyse the total abundances obtained from RLs and CELs, in the second case both with and without the assumption of the existence of temperature fluctuations ($t^2>0$ and $t^2=0$, respectively). We also discuss the increase in the gaseous Fe abundance due to dust destruction in HH~529~II and HH~529~III. In Section~\ref{sec:kin_analisys} we describe the radial velocity structure of each component, both the nebular and the high-velocity ones. We also derive the electron temperature from the thermal broadening of the line profiles. In Sections~\ref{sec:proper-motions-hh} and \ref{sec:dym_impl} we calculate the proper motions of HH~529 and discuss some physical properties of the shock, such as the pre-shock density. Finally, in Section~\ref{sec:summary} we summarize our main conclusions. In the appendix, some extra information, tables and figures are attached as supporting material.

\section{Observations and data reduction}
\label{sec:data}

The observations were made under photometric conditions during the night of November 28 and 29, 2013 using UVES in the UT2 of the Very large Telescope (VLT) in Cerro Paranal, Chile. The slit position was centred at the coordinates RA(J2000)=05$^h$35$^m$16$^s$.80, DEC(J2000)=$-$05$^{\circ}$23$'$57.48$''$, with a slit length of 10 arcsec in the blue arm and 12 arcsec in the red arm in order to give an adequate interorder separation. Table~\ref{tab:observations} shows the main parameters of UVES observations. The slit width was set to 1 arcsec, which provides an effective spectral resolution $\lambda$/$\Delta \lambda \approx$ 40000 (6.5 km s$^{-1}$). To perform the flux calibration of the data, three exposures of 150s of the standard star GD71 \citep{Moehler14a,Moehler14b} were taken under similar conditions of seeing and airmass than the science observations during the same night. The spatial coverage of the slit is shown in Fig.~\ref{fig:hst}.

Our observations cover the spectral range between 3100-10420~\AA, using two standard dichroic settings of UVES.  Dichroic~\#1 setting split the light in two wavelengths ranges: from 3100 to 3885~\AA$\text{}$ in the blue arm and from 4785 to 6805~\AA$\text{}$ in the red one, while the dichroic~\#2 setting covers from 3750 to 4995 \AA$\text{}$ in the blue arm and from 6700 to 10420~\AA$\text{}$ in the red one. However, in our high resolution and wide spectral range observations, there are some observational gaps. The red arm use two CCDs, and due to their physical separation, spectral ranges 5773--5833 \AA$\text{}$ and 8540--8650 \AA$\text{}$ could not be observed. Additionally there are some narrow gaps that could not be observed in the redmost part of the red arm in the dichroic~\#2 setting because the spectral orders could not fit entirely within the CCD. These ranges are $\sim$ 8911--8913 \AA, 9042--9046 \AA, 9178--9182 \AA, 9317--9323 \AA, 9460--9469 \AA, 9608--9619 \AA, 9760--9774 \AA, 9918--9935 \AA, 10080--10100 \AA$\text{}$ and 10248--10271 \AA. 

\begin{table}
\caption{Main parameters of UVES spectroscopic observations.}
\label{tab:observations}
\begin{tabular}{ccccc}
\hline
Date & $\Delta \lambda$& Exp. time  &Seeing &Airmass\\
 & (\AA) &  (s) & (arcsec)&\\
\hline
2013-11-29 & 3100-3885 & 5, 3$\times$180 &0.79&1.20\\
2013-11-29 & 3750-4995 & 5, 3$\times$600 &0.65&1.14\\
2013-11-29 & 4785-6805 & 5, 3$\times$180 &0.79&1.20\\
2013-11-29 & 6700-10420 & 5, 3$\times$600 &0.65&1.14\\
\hline
\end{tabular}
\end{table}

\begin{table}
  \caption{\textit{HST} observations used in proper motion study.}
  \label{tab:programs}
  \setlength\tabcolsep{1ex}
  \begin{tabular}{llll}
    \toprule
    Date & Program & Camera, CCD, Filter & Reference \\
    \midrule
    1995-03 & 5469 & WFPC2, PC, F656N & \citet{Bally:1998a} \\
    2005-04 & 10246 & ACS, WFC, F658N & \citet{Robberto:2013a} \\
    2015-01 & 13419 & WFC3, UVIS, F656N & \citet{Bally:2018c} \\
    \bottomrule
  \end{tabular}
\end{table}

We reduced the spectra using a combination of tasks from the public ESO UVES pipeline  \citep{Ballester00} under the {\sc gasgano} graphic user interface, and tasks built by ourselves based on IRAF\footnote{IRAF is distributed by National Optical Astronomy Observatory, which is operated by Association of Universities for Research in Astronomy, under cooperative agreement with the National Science Foundation} \citep{Tody93} and several python packages. Firstly, we used IRAF taks FIXPIX and IMCOMBINE to mask known bad pixels in our images and to combine all the images with the same exposure time. Then, we used the ESO UVES pipeline for bias subtraction, background subtraction, aperture extraction, flat-fielding and wavelength calibration. As a product, we obtained a 2D science spectrum for each arm in each dichroic setting without flux calibration. We followed the same procedure for GD\,71 but extracting both a 2D and a 1D spectrum. The 2D spectrum of the calibration star helps us to note the presence of faint sky lines which are also present in the science spectra.
 

\begin{figure*}
  \begin{subfigure}{6cm}
    \centering\includegraphics[height=2cm,width=\columnwidth]{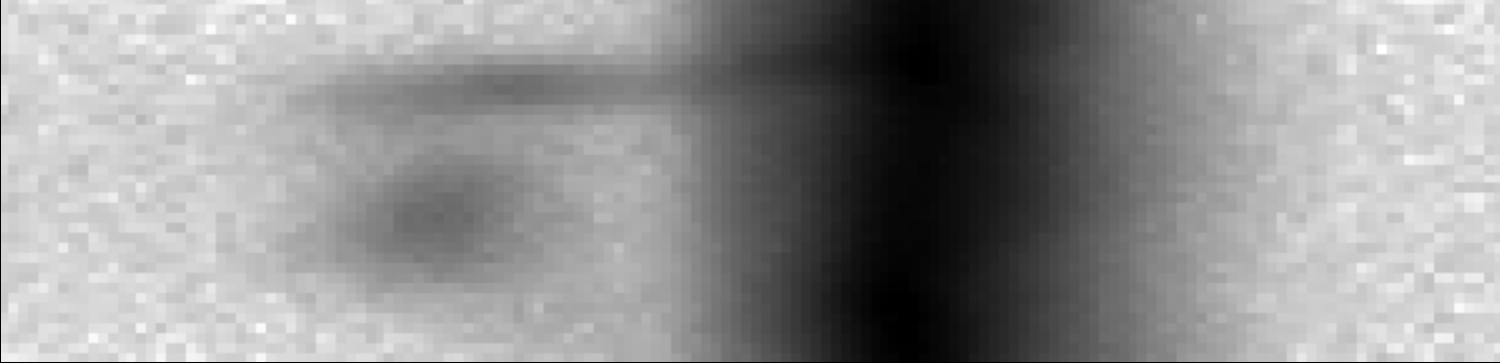}
    \caption{[O\thinspace II] $\lambda 3729$.}
  \end{subfigure}
  \begin{subfigure}{6cm}
    \centering\includegraphics[height=2cm,width=\columnwidth]{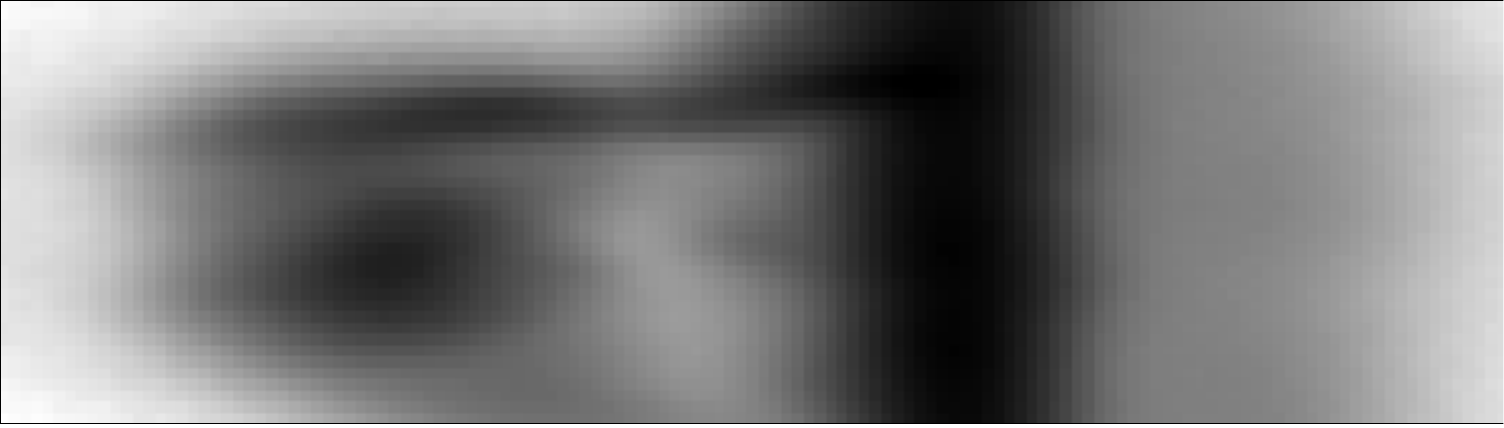}
    \caption{[O\thinspace III] $\lambda 4959$.}
  \end{subfigure}
 
  \begin{subfigure}{6cm}
    \centering\includegraphics[height=2cm , width=\columnwidth]{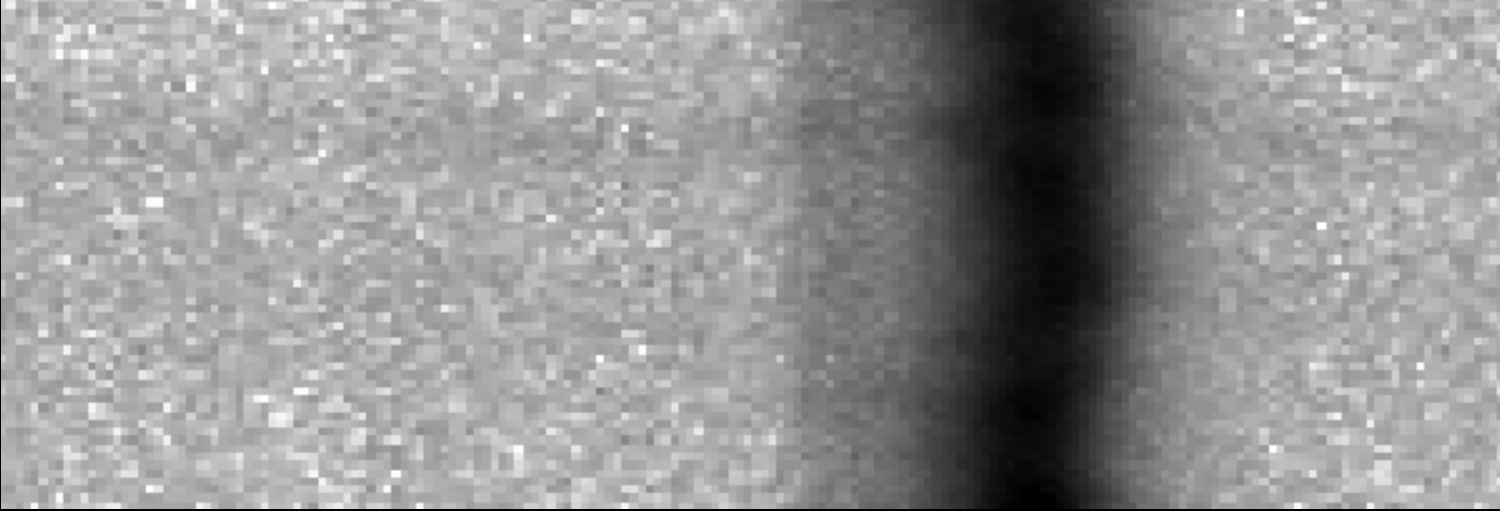}
    \caption{[O\thinspace I] $\lambda 6300$.}
  \end{subfigure}
  \begin{subfigure}{6cm}
    \centering\includegraphics[height=2cm , width=\columnwidth]{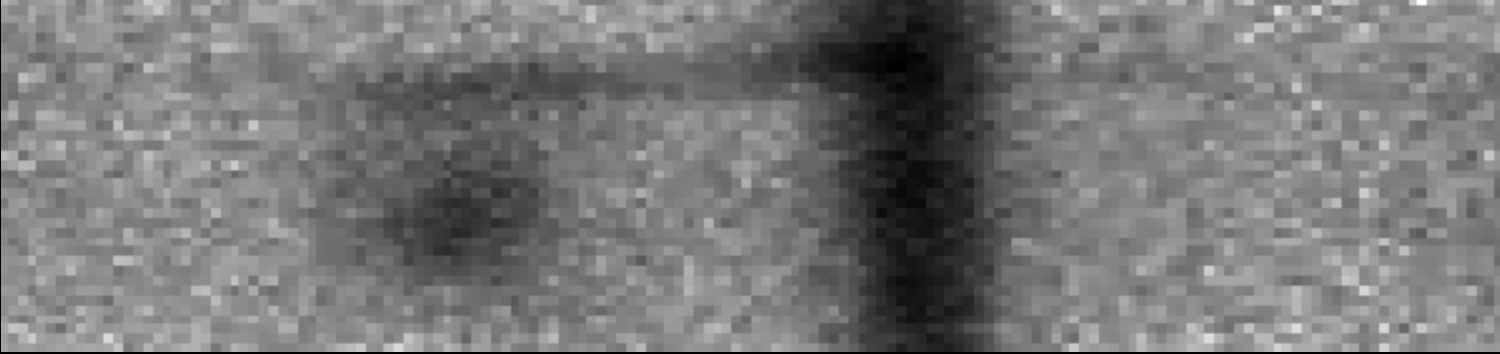}
    \caption{O\thinspace II $\lambda 4649$.}
  \end{subfigure}
\begin{subfigure}{10cm}
\centering\includegraphics[height=4cm, width=\columnwidth]{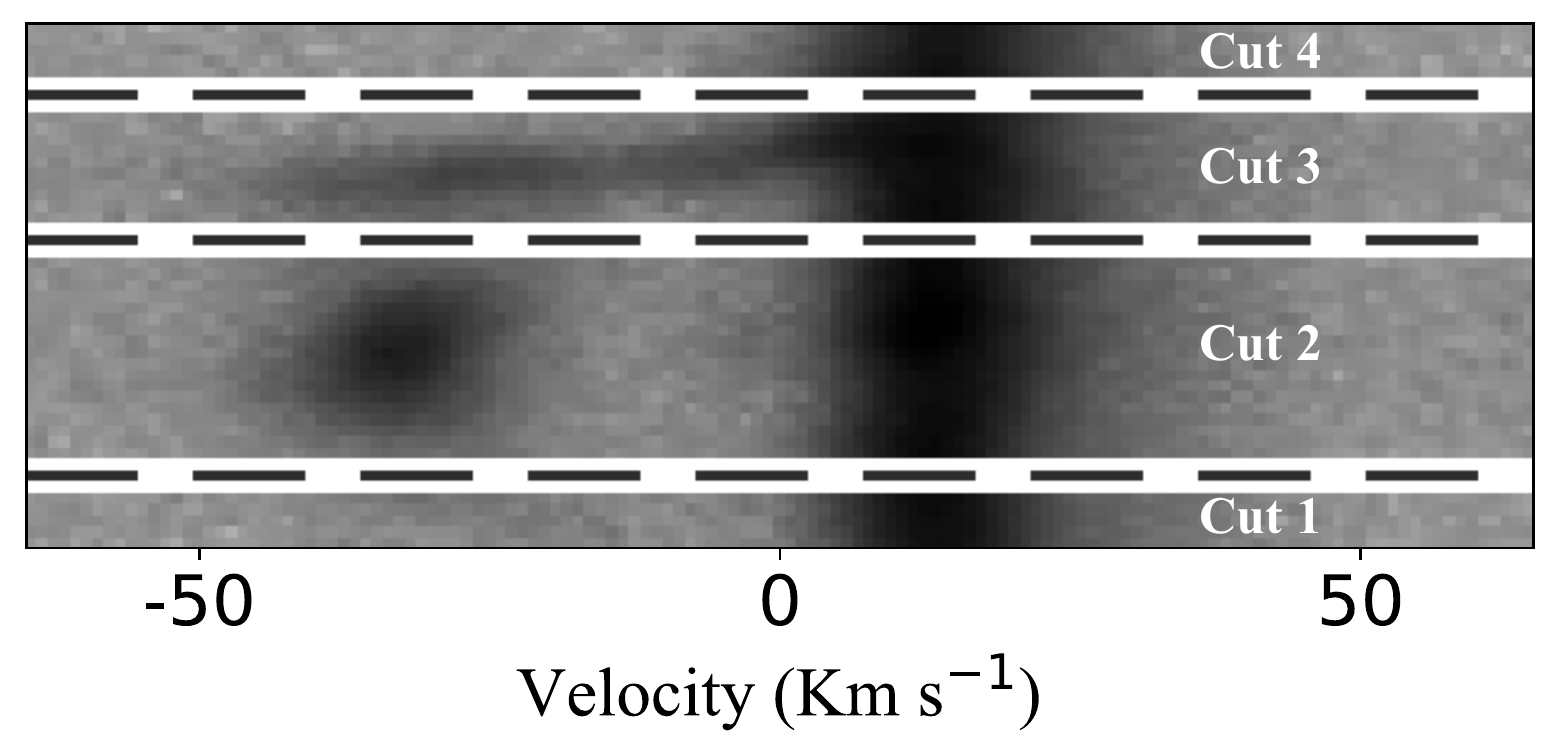}
\caption{[Fe\thinspace III] $\lambda 4658$. The 4 cuts are identified. Bottom: Cut 1, Top: Cut 4.} 
\end{subfigure}
\caption{\textit{Upper panels:} Sample of representative lines in the bi-dimensional spectrum. The Y axis corresponds to the spatial direction (up east, down west, see Fig.\ref{fig:hst} for the spatial location of the slit) while the X axis is the spectral axis. All figures are centered at $\lambda_0$, the rest-frame reference wavelength of each line. The ``ball-shaped'' emission corresponds to HH~529~II while the elongated one to HH~529~III. The blueshifted high-velocity components do not present emission from neutral elements as [O\thinspace I] and rather faint emission from low ionization ions such as [O\thinspace II]. \textit{Bottom panel:} Emission of the [Fe\thinspace III] $\lambda 4658.17$ line as well as the limits and extension of the different spatial cuts selected to analyse each velocity component. Cut 1 is at the bottom, which corresponds to the westernmost one. 
The spatial coverage is 1.23 arcsec, 4.43 arcsec, 2.46 arcsec and 1.23 arcsec for cuts 1, 2, 3 and 4, respectively.
The velocity scale is heliocentric.}
\label{fig:cuts}
\end{figure*}

One crucial step of the data reduction is to perform adequate cuts in the spatial direction of the slit to extract 1D spectra. We chose these spatial cuts in order to study in detail each observed velocity component and trying to maximise the shock/nebular emission ratio. We relied on the [Fe\thinspace III] $\lambda$4658 line, which is relatively bright in the high-velocity components, to delimit the cuts. In the bi-dimensional spectrum shown in Fig.~\ref{fig:cuts} for some representative lines, we show that the seeing conditions permit us to spatially separate HH~529~II from HH~529~III. HH~529~II has a ``ball shape'' while HH~529~III presents an elongated distribution along the spectral axis. This is related to the morphology of the outflow system of HH~529 \citep[firstly identified by][]{bally00}. This system shows three prominent bright arcs, identified by the numbers I, II and III, being numbered by their position from west to east \citep{odellyhenney08}. However, the system is more complex than just three homogeneous arcs as we will analyse in Section~\ref{sec:proper-motions-hh}. The length covered by each cut in the spatial direction is 1.23 arcsec, 4.43 arcsec, 2.46 arcsec and 1.23 arcsec for cuts 1, 2, 3 and 4, respectively. The numbering of the cuts has been defined from west to east. The high-velocity component of cut 3 corresponds to HH~529~III, while that of cut 2 is HH~529~II. We have also defined an additional 1D spectrum, labelled as ``combined cuts''. This was created by adding the flux of the lines in all the velocity components when they were detected at least in the nebular emission of all cuts. The spectrum of the combined cuts is useful for analysing the effect that a non-resolved shock component would have in the properties of a low-resolution spectrum.
We used the Python-based Astropy package \citep{astropy_a,astropy_b} to obtain 1D spectra for each cut, doing the conversion between the different pixel scale of the CCDs in the blue and the red arm. Each spatial cut covers an area larger than the seeing size during the observations, as is shown in Table~\ref{tab:observations}. We used the IRAF tasks STANDARD, SENSFUNC and CALIBRATE to perform the flux calibration of each 1D spectra of all cuts. The radial velocity correction was made using Astropy. 

For the determination of the proper motion of HH~529, we take advantage of the 20~years of archival $HST$ imaging that is now available. We employ three epochs of observations, as detailed in Table~\ref{tab:programs}.
All data were downloaded from the Barbara~A. Mikulski Archive for Space Telescopes%
\footnote{MAST, \url{https://archive.stsci.edu/}}.

\section{Line intensities and reddening}
\label{sec:line_inten}

We used the SPLOT task from IRAF to measure line intensities and estimate their uncertainties. We applied a double Gaussian profile fit for the nebular and the high-velocity component, delimiting the continuum by eye. The error estimations are carried out by SPLOT by Monte-Carlo simulations around a gaussian sigma defined as the average {\it rms} measured on the continuum on both sides of each line with 100 iterations. The error estimates are one sigma estimates. 
We also consider an error of the absolute flux calibration of 2\%, added quadratically. In case of evident line blending, we applied as many Gaussians as necessary to properly reproduce the line profile. As was mentioned in Section~\ref{sec:data}, the observed wavelength range (3100--10420 \AA) was covered in 4 sections (two dichroic settings splitting the light into two spectrograph arms). Between each section, there is an overlapping zone from where we used the most intense lines to normalize the entire spectrum with respect to  H$\beta$. The measured flux of H~{\sc i} $\lambda$3835, [O~{\sc iii}] $\lambda$4959 and [S~{\sc ii}] $\lambda$6731 lines were used to normalize the spectra from the blue arm of dichroic~\#1, the red arm of dichroic~\#1 and the red arm of dichroic~\#2 settings respectively (H$\beta$ is in the blue arm of dichroic~\#2 setting). This normalization eliminates the differences in flux between each part of the spectrum due to the different pixel scale between the blue and the red arms. 

The emission lines were corrected for reddening using Eq.~(\ref{eq:red}), where $f(\lambda)$ is the adopted extinction curve from \citet{Blagrave07}, normalized to $\text{H}\beta$. We calculate the reddening coefficient, $\text{c}(\text{H}\beta)$, by using the ratios of $\text{H}\varepsilon$, $\text{H}\delta$, $\text{H}\gamma$ and $\text{H}\alpha$ Balmer lines and the P12, P11, P10, P9 Paschen lines with respect to $\text{H}\beta$ and the emissivity coefficients of \citet{Storey95}.

\begin{equation}
\label{eq:red}
\frac{I(\lambda)}{I(\text{H}\beta)}=\frac{F(\lambda)}{F(\text{H}\beta)} \times 10^{\text{c}(\text{H}\beta) f(\lambda)}.
\end{equation}

\noindent  The final adopted $\text{c}(\text{H}\beta)$ value is the weighted average value obtained from the aforementioned Balmer and Paschen lines and is shown in Table~\ref{tab:c_extin} for each component. The selected H\thinspace I lines are free of line-blending or telluric absorptions that may affect the determination of $\text{c}(\text{H}\beta)$. Despite the existence of further isolated and bright Balmer and Paschen lines in the spectra, we did not use them since their emission depart from the case B values. This behaviour was reported previously  in the Orion Nebula \citep{mesadelgado09}, the Magellanic Clouds \citep{dominguezguzman19} and in several planetary nebulae \citep[PNe, see][]{rodriguez20}.

\begin{table}
\caption{Reddening coefficients for each component.}
\label{tab:c_extin}
\begin{tabular}{lccccc}
\hline
 & \multicolumn{2}{c}{$\text{c}(\text{H}\beta)$} \\
 & High-velocity & Nebula\\
\hline
Cut 1 & - & $0.82 \pm  0.02$\\
Cut 2 & $0.90 \pm 0.03$ &$0.83 \pm 0.02$\\
Cut 3 &$0.89 \pm 0.05 $&$0.84 \pm 0.03$\\
Cut 4 & - & $0.83 \pm  0.02$\\
Combined cuts &-&$0.85 \pm 0.02$\\
\hline
\end{tabular}
\end{table}

\begin{figure}
\includegraphics[width=\columnwidth]{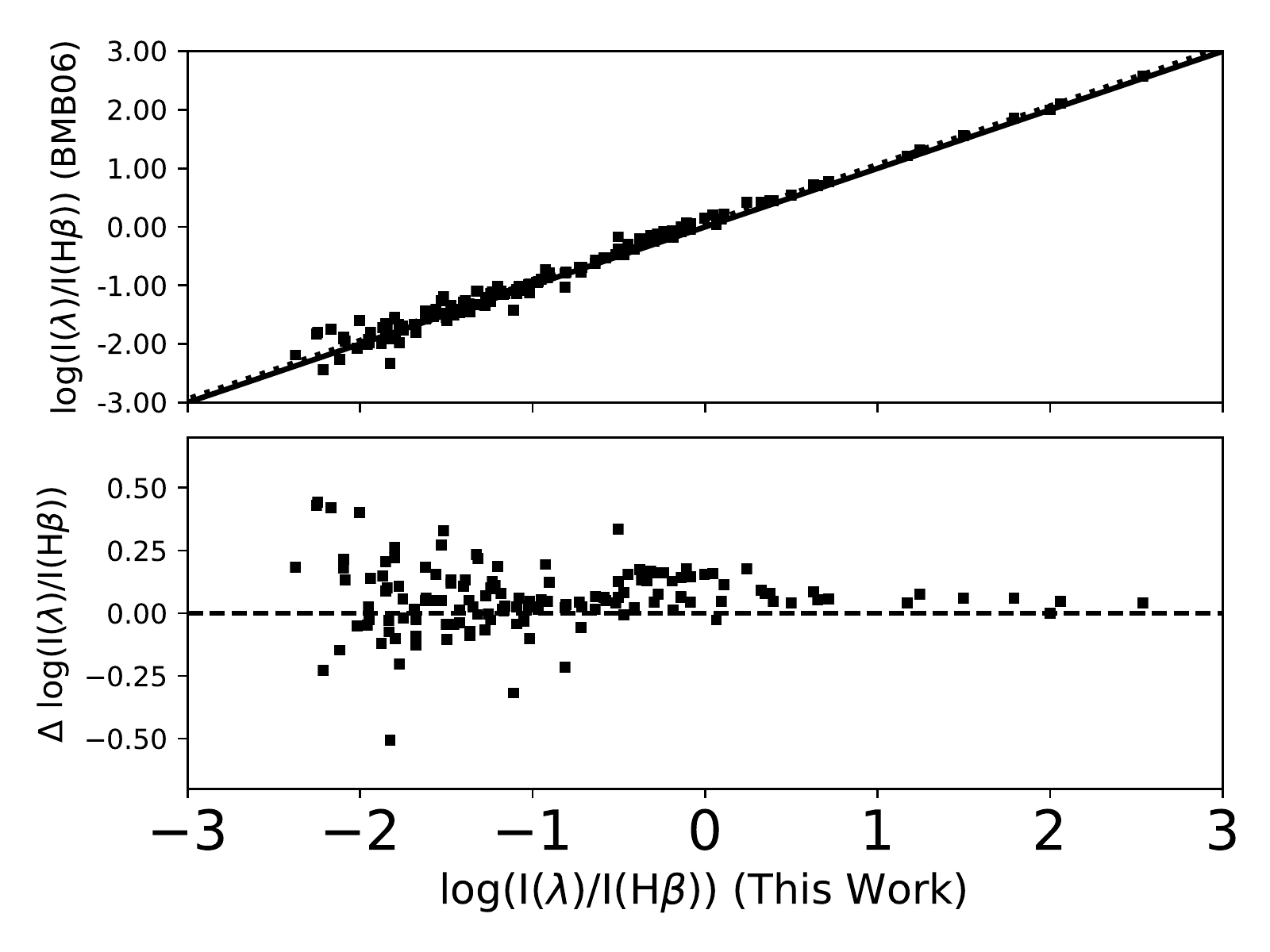}
\caption{\textit{Upper panel:} Comparison between the reddening corrected nebular spectrum (log(I($\lambda$)/I(H$\beta$))) from \citet{Blagrave06} and this work. The dotted line represents the linear fit $y = 1.00x + 0.05$, while the solid line represents $y = x$. \textit{Bottom panel:} Difference of the logarithm of line intensity ratios with respect to I(H$\beta$) in the spectrum of BMB06 and ours. }
\label{fig:compari}
\end{figure}

\citet{Blagrave06} (hereinafter BMB06) observed a zone of the Orion Nebula that includes HH~529~II+III using the 4m Blanco telescope at the Cerro Tololo Inter-American Observatory, covering the 3500-7500\AA$\text{ }$ spectral region. Fig.~\ref{fig:compari} shows a comparison between their reddening corrected nebular spectrum and ours (from cut 2). For the comparison, we have excluded lines flagged with notes of ``Avg'', ``blend'' or ``small FWHM'' in Table 1 of BMB06, due to their uncertain fluxes. For example, [Ne\thinspace III] $\lambda 3967.46$ line, marked with an ``Avg'', is inconsistent with the measured intensity of [Ne\thinspace III]$\lambda 3868.75$, since their observed ratio is 2.02, quite far from the theoretical one of 3.29 \citep{McLaughlin11}. A least squares linear fit of the data represented in Fig.~\ref{fig:compari} yields the relationship $y = 1.00\left( \pm 0.01 \right)\text{x} + 0.05\left(\pm 0.02\right)$, indicating that the BMB06's spectrum ($y$ values) presents systematically larger (by a factor of $\sim$ 1.12) line ratios (relative to H$\beta$) than ours ($x$ values). This is very noticeable in the spectral region of the high-level Balmer lines (3660-3720\AA), where this difference can reach up to 50\%. This may be due to the relative weakness of these lines, coupled with the abrupt change in the continuum level due to the closeness to the Balmer discontinuity. In Table~\ref{tab:comparison_balmer}, we compare our values of some selected reddening-corrected Balmer line ratios with those obtained by BMB06. The Balmer line ratios with respect to H$\beta$ obtained by BMB06 for both components differ significantly from the theoretical values. However, this does not seem to be the case when we use ratios of Balmer lines excluding H$\beta$. An underestimation of around 10\% in the flux of H$\beta$ in the BMB06's spectrum explains the systematic trend observed in Fig.~\ref{fig:compari}. We do not compare the high-velocity component of BMB06 with our data of HH~529~II and HH~529~III since their slit position and spatial coverage is slightly different than ours.

\begin{figure}
\includegraphics[width=\columnwidth]{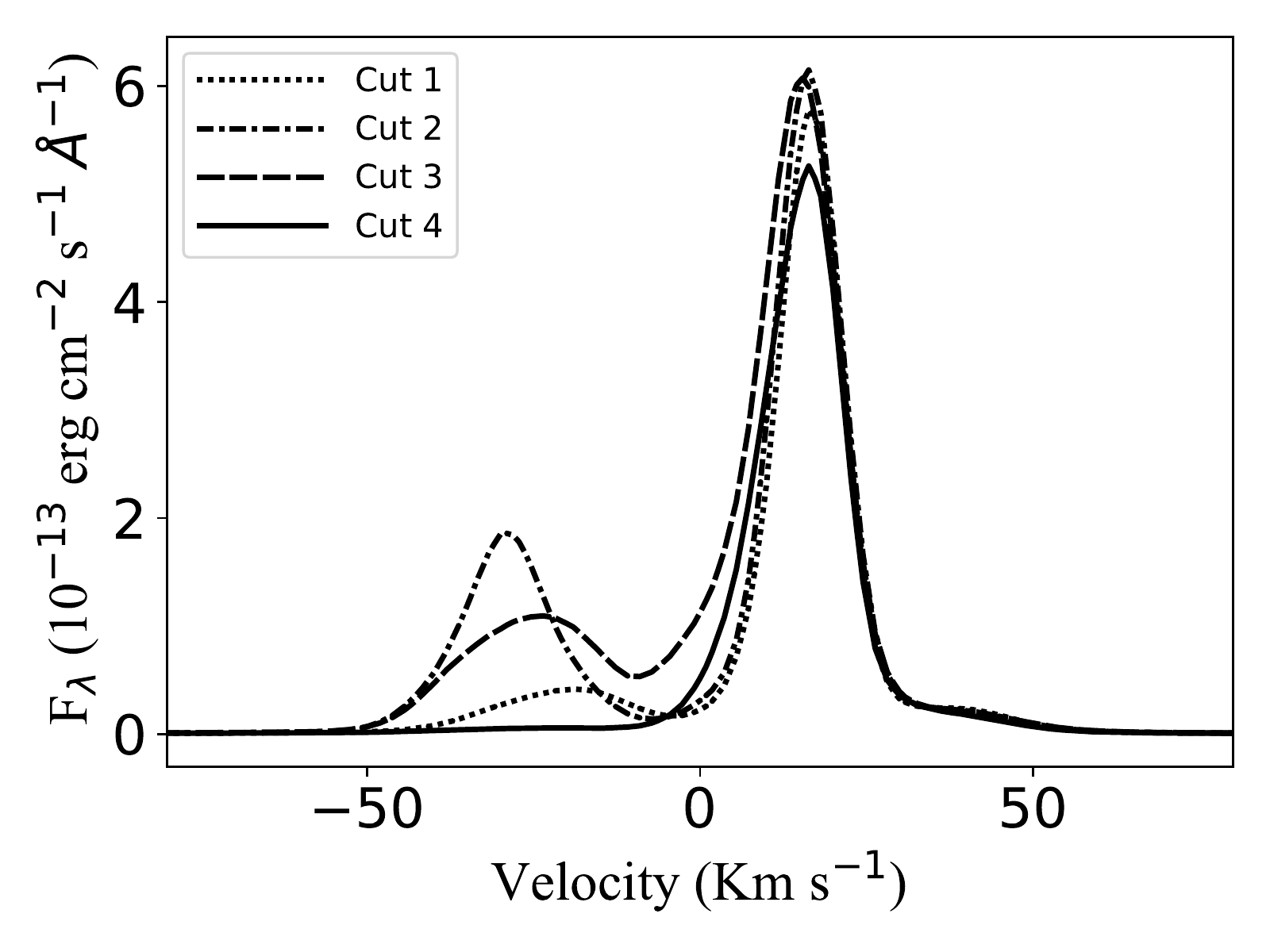}
\caption{Profile of [O\thinspace III] $\lambda 4958.91$ line in each of the analysed spatial cuts.
The velocity scale is heliocentric.}
\label{fig:4959_vel}
\end{figure}

Fig.~\ref{fig:4959_vel} shows the [O\thinspace III] $\lambda 4958.91$ line profile in the different cuts. As can be seen, the reddest component of each profile (corresponding to the nebular component) shows practically the same shape in all cuts except in cut 3, where the line is broadened by the presence of a larger velocity dispersion in the high-velocity component. The complexity of the velocity components of HH~529 is discussed in more detail in Section~\ref{sec:proper-motions-hh}.

Line identifications were consistently made by adopting the theoretical wavelengths of Peter Van Hoof's latest Atomic Line List v2.05b21 \footnote{\url{https://www.pa.uky.edu/~peter/newpage/}} \citep{vanhoof18} for all ions except for \mbox{Cl}\thinspace \mbox{III}, \mbox{Cl}\thinspace \mbox{IV} and \mbox{Ne}\thinspace \mbox{III} due to some inconsistencies found (see Section~\ref{sec:kin_analisys} for a detailed discussion). The number of lines we have identified in our spectra is very large. Line identifications and observed and dereddened flux line ratios  are presented in 4 online tables, one for each analysed cut. Tables of cut 2 and cut 3 contain, in addition to the observed nebular component, the spectra of HH~529~II and HH~529~III, respectively. These tables contain, for each measured line, the identified rest-frame wavelength ($\lambda_0$), the identified ion, the observed wavelength ($\lambda$), the radial velocity with respect to $\lambda_0$ ($v_r$), the full width at half maximum (FWHM), the observed flux relative to F(H$\beta$)=100 (F$\left( \lambda \right)$/F$\left( \mbox{H}\beta \right)$), the reddening corrected intensity relative to I(H$\beta$)=100 (I$\left( \lambda \right)$/I$\left( \mbox{H}\beta \right)$), the estimated error of the reddening corrected intensity and some notes. In the nebular components, 514, 633, 579 and 522 lines were measured in cuts 1, 2, 3 and 4, respectively. For HH~529~II and HH~529~III, 376 and 245 lines were detected, respectively. Multi-line blends were counted as single detections. As an example of our line tables, in Table~\ref{tab:sample_spectra} we show a sample of 15 lines of the spectra of cut 2.

\section{Physical conditions}
\label{sec:physical_conditions}

The estimation of the physical conditions and chemical abundances in the different components analyzed in this work are based on their photoionization equilibrium state. However, since HH~529~II and HH~529~III are produced by the interaction of high-velocity flows within the photoionized gas of the Orion Nebula, some contribution of the shock in the energy balance of the ionized gas may be expected. In Sec.~\ref{sec:dym_impl}, we demonstrate that the shock contribution in the observed optical spectra of the high-velocity components is very small and, therefore, the physical conditions and ionic abundances of these objects can be determined by means of the usual tools for analyzing ionized nebulae.

\begin{table*}
\centering
\caption{Physical conditions.}
\label{tab:pc}
\begin{adjustbox}{width=\textwidth}
\begin{tabular}{lcccccccc}
\hline 
 & \multicolumn{1}{c}{Cut 1} & \multicolumn{2}{c}{Cut 2} & \multicolumn{2}{c}{Cut 3} & \multicolumn{1}{c}{Cut 4} & \\
Diagnostic &  Nebula & HH~529~II & Nebula & HH~529~III & Nebula & Nebula &  Combined Cuts\\
\hline \noalign{\vskip3pt}
{\bf Density} & \multicolumn{7}{c}{\boldmath $n_{\rm e}$  {\bf (cm$^{-3}$)}}\\
\noalign{\vskip3pt}

[O\thinspace II] $\lambda$3726/$\lambda$3729 & $5460^{+1000} _{-750}$& $10570^{+3680} _{-2420}$& $5220^{+960} _{-720}$& $18020^{+17170} _{-6930}$ & $5530^{+1000} _{-800}$& $5070^{+880} _{-710}$& $5530^{+990} _{-810}$\\

[S\thinspace II] $\lambda$6731/$\lambda$6716 & $4230^{+1500} _{-980}$& $9390^{+10170} _{-3950}$&$4160^{+1570} _{-1040}$& $13130^{+15820} _{-6550}$ & $4130^{+2020} _{-1250}$& $4160^{+1400} _{-960}$& $4510^{+2270} _{-1330}$\\

[Cl\thinspace III] $\lambda$5538/$\lambda$5518 & $7020^{+960} _{-900}$& $8170^{+1810} _{-1610}$& $6670^{+920} _{-860}$& $15040^{+5620} _{-4490}$ & $7370^{+1190} _{-1120}$& $7000^{+960} _{-900}$& $7420^{+1170} _{-1100}$\\

[Fe\thinspace III] $\lambda$4658/$\lambda$4702 & $9260^{+3700} _{-2890}$& $12390^{+5010} _{-3460}$& $8990^{+2840} _{-2350}$& $33800^{+13820} _{-10530}$& $10490^{+4240} _{-3090}$& $8340^{+3620} _{-2530}$& $9510^{+3790} _{-2750}$\\

$\mbox{[Ar}\thinspace \mbox{IV]}$ $\lambda$4740/$\lambda$4711 & $4480^{+1700} _{-1640}$& $6410^{+1900} _{-1880}$& $5920^{+980} _{-940}$&$15050^{+13300} _{-9240}$& $6400^{+1690} _{-1660}$& $5460^{+1000} _{-1050}$& $6580^{+2000} _{-1870}$\\

$\mbox{O}\thinspace \mbox{II}^{*}$ &$4710 \pm 710$ & $3490 \pm 340$ & $4390 \pm 400$&$3600 \pm 850$ & $4920 \pm 550$&$4350 \pm 610$&$5420 \pm 690$\\

$\mbox{[Fe}\thinspace \mbox{III]}^{*}$ & $8530 \pm 1050$ & $11880 \pm 1860$ & $9430 \pm 1010$ & $30200 \pm 8080$ &$10330 \pm 1700$ & $9020 \pm 1170$ & $10360 \pm 1410$ \\

\textbf{Adopted} &   \boldmath${ 5830 \pm 1210 }$ &   \boldmath${ 11880 \pm 1860 }$& \boldmath${ 5870 \pm 970}$  &   \boldmath${30200 \pm 8080 }$   &   \boldmath${6180 \pm 1220 } $ &   \boldmath${5650 \pm 1030 } $&   \boldmath${6290 \pm 1130 } $\\

\noalign{\vskip3pt}
{\bf Temperature} & \multicolumn{7}{c}{\boldmath $T_{\rm e}$ \bf (K)}\\
\noalign{\vskip3pt}

T$\left(\mbox{H}\thinspace \mbox{I} \right)_{\text{Balmer}}$ &-&-&-&-&-&-&$7520 \pm 790$\\ 
T$\left(\mbox{H}\thinspace \mbox{I} \right)_{\text{Paschen}}$ &-&-&-&-&-&-&$7550 \pm 1160$\\ 

T$\left(\mbox{He}\thinspace \mbox{I} \right)$ &$8280 ^{+520} _{-570}$&$7200 \pm 550$& $8060 ^{+540} _{-510}$&  $7340 \pm 710$  &$8090 \pm 530$&$7390 ^{+470} _{-580}$&$7690 ^{+500} _{-510}$\\

[N\thinspace II] $\lambda$5755/$\lambda$6584 & $9910 \pm 250$ & $10150^{+570} _{-510}$ & $9850\pm 240$ & $11040^{+920} _{-970}$  & $10060^{+260} _{-280}$& $9860^{+250} _{-270}$& $9990^{+250} _{-270}$\\

[O\thinspace II] $\lambda \lambda$ 3726+29/$\lambda \lambda$7319+20+30+31 &$10340^{+1330} _{-940}$&-&-&-&-& $11230^{+1330} _{-1110}$& $10910^{+1320} _{-950}$\\

[S\thinspace II] $\lambda \lambda$4069+76/$\lambda \lambda$ 6716+31&$11430^{+3290} _{-1870}$&-& $11070^{+2420} _{-1450}$&-& $10550^{+2450} _{-1570}$& $10790^{+2340} _{-1470}$& $11000^{+2510} _{-1600}$\\

[O\thinspace III] $\lambda$4363/$\lambda \lambda$4959+5007 & $8430 \pm 90$ & $8240 \pm 80$& $8410^{+80} _{-90}$& $8600^{+110} _{-120}$ & $8510 \pm 90$& $8320\pm 90$& $8450^{+80} _{-90}$\\

$\mbox{[S}\thinspace \mbox{III]}$ $\lambda$6312/$\lambda \lambda$9069+9531 & $9220^{+290} _{-330}$ & $8670\pm 310$ & $8990^{+290} _{-330}$&  $9040^{+410} _{-420}$& $8920^{+330} _{-300}$ & $8850^{+290} _{-320}$& $8970^{+280} _{-290}$\\

$\mbox{[Ar}\thinspace \mbox{III]}$ $\lambda$5192/$\lambda \lambda$7136+7751  & $8280^{+280} _{-310}$ & $8620^{+500} _{-540}$ & $8390^{+220} _{-280}$&-& $8250^{+260} _{-290}$& $8280^{+380} _{-420}$& $8270^{+280} _{-320}$\\

$\mbox{O}\thinspace \mbox{II}^{*}$ & - & - & $9350 \pm 1090$ & - & - & - &-\\

$\mbox{[Fe}\thinspace \mbox{III]}^{*}$ & $7800 \pm 800$ & $8500 \pm 1050$ & $8450 \pm 730$ & $7900 \pm 1910$ &$7970 \pm 920$ & $7350 \pm 590$ & $8440 \pm 710$  \\

Thermal broadening  & - & $8670 \pm 50$ & $8340 \pm 410$ & 10470: & -&-&-\\

\textbf{\boldmath${T_e}$ (low) Adopted} &\boldmath${ 9930 \pm  140 }$ & \boldmath${ 10150^{+570} _{-510} }$& \boldmath${ 9860 \pm  240 }$ & \boldmath${11040^{+920} _{-970}}$&\boldmath${10070 \pm 270}$&\boldmath${9920 \pm 280}$&\boldmath${10040 \pm 210}$\\

\textbf{\boldmath${T_e}$ (high) Adopted} &\boldmath${8470 \pm 200 }$&\boldmath${8270 \pm 110 }$&\boldmath${8440 \pm 140 }$&\boldmath${ 8630 \pm 120 }$&\boldmath${8510 \pm 120 }$&\boldmath${8360 \pm 140 }$&\boldmath${8480 \pm 150 }$\\

\hline
\end{tabular}
\end{adjustbox}
\begin{description}
\item $^*$ A maximum likelihood method was used. \\
\end{description}
\end{table*}

\subsection{Physical conditions based on CELs}
\label{subsec:physical_conditions_cels}

We use PyNeb (version 1.1.10) \citep{Luridiana15} and the updated atomic dataset listed in Table~\ref{tab:atomic_data} to calculate physical conditions based on the intensity ratios of CELs from different ions. 

The first step was to test all the intensity ratios of CELs that can serve as a temperature or density diagnostic using the PyNeb task {\it getCrossTemDen}. This task uses two line ratios at the same time: one as density diagnostic and the other one for temperature, giving their convergence to a pair $n_{\rm e},\text{}T_{\rm e}$ as a result. We tried all possible permutations for all the available diagnostics in all components from all cuts. We only discarded the use of lines strongly affected by blends, telluric emissions and/or absorptions or reflections in the optical system of the spectrograph. We did not consider the $n_{\rm e}$ diagnostic based on $ \mbox{ [N}\thinspace \mbox{I] }\lambda \lambda 5198/5200$ owing to a significant fluorescent contribution in the Orion Nebula \citep{Ferland12}.  

Diagnostics based on [Ni\thinspace III] $\lambda_1/\lambda_2$, where $\lambda_1, \lambda_2 \in [6000,6534,6682,6797,6946,7890]$ do not give any useful physical information since they either did not converge or showed convergences at values highly discordant with the other diagnostics. This will be discussed in Section~\ref{sec:ni2_ab_comment}.  Another interesting diagnostics are based on \mbox{[Fe}\thinspace \mbox{III] }$\lambda_1/\lambda_2$, where $\lambda_1, \lambda_2 \in [4658,4702,4734,4881,5011,4925,4987,5271]$. With the exception of \mbox{[Fe}\thinspace \mbox{III] }4658/4702, all the diagnostics converge in a fairly wide range of physical conditions. This is due to the ambivalence and/or high dependence of these ratios on both density and temperature. This will be discussed in Section~\ref{subsec:iron_conditions}.

After the initial exploration, we define the ratios we consider good indicators of electron density and temperature. Then we use Monte Carlo simulations with 1000 points to estimate uncertainties in the physical conditions given by the {\it getCrossTemDen} task of PyNeb. For example, using $\mbox{[O}\thinspace \mbox{III]}  \lambda 4363/\lambda \lambda$4959+5007 as a temperature indicator and the following density diagnostics: $\mbox{[Cl}\thinspace \mbox{III]} \lambda5538/\lambda5518$, $\mbox{[Fe}\thinspace \mbox{III]} \lambda4658/\lambda4702$, $\mbox{[O}\thinspace \mbox{II]}\lambda3726/\lambda3729 $, $\mbox{[S}\thinspace \mbox{II]} \lambda6731/\lambda6716 $ and $\mbox{[Ar}\thinspace \mbox{IV]} \lambda4740/\lambda4711$, we estimate the convergence in $T_{\rm e}$ and $n_{\rm e}$ and their uncertainties in every case. Analogously, we use the rest of $T_{\rm e}$-diagnostics. The central value of $T_{\rm e}$ or $n_{\rm e}$ corresponds to the median of the Monte Carlo distribution and the errors are represented by the deviations to 84th and 16th percentiles, corresponding to $\pm 1 \sigma$ in the case of a Gaussian. After this procedure, all diagnostics (either $T_{\rm e}$ or $n_{\rm e}$), will have a result for each cross-comparison. 

For the nebular components on each cut, we define the representative $n_{\rm e}$ as the weighted mean\footnote{The weights were defined as the inverse of the square of the error associated to each density diagnostic.} in each cross-comparison with all the the temperature indicators. In the case of high-velocity components, the treatment is more complex since all the density diagnostics based on CELs reveal considerably higher densities than in the nebular components, reaching values at or above the critical densities of the atomic levels involved in some diagnostics as shown in Table~\ref{tab:critical_densities}. At densities of $10^4-10^6 \text{ cm}^{-3}$, diagnostics based on [Fe \thinspace III] lines are more reliable than other classic ones such as [O\thinspace II] $\lambda$3726/$\lambda$3729 or [S\thinspace II] $\lambda$6731/$\lambda$6716. In addition, dust destruction processes release gaseous Fe in the shock front that should favor the larger contribution of the emission of [Fe \thinspace III] lines of the post-shock gas and, therefore, the derived physical conditions would be biased to those of the post-shock zones. We adopted a maximum-likelihood method to determine the density from [Fe\thinspace III] lines for the high-velocity components. This procedure and its interpretation is described in detail in Section~\ref{subsec:iron_conditions}.

Finally, using the adopted representative $n_{\rm e}$, we calculate $T_{\rm e}$ with the available diagnostics using the \textit{getTemDen} task of PyNeb. Assuming the scheme of two ionization zones, we define \textbf{\boldmath${T_{\rm e}}$(high) } as the weighted mean $T_{\rm e}$ obtained from  $\mbox{[Ar}\thinspace \mbox{III]} \lambda 5192/ \lambda \lambda$7136+7751, $\mbox{[O}\thinspace \mbox{III]} \lambda4363/\lambda \lambda$4959+5007 and $\mbox{[S}\thinspace \mbox{III]} \lambda6312/\lambda \lambda$9069+9531 line ratios. Similarly, we define \textbf{\boldmath${T_{\rm e}}$(low) } based on the resulting $T_{\rm e}$ obtained from $\mbox{[S}\thinspace \mbox{II]} \lambda \lambda$4069+76/$\lambda \lambda$ 6716+31, $\mbox{[N}\thinspace \mbox{II]}\lambda5755/\lambda6584$ and $\mbox{[O}\thinspace \mbox{II]} \lambda \lambda$3726+29/$\lambda \lambda$7319+20+30+31 line ratios.

We note that in the nebular component of all cuts the observed \mbox{[S}\thinspace \mbox{III]} $\lambda9531/\lambda9069$ line intensity ratio does not agree with the theoretical value. This is owing to strong telluric absorptions that affect the \mbox{[S}\thinspace \mbox{III]} $\lambda 9069$ line that, on the other hand, do not affect the blueshifted lines of the high velocity components. After an inspection in the 2D spectra of the calibration star and in the science object, we concluded that \mbox{[S}\thinspace \mbox{III]} $\lambda 9531$ is not affected by telluric absorptions or emissions at the earth velocities at which the observations were taken. In the nebular component of all cuts, we assumed the theoretical ratio $I$(\mbox{[S}\thinspace \mbox{III]} 9531)/$I$(\mbox{[S}\thinspace \mbox{III]} 9069) = 2.47 obtained from the atomic data given in Table~\ref{tab:atomic_data} to estimate $T_{\rm e}$(\mbox{[S}\thinspace \mbox{III]}).

Plasma diagnostic plots shown in Fig.~\ref{fig:plasma}, indicate that the resulting values of each diagnostic are consistent with each other. The numerical values in each case are presented in Table~\ref{tab:pc}. 

\subsection{Physical conditions based on [Fe~III] lines.}
\label{subsec:iron_conditions}

\begin{figure}
\includegraphics[width=\columnwidth]{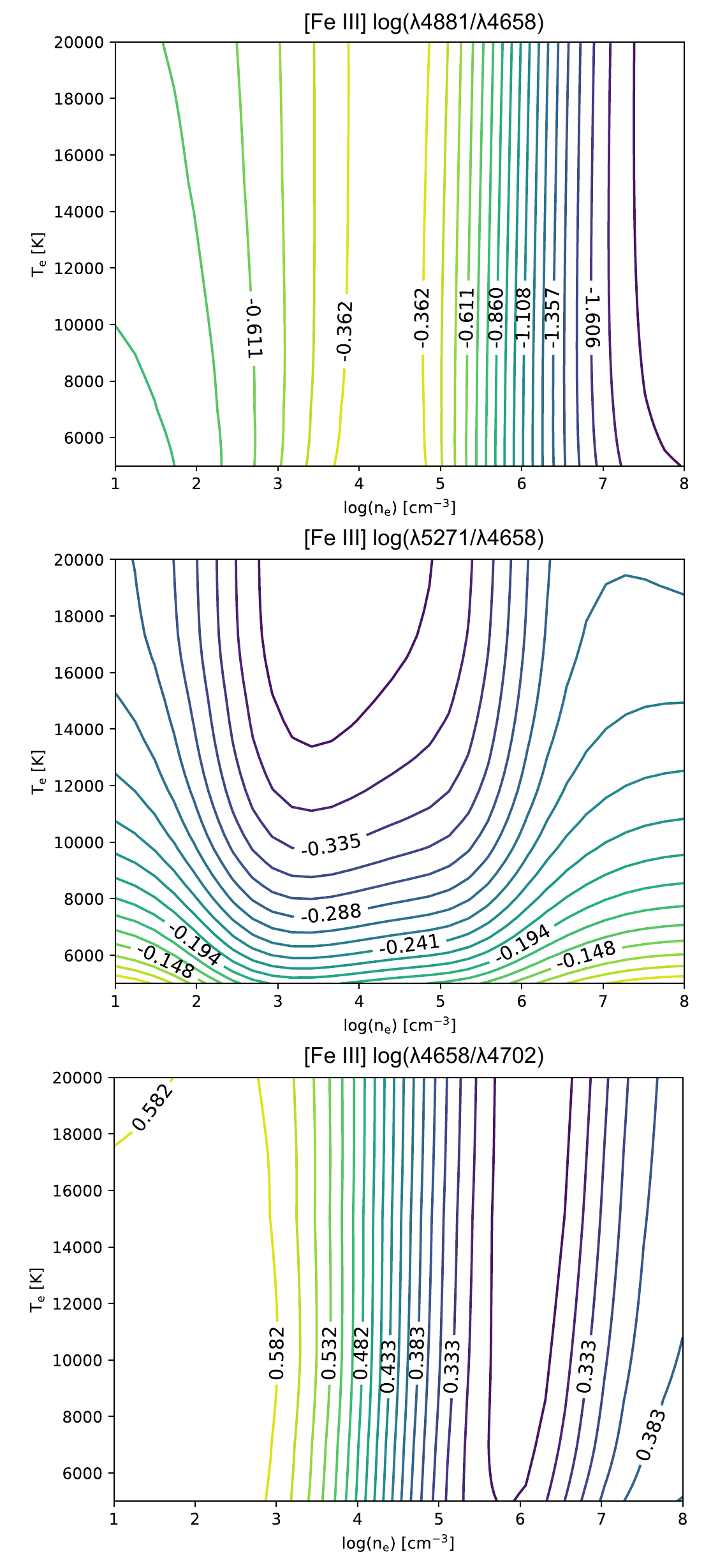}
\caption{Predicted dependence of the [Fe\thinspace III] $\lambda4881/\lambda4658$, $\lambda5271/\lambda4658$ and $\lambda4658/\lambda4702$
line intensity ratios with physical conditions.}
\label{fig:predicted_ratios_feiii}
\end{figure}

As mentioned in Section~\ref{subsec:physical_conditions_cels}, density diagnostics based on different [Fe\thinspace III] line intensity ratios give apparently discordant results. This is mainly due to the ambivalence in the density dependence of some observed intensity ratios and/or due to their high dependence on $T_{\rm e}$ as well as on $n_{\rm e}$. These two scenarios are exemplified in Fig.~\ref{fig:predicted_ratios_feiii} for [Fe\thinspace III] $\lambda4881/\lambda4658$ and $\lambda5271/\lambda4658$ line ratios, upper and middle panels, respectively. $\lambda 4881/\lambda 4658$ has a broad maximum around $n_{\rm e} \sim 2\times 10^4 \text{ cm}^{-3}$, so only becomes an accurate density diagnostic for $n_{\rm e}<5\times 10^3 \text{ cm}^{-3}$ or $n_{\rm e}>10^5\text{ cm}^{-3}$. In the case of $\lambda 5271/\lambda 4658$, the $T_{\rm e}$ dependence is always important except for some narrow density ranges between $\sim$10$^2$ cm$^{-3}$ and $\sim$10$^3$ cm $^{-3}$ and between $\sim$10$^5$ cm$^{-3}$ and $\sim$10$^6$ cm $^{-3}$. For the expected densities in the different components observed in this work ($n_{\rm e}$ between $\sim$10$^3$ cm$^{-3}$ and $\sim$10$^5$ cm $^{-3}$), these diagnostics are not very enlightening on their own. On the other hand, for $10^3 \text{ cm}^{-3}<n_{\rm e}<10^6 \text{ cm}^{-3}$, $\lambda 4658/\lambda 4702$ (see bottom panel of Fig.~\ref{fig:predicted_ratios_feiii}) varies monotonically with $n_{\rm e}$ and is insensitive to $T_{\rm e}$. Thus, it is the most reliable diagnostic in our case.

We consider that the option to determine the physical conditions based on the observed intensity ratios of  [Fe\thinspace III] lines is using a maximum-likelihood process. This method is based on a $\chi^{2}$ minimization by testing a wide range of parameters. The value of $\chi^{2}$ is defined in Eq.~\eqref{eq:chi2}, as the sum of the quadratic differences between the abundance of ion X$^i$ (in this case Fe$^{2+}$) determined with each emission line included in the procedure and the weighted average of the abundance defined in Eq.~\eqref{eq:average}. 

\begin{equation}
    \label{eq:chi2}
    \chi^2=\sum_{\lambda} \frac{\left(n\left( \frac{\text{X}^{i}}{\text{H}^{+}} \right)_{\lambda}-\overline{n\left( \frac{\text{X}^{i}}{\text{H}^{+}} \right)} \right)^2}{  \Delta n\left( \frac{\text{X}^{i}}{\text{H}^{+}} \right)_{\lambda} ^{2} },
\end{equation}

\begin{equation}
    \label{eq:average}
    \overline{n\left( \frac{\text{X}^{i}}{\text{H}^{+}} \right)}= \frac{ \sum_{\lambda} \left( n\left( \frac{\text{X}^{i}}{\text{H}^{+}} \right)_{\lambda}/ \Delta n\left( \frac{\text{X}^{i}}{\text{H}^{+}} \right)_{\lambda}^2  \right)   }{ \sum_{\lambda} \left( 1/ \Delta n\left( \frac{\text{X}^{i}}{\text{H}^{+}} \right)_{\lambda}^2  \right) }.
\end{equation}

This self-consistent procedure gives the physical parameters that minimize $\chi^2$ with an associated uncertainty based on the resulting values within $\chi^2-\chi^2_{\text{min}}\leq 1$. This method requires a strict control on the variables that affect the line fluxes, otherwise a spurious contribution appears, and can change the resulting parameters that minimize $\chi^2$. For example, undetected blends in the studied lines can result in incorrect density and/or temperature determinations.

We have considered several aspects to choose the set of [Fe\thinspace III] lines that should be included in the maximum-likelihood process. We discard lines with evident line blending or contamination by telluric emission or ghosts. To test unnoticed line blends or inaccuracies in flux estimations, we use ratios of observed lines that should depend only on transition probabilities and not on physical conditions. The results are shown in Table~\ref{tab:fe3_ratios_theo}. As can be seen, there are some deviations between the theoretical and the observed values in the cases of [Fe\thinspace III] $\lambda 4667/\lambda 4734$, $\lambda 4778/\lambda 4667$, $\lambda 4607/\lambda 4702$, $\lambda 4607/\lambda 4770$ and $\lambda 4881/\lambda 4987$ due to the contamination of [Fe\thinspace III] $\lambda 4667$ by a ghost, by the blend of [Fe\thinspace III] $\lambda 4607$ with N\thinspace II $\lambda 4607.15$ and by the blend of [Fe\thinspace III] $\lambda 4987.29$ with N\thinspace II $\lambda 4987.38$. On the other hand, we detect that $\lambda 4778$ line is 35\% wider than the rest of the [Fe\thinspace III] bright lines in HH~529~II, although, this is not observed in HH~529~III. This suggests that, due to the higher signal to noise ratio in the cut 2 spectrum, a line blend with an emission feature is detected in HH~529~II while for HH~529~III it remains below the detection level. The line ratio with the largest deviation is [Fe\thinspace III] $\lambda 5011/\lambda 4085$. This could be mainly due to the low signal-to-noise ratio of the [Fe\thinspace III] $\lambda 4085$ line. However, [Fe\thinspace III] $\lambda 5011$ is located close to [O\thinspace III] $\lambda 5007$, which presents broad wings in its line profile that affects the shape of the continuum close to  [Fe\thinspace III] $\lambda 5011$ and perhaps the measurement of its line flux. The presence of bright lines affecting the continuum shape in areas close to relatively weak [Fe\thinspace III] lines may also contribute to some differences between the observed and predicted line ratios shown in Table~\ref{tab:fe3_ratios_theo}. This problem is reduced by using the brightest line of each ratio in the maximum-likelihood process.

\begin{figure*}
    \begin{subfigure}{6cm}
    \centering\includegraphics[height=5cm,width=\columnwidth]{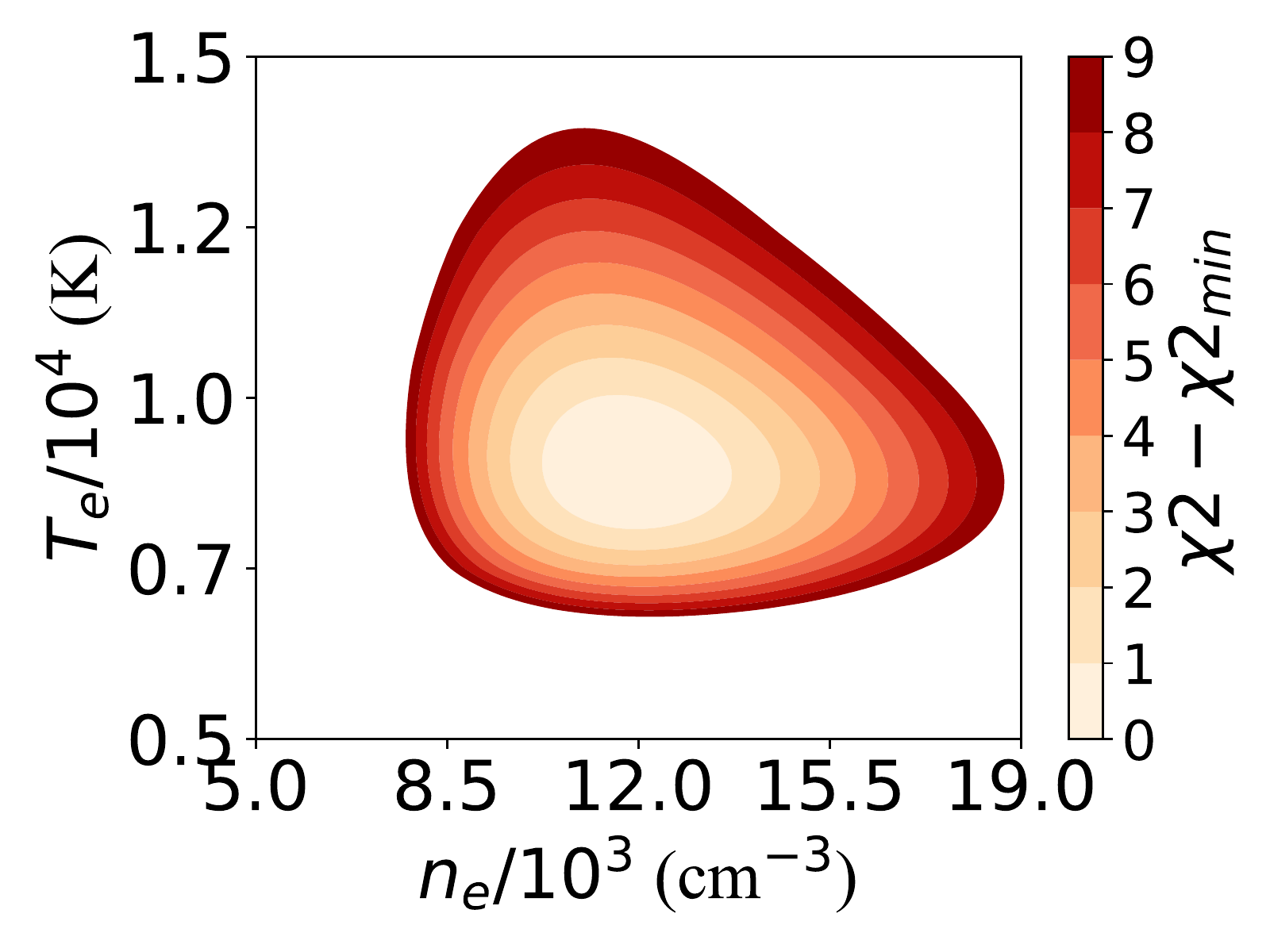}
    \caption{HH~529~II}
    \label{fig:fe3_physical_cond_cut2_velyneb_a}
  \end{subfigure}
  \begin{subfigure}{6cm}
    \centering\includegraphics[height=5cm,width=\columnwidth]{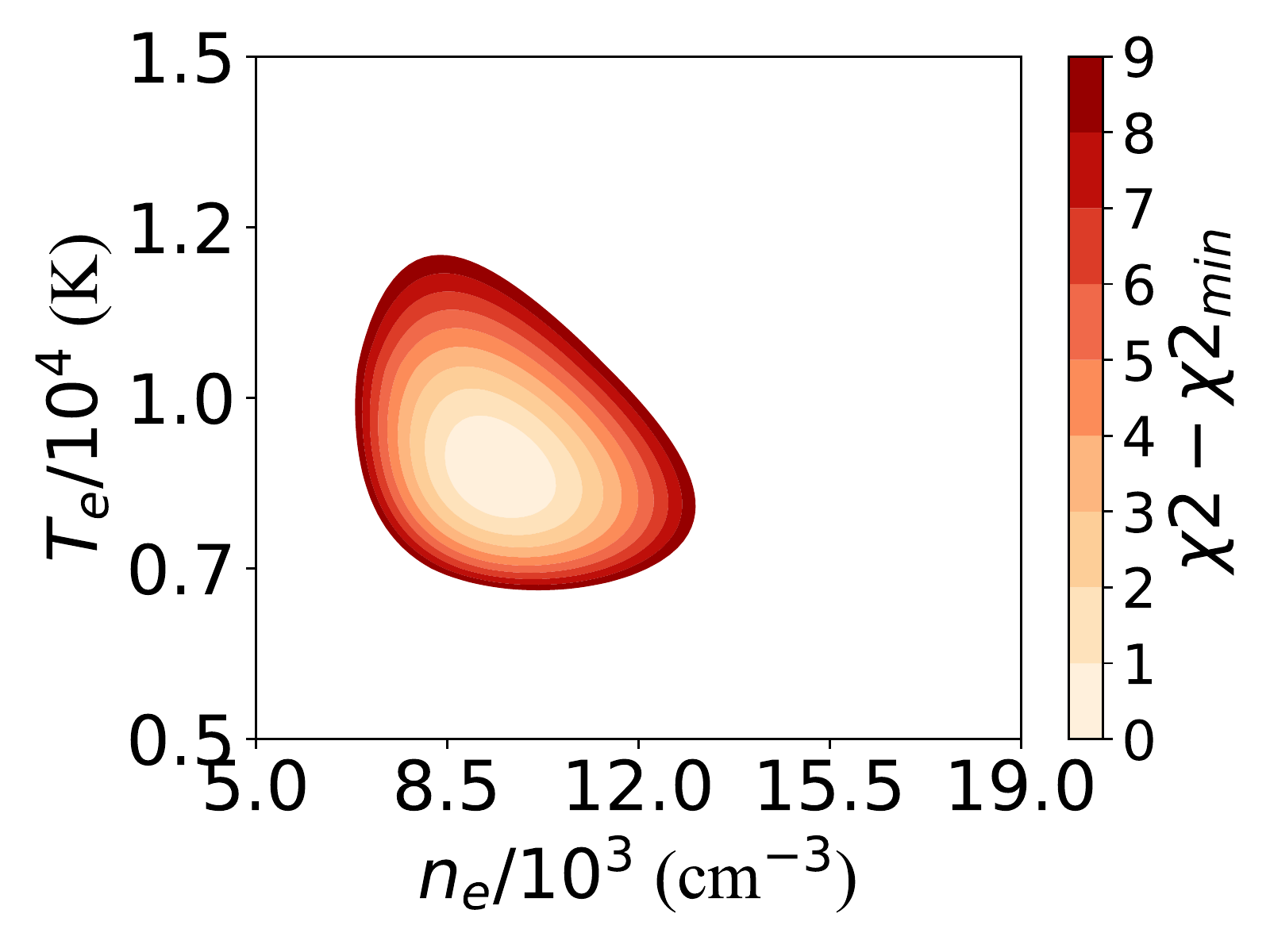}
    \caption{Nebular component}
    \label{fig:fe3_physical_cond_cut2_velyneb_b}
  \end{subfigure}
 \caption{$\chi^2$ in the space of $T_{\rm e}$ and $n_{\rm e}$ in the maximum-likelihood procedure for [Fe\thinspace III] lines. Both in the case of HH~529~II (left panel) and the nebular component of cut 2 (right panel) there is a convergence to well defined physical conditions.}
\label{fig:fe3_physical_cond_cut2_velyneb}

\end{figure*}

We select the following [Fe\thinspace III] lines for the maximum-likelihood process: $\lambda\lambda$4658.17, 4701.64, 4734.00, 4881.07 and 5270.57. This selection includes the brightest [Fe\thinspace III] lines that are free of blends or telluric emissions and/or absorptions.  Moreover, these lines lie in a relatively small spectral range and hence, uncertainties in the reddening correction would have a negligible effect. This allows us to restrict the parameter space to electron density and temperature to test  $\chi^2$. Studies of the primordial helium abundance have used similar maximum-likelihood procedures to calculate the He$^{+}$ abundance and have found that this procedure can lead to degeneracies in the fitted parameters and $\chi^2$  \citep[see][and references therein]{olive04,Aver11}. Because of this, it is important to have an overview of the behavior of $\chi^2$ in the complete  parameter space. In Fig.~\ref{fig:fe3_physical_cond_cut2_velyneb} we present the convergence of $\chi^2$ in the $n_{\rm e} - T_{\rm e}$ space for both high-velocity and nebular components of cut 2. As it can be seen, $\chi^2$ falls into a single minimum in each case, corresponding to $T_{\rm e}=8500 \pm 1050$ K and $n_{\rm e}=11880 \pm 1860$ $\text{cm}^{-3}$ for HH~529~II and $T_{\rm e}=8450 \pm 730$ K and $n_{\rm e}=9430 \pm 1010$ $\text{cm}^{-3}$ for the nebular component. The $T_{\rm e}$ and $n_{\rm e}$ values obtained for the rest of cuts using this approach are presented in Table~\ref{tab:pc}. The convergence to the resulting $n_{\rm e}$ is consistent with the diagnostic based on [Fe\thinspace III] $\lambda 4658/\lambda 4702$ ratio but with a smaller uncertainty due to the application of the $\chi^2$ maximum-likelihood procedure. 
It is notable that in all cases, [Fe\thinspace III] lines give $n_{\rm e}$ values higher than the usual diagnostics based on CELs such as [S\thinspace II] $\lambda 6731/ \lambda6716 $ or [O\thinspace II] $\lambda 3726/ \lambda3729 $. The largest difference is found in the high-velocity components, in particular in HH~529~III. In the case of nebular components, the low dependence on density of some ratios such as $\lambda 4734/\lambda 4658$, $\lambda 4658/\lambda 4702$ or $\lambda 4734/\lambda 4702$ at density values smaller than $\sim10^3$ $\text{cm}^{-3}$, gives more weight to the higher-density zones within the line of sight. On the other hand, in the high-velocity components, the larger differences suggest the presence of high densities in the range of $10^4-10^5 \text{ cm}^{-3}$, where the usual density diagnostics, such as [S\thinspace II] $\lambda 6731/ \lambda6716 $ or [O\thinspace II] $\lambda 3726/ \lambda3729 $, are uncertain, being well above the critical densities, as is shown in Table~\ref{tab:critical_densities}. In addition, as found in HH~202 \citep{mesadelgado09,espiritu17} and in this work (see Section~\ref{subsubsec:total_abun_fe}), the gaseous Fe abundance is higher in the high-velocity components due to the dust destruction where the Fe is commonly depleted, thus, the flux of [Fe\thinspace III] lines increase in the shock front, where the gas is being compressed.  Therefore, the $n_{\rm e}$ determinations based on [Fe\thinspace III] lines will be biased to the higher values of the density at the head of the shock although the volume of gas integrated in the high-velocity components include not only the denser gas of the head but also some contribution of the jet beam gas behind since it is fully photoionized and flowing towards the observer (see Section~\ref{sec:dym_impl}).

The results indicate a closer similarity between $T_{\rm e}$([Fe\thinspace III]) and $T_{\rm e}$(high), contrary to what the ionization potential of Fe$^{2+}$ would suggest, closer to N$^+$ than to O$^{2+}$, which are ions representative of the low and high ionization zones, respectively. In the case of the nebular components, the fact that the [Fe\thinspace III] density diagnostics give more weight to the high-density zones in the line of sight, as we mentioned previously, may bias the results towards lower temperatures, which are not representative for all the Fe$^{2+}$. On the other hand, in the high-velocity components, this indicates that in the shock front, where further dust destruction and incorporation of Fe into the gas phase is expected, the high-ionization gas dominates over the remaining low-ionization one, which may be flowing behind of the shock front. This suggests that the optimal temperature to calculate the Fe$^{2+}$ abundance in the high-velocity components is $T_{\rm e}$(high). Estimates of Fe$^{2+}$ abundances based on both $T_{\rm e}$(low) and $T_{\rm e}$(high) will be discussed separately in Section~\ref{subsubsec:total_abun_fe}.

\subsection{Physical conditions based on RLs.}
\label{subsec:physical_conditions_rls}

\subsubsection{Physical conditions based on \mbox{O}\thinspace \mbox{II} RLs}
\label{subsubsec:oii_pc}

\begin{figure}
\includegraphics[width=\columnwidth]{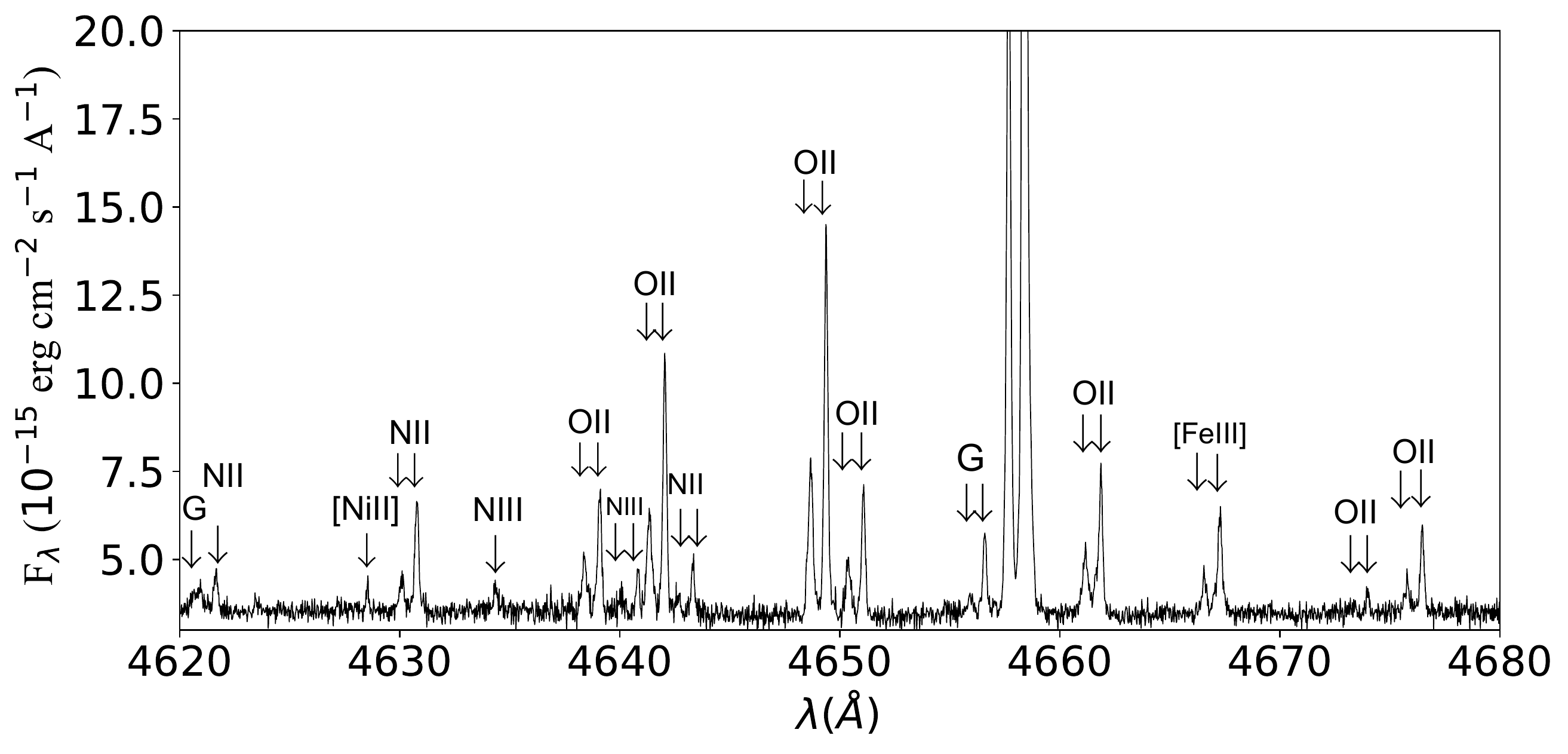}
\caption{Section of the spectrum of the spatial cut 2 covering the spectral range 4620-4680\AA. Several couples of O\thinspace II lines from multiplet 1 are present, showing the quality of the spectra of HH~529~II and the nebular component. Emissions marked with a G are ghosts (see Section~\ref{sec:siiv_coment}). }
\label{fig:oii_lines}
\end{figure}

To estimate physical conditions based on O\thinspace II RLs, we use the effective recombination coefficients from \citet{Storey17}. These coefficients fully account the dependence on electron density and temperature of the population distribution among the ground levels of O\thinspace II. We follow a similar maximum-likelihood procedure as described in Section~\ref{subsec:iron_conditions} to derive the  physical conditions. For this case, we chose the observed lines from multiplet 1 and $\lambda \lambda 4089.29, 4275.55$ from 3d-4f transitions, due to the following reasons: (1) lines from multiplet 1 are the brightest O\thinspace II RLs and are comparatively less affected by line blending or instrumental reflections as is illustrated in Fig.~\ref{fig:oii_lines} for cut 2. (2) The line ratios within multiplet 1 deviate from the local thermodynamic equilibrium (LTE) values for $n_{\rm e}\leq 10^5 \text{ cm}^{-3}$ \citep{Storey17}, providing a density diagnostic. (3) O\thinspace II $\lambda \lambda 4089.29, 4275.55 $ RLs corresponding to 3d-4f transitions depend slightly stronger on $T_{\rm e}$ than the lines from multiplet 1 and their ratio with O\thinspace II $\lambda 4649.13$ is practically insensitive to $n_{\rm e}$, since the population of the levels that arise these lines depend on the population of the same $^{3}P_2$ ground level \citep{Storey17}, giving a $T_{\rm e}$ diagnostic. Nevertheless, $\lambda \lambda 4089.29, 4275.55 $ are relatively weak and we expect comparatively larger uncertainties in the $T_{\rm e}$ determinations than using diagnostics based on CELs. 

\begin{figure*}
 
    \begin{subfigure}{6cm}
    \centering\includegraphics[height=5cm,width=\columnwidth]{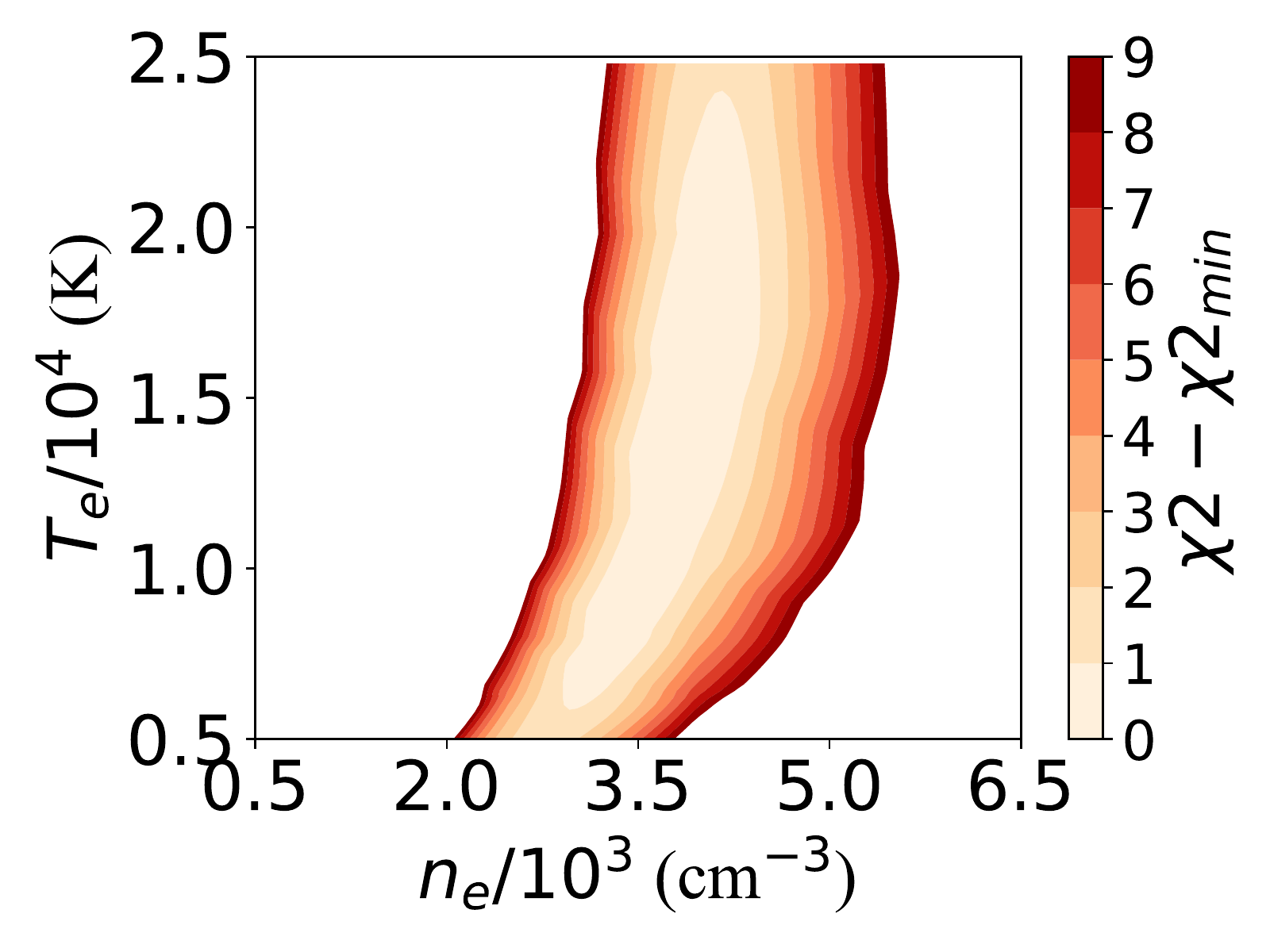}
    \caption{HH~529~II}
    \label{fig:OII_physical_cond_cut2_velyneb_a}
  \end{subfigure}
  \begin{subfigure}{6cm}
    \centering\includegraphics[height=5cm,width=\columnwidth]{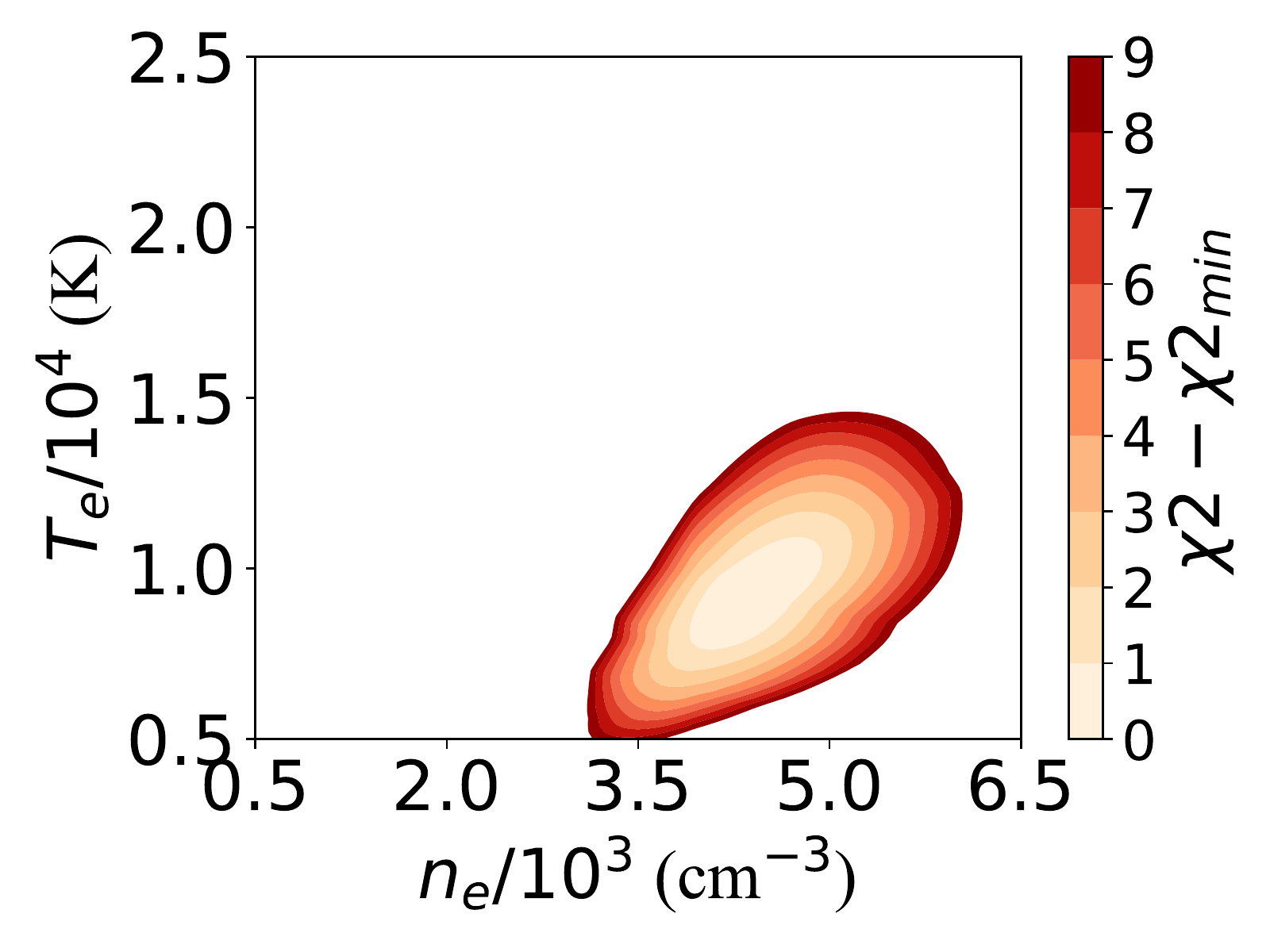}
    \caption{Nebular component}
    \label{fig:OII_physical_cond_cut2_velyneb_b}
  \end{subfigure}
 \caption{The same as in Fig.~\ref{fig:fe3_physical_cond_cut2_velyneb} but using O\thinspace II lines. In this case, there is an evident  $T_{\rm e}$ degeneracy in HH~529~II (left panel) because that the temperature-dependent O\thinspace II RL $\lambda4089.29$ cannot be measured in its spectrum. A clear convergence can be seen in the nebular component (right panel). The uncertainties are somewhat large owing to the relative weakness of the temperature-dependent O\thinspace II lines ($\lambda \lambda 4089.23,  4275.55$).}
\label{fig:OII_physical_cond_cut2_velyneb}

\end{figure*}

Fig.~\ref{fig:OII_physical_cond_cut2_velyneb} shows $\chi^2$-maps in the space of $T_{\rm e}$ and $n_{\rm e}$ for both components of cut 2. As it can be seen, HH~529~II has a temperature degeneracy. This is not surprising, due to the fact that multiplet 1 is rather independent of $T_{\rm e}$ and the weak line $\lambda 4275.55$ is the only one that can break the degeneracy in this component since O\thinspace II $\lambda 4089.29$ is blended with a ghost feature (see Section~\ref{sec:siiv_coment}). However, it is clear that the density dependence is well limited within a range of 3000-3700 cm$^{-3}$. Fixing the temperature to the adopted one for the high ionization zone using CELs, we obtain $n_{\rm e} = 3490 \pm 340$ cm$^{-3}$ for HH~529~II. On the other hand, since we were able to use the O\thinspace II $\lambda 4089.29$ together with $\lambda 4275.55$ in the nebular component of cut 2, we have a convergence within a more limited interval of values. The physical conditions that minimizes $\chi^2$ in this case are $n_{\rm e} = 4390 \pm 400$ cm$^{-3}$ and $T_{\rm e} = 9350 \pm 1090$ K. This result is compatible with $T_{\rm e}$([O\thinspace III]) within the uncertainties,  indicating that the emission of CELs and RLs of O$^{2+}$ comes basically from the same gas (see Section~\ref{subsec:overmetal}). 

In Table~\ref{tab:pc}, we can see that the density values obtained from O\thinspace II lines are similar to those obtained from other diagnostics in the nebular components but lower in the high-velocity ones. This may be because, although formally the population of the $^3P_J$ levels from O$^{2+}$ do not reach the statistical equilibrium until densities of $\sim 10^5 \text{ cm}^{-3}$, the density dependence becomes rather weak from values above $\sim 10^4 \text{ cm}^{-3}$, as it is shown in Fig. 4 from \citet{Storey17}. Therefore, the values obtained by this diagnostic may not be representative of the shock front, where the density is expected to be higher than $10^4 \text{ cm}^{-3}$ but from a lower density component flowing in the jet beam. A gas component with a density around $\sim 10^3 \text{ cm}^{-3}$ has larger deviations from LTE in the populations of the levels from which multiplet 1 arise. This might bias the results to lower values. However, this discrepancy may have a different origin, which requires further investigation.

\subsubsection{Electron temperature from He\thinspace I recombination line ratios}
\label{subsubsec:helium_temperature}

Following the procedure used by \citet{Zhang05} for PNe, we have used the $I$(\mbox{He}\thinspace \mbox{I} $\lambda$7281)/$I$(\mbox{He}\thinspace \mbox{I} $\lambda$6678) ratio for deriving $T_{\rm e}\left(\mbox{He}\thinspace \mbox{I} \right)$ in our spectra. The use of those particular lines have se\-ve\-ral advantages. First, $\lambda7281$ and $\lambda6678$ are among the brightest He\thinspace I RLs and their use minimizes observational errors. Second, they are produced in transitions between singlet levels, ensuring that they are free of significant self-absorption effects. We have explored the temperature dependence of other intensity ratios of \mbox{He}\thinspace \mbox{I} $\lambda$7281 with respect to other relevant singlet lines ($\lambda\lambda$4388, 4922, 4438, 3614, 3965 and 5016) using the recombination coefficients of \citet{Porter12,Porter13}. Intensity ratios of transitions coming from $5^1\text{D}$, $4^1\text{D}$, $3^1\text{D}$ levels to $2^1\text{P}$ show the the strongest dependence on $T_{\rm e}$ (see Fig.~\ref{fig:grotrian_he}). $2^1\text{P}$ is the same lower level of the transition producing the \mbox{He}\thinspace \mbox{I} $\lambda 7281$ line, which comes from  the $3^1\text{S}$ level. On the other hand, comparatively, the  $I$(\mbox{He}\thinspace \mbox{I} $\lambda$7281)/$I$(\mbox{He}\thinspace \mbox{I} $\lambda$6678) ratio has the weakest $n_{\rm e}$ dependence, in agreement with the conclusion of \citet{Zhang05}, despite using different recombination coefficients.

Fig.~\ref{fig:7281_temperature_helium_dependence} shows that the $T_{\rm e}$ dependence of $I$(\mbox{He}\thinspace \mbox{I} $\lambda$7281)/$I$(\mbox{He}\thinspace \mbox{I} $\lambda$6678) ratio is practically linear in the interval 5000 K $\leq T_{\rm e}$(K) $\leq$ 10000 K. The deviation between the determination of $T_{\rm e}$(\mbox{He}\thinspace \mbox{I}) using a linear fit (as in Eq.~(\ref{eq:helium_fit})) and a more complex interpolation of the recombination coefficients of \citet{Porter12,Porter13} is always smaller than $35 \text{ K}$. At $T_{\rm e}$ $>$ 10000 K, any linear fit will fail for almost all $n_{\rm e}$ values except for the lowest ones ($n_{\rm e}\leq$ 100 cm$^{-3}$). In these cases, a more complex treatment is necessary to estimate $T_{\rm e}$(He\thinspace I).  The linear fit (slope and intercept) varies significantly in the lower density ranges, and tends to remain almost constant for densities $n_{\rm e}\geq$ 10000 cm$^{-3}$.

\begin{equation}
    \label{eq:helium_fit}
    T_{\text e}\left(\mbox{He}\thinspace \mbox{I} \right) ({\text K})=\alpha \left[ \frac{I\left(\lambda7281\right)}{I\left(\lambda6678 \right)} \right]+\beta.
\end{equation}

In Table~\ref{tab:slopes}, we present the slope and intercept values given by Eq.~(\ref{eq:helium_fit}) for a density range representative for H\thinspace II regions and some PNe. The resulting $T_{\rm e}$(\mbox{He}\thinspace \mbox{I}), using the average values  obtained with $I(\lambda7281)/I(\lambda6678 )$, $I(\lambda7281)/I(\lambda4922 )$ and $I(\lambda7281)/I(\lambda4388)$ ratios, are all consistent with each other and are included in Table~\ref{tab:pc} for all components.

\subsection{Electron temperature determinations from nebular continuum.}
\label{subsec:balmer_paschen_jumps}

\begin{figure*}
\includegraphics[width=\columnwidth]{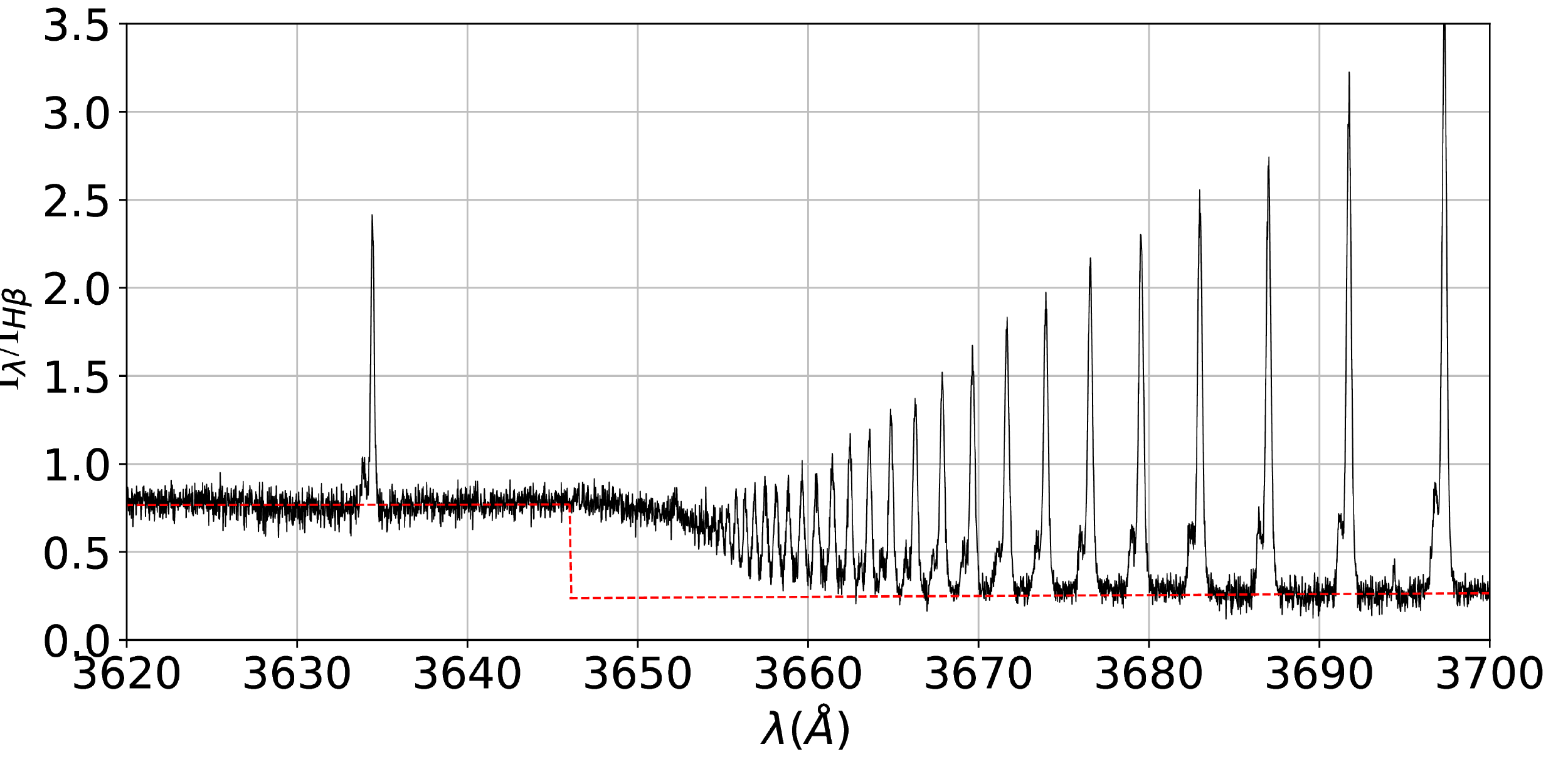}
 \includegraphics[width=\columnwidth]{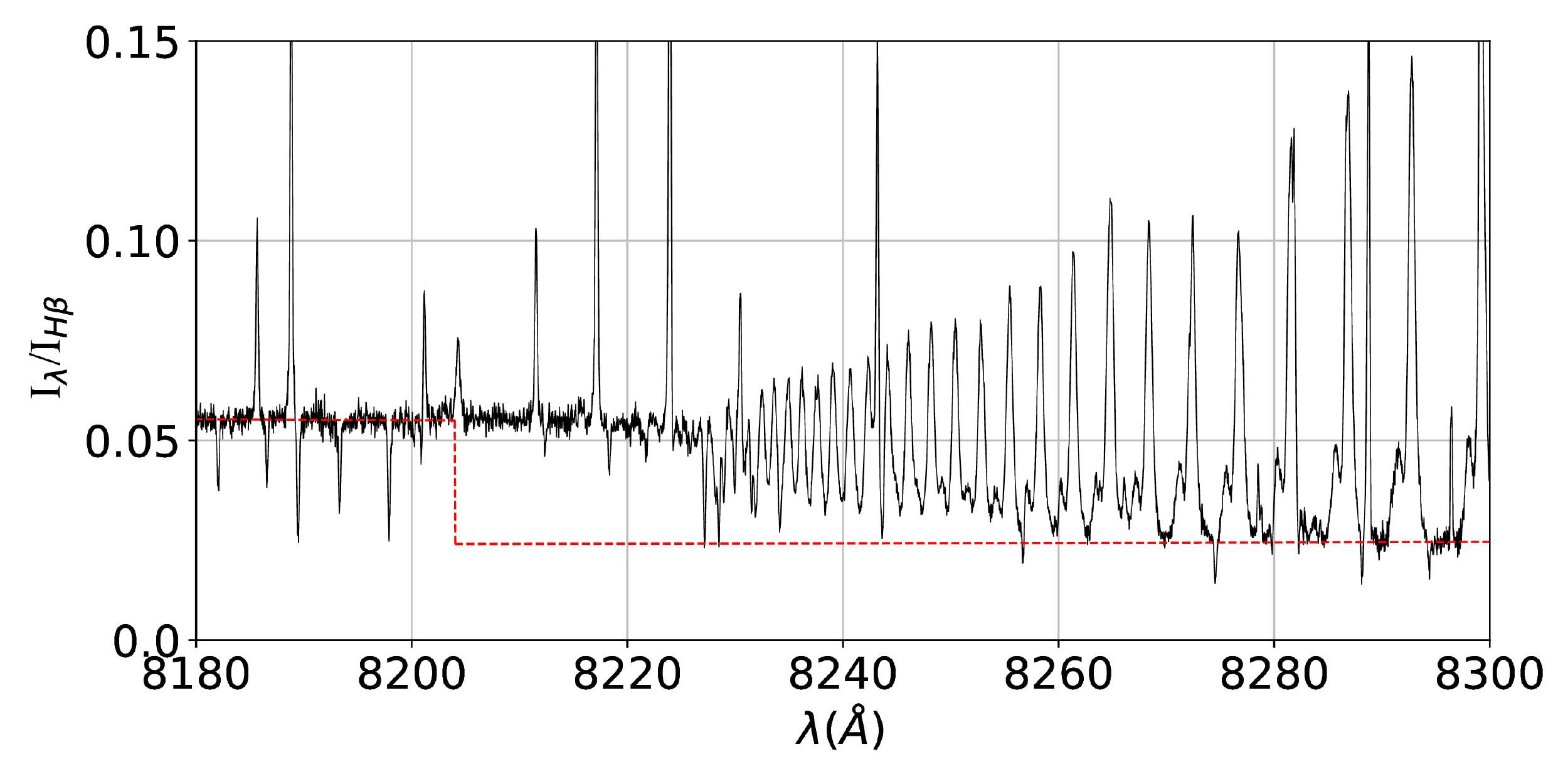}
 \caption{Reddening-corrected spectrum resulting after adding all cuts, showing the Balmer (left panel) and Paschen (right panel) discontinuities. Both jump estimations are shown in red.}
\label{fig:jumps}
\end{figure*}

 Thanks to the high signal-to-noise ratio of our spectra, we can obtain a good determination of the Balmer and Paschen discontinuities of the nebular continuum in the spectrum obtained adding all the cuts (see Fig.~\ref{fig:jumps}). We used Eq.~(\ref{eq:balmer_fit}), taken from \citet{liu01} for He$^{2+}$/H$^+$ = 0 to estimate $T_{\rm e}$(H\thinspace I)$_{\rm Balmer}$. This formula is based on theoretical continuum emission of H\thinspace I, He\thinspace I and He\thinspace II calculated by \citet{brownymathews70} and the theoretical line emission of \mbox{H}\thinspace \mbox{I} $\lambda 3770.63$ (H11) from \citet{Storey95}. 
Analogously, we used Eq.~(\ref{eq:paschen_fit}), taken from \citet{fangyliu11} to estimate $T_{\rm e}$(H\thinspace I)$_{\rm Paschen}$ using the measured Paschen discontinuty and the intensity of  \mbox{H}\thinspace \mbox{I }$\lambda 8862.78$ (P11) line.

\begin{equation}
\label{eq:balmer_fit}
T_{\rm e}{\rm (H\thinspace I)}_{\rm Balmer} ({\rm K})= 368\times\left(1+0.259\frac{\text{He}^{+}}{\text{H}^{+}}  \right)\left(\frac{\text{BJ}}{\text{H11}}\right)^{-3/2}.
\end{equation}

\begin{equation}
\label{eq:paschen_fit}
T_{\rm e}{\rm (H\thinspace I)}_{\rm Paschen} ({\rm K})=8.72\times \left(1+0.52\frac{\text{He}^{+}}{\text{H}^{+}} \right) \left( \frac{\text{PJ}}{\text{P11}}\right)^{-1.77}.
\end{equation}

The estimation of the temperature requires a precise fit to the continuum emission at both sides of 3646 \AA\ and 8204 \AA, the approximate wavelengths of the Balmer and Paschen discontinuities, respectively, since both estimations are very sensitive to changes in the jump value. We do not determine $T_{\rm e}$(H\thinspace I)$_{\rm Balmer}$ and $T_{\rm e}$(H\thinspace I)$_{\rm Paschen}$ in the remaining cuts because of the much larger noise level of the continuum in their spectra. However, using the spectrum of the combined cuts has the drawback of mixing the emission of the nebular and the high-velocity components in the continuum. In any case, as \citet{Bohigas15} suggests, the total $T_{\rm e}$(H\thinspace I) would be the weighted average of the individual values of the mixed components, where the weight would be the H$^{+}$ mass of each component. Thus, given that the high-velocity component should contain a much smaller mass, we can assume that the contribution of the high-velocity component to the continuum should be small, not affecting the $T_{\rm e}$(H\thinspace I) determination in a substantial manner.

Fig.~\ref{fig:jumps} shows the discontinuities and the fitted Balmer and Paschen continua in the normalized and reddening corrected spectrum. The best fit is achieved with $\text{BJ}/\text{I}_{H\beta} = 0.532 \pm 0.036$ and $\text{PJ}/\text{I}_{H\beta} = 0.031 \pm 0.002$.

\section{Chemical abundances}
\label{sec:chemical_abundances}

\subsection{Ionic abundances from CELs}
\label{subsec:ionic_from_CELS}

 We determine the ionic abundances based on the observed CELs using the PyNeb routines and the transition probabilities and collision strengths given in Table~\ref{tab:atomic_data}. Abundances for O$^{+}$, N$^{+}$, S$^{+}$, Ni$^{2+}$ and Cl$^{+}$ were derived using the $n_{\rm e}$ and $T_{\rm e}$(low) adopted for each component of each cut, while abundances for O$^{2+}$, Ne$^{2+}$, Cl$^{3+}$, Fe$^{3+}$, Ar$^{2+}$ and Ar$^{3+}$ rely on the adopted $T_{\rm e}$(high). S$^{2+}$ and Cl$^{2+}$ abundances were derived using $T_{\rm e}$([S\thinspace III]) (see Section~\ref{sec:cl_comment}). In the case of Fe$^{2+}$, estimations of its abundance are presented using both $T_{\rm e}$(high) and $T_{\rm e}$(low) (see Table~\ref{tab:Fe3_abundances_low} and Table~\ref{tab:Fe3_abundances_high}). This will be analysed in Section~\ref{subsubsec:total_abun_fe}. General results are presented in Table~\ref{tab:cels_abundances}.

\begin{table*}
\centering
\caption{Chemical abundances based on CEL's without considering the temperature fluctuations scenario ($t^2=0$). The units are logarithmic with $n(\text{H})=12$.}
\label{tab:cels_abundances}
\begin{adjustbox}{width=\textwidth}
\begin{tabular}{ccccccccccccc}
\hline
 & \multicolumn{1}{c}{Cut 1} & \multicolumn{2}{c}{Cut 2} & \multicolumn{2}{c}{Cut 3} & \multicolumn{1}{c}{Cut 4} \\
Ion &  Nebula & HH~529~II &  Nebula & HH~529~III &  Nebula &  Nebula & Combined cuts\\
\hline

O$^{+}$ & $7.88 \pm 0.04$ & $7.36^{+0.12} _{-0.09}$  & $7.83^{+0.06} _{-0.05}$ & $7.51^{+0.22} _{-0.14}$ & $7.76^{+0.07} _{-0.06}$ & $7.81^{+0.07} _{-0.06}$& $7.75 \pm 0.05$ \\

O$^{2+}$ & $8.32 \pm 0.03$ & $8.54^{+0.03} _{-0.02}$ & $8.35 \pm 0.03$ & $8.48^{+0.03} _{-0.02}$ & $8.35^{+0.03} _{-0.02}$  & $8.36 \pm 0.03$ & $8.37 \pm 0.03$   \\

N$^{+}$  & $7.00 \pm 0.02$ & $6.19^{+0.06} _{-0.05}$  & $6.99 \pm 0.03$ &  $6.45^{+0.12} _{-0.08}$  & $6.91^{+0.04} _{-0.03}$ & $6.98^{+0.04} _{-0.03}$ & $6.89 \pm 0.03$  \\

Ne$^{2+}$ & $7.67 \pm 0.04$ & $7.91 \pm 0.03$ & $7.70^{+0.04} _{-0.03}$ &  $7.80 \pm 0.03 $  & $7.73 \pm 0.03$ & $7.75^{+0.04} _{-0.03}$ & $7.73^{+0.04} _{-0.03}$  \\

S$^{+}$&  $5.58 \pm 0.05$ & $4.88^{+0.08} _{-0.07}$ & $5.57^{+0.05} _{-0.04}$ &  $5.17^{+0.15} _{-0.12}$ & $5.53^{+0.06} _{-0.05}$ & $5.59 \pm 0.05$ & $5.49 \pm 0.05$  \\

S$^{2+}$&  $6.79 \pm 0.04$ & $6.92^{+0.05} _{-0.04}$ & $6.85^{+0.09} _{-0.07}$& $6.89^{+0.06} _{-0.05}$  & $6.86^{+0.05} _{-0.04}$& $6.82^{+0.05} _{-0.04}$& $6.85 \pm 0.04$\\

Cl$^{+}$ & $3.72 \pm 0.04$ & $2.95^{+0.14} _{-0.13}$ & $3.71 \pm 0.04$ & <3.34& $3.68 \pm 0.04$ & $3.75 \pm 0.04$ & $3.63 \pm 0.04$  \\

Cl$^{2+}$ & $4.88^{+0.06} _{-0.05}$ & $5.01^{+0.06} _{-0.05}$ & $4.93^{+0.06} _{-0.05}$ & $5.03^{+0.08} _{-0.07}$   &$4.96^{+0.06} _{-0.05}$&$4.95^{+0.06} _{-0.05}$&$4.94^{+0.05} _{-0.04}$ \\

Cl$^{3+}$ & $3.28 \pm 0.06$ & $3.60 \pm 0.05$  & $3.28^{+0.04} _{-0.03}$  & $3.40 \pm 0.10 $ & $3.33 \pm 0.04$ & $3.45 \pm 0.04$ & $3.38 \pm 0.05$ \\

Ar$^{2+}$ & $6.31 \pm 0.03$ & $6.39 \pm 0.02$ & $6.31 \pm 0.03$ &  $6.36 \pm 0.03 $  & $6.33 \pm 0.02$ & $6.29 \pm 0.03$ & $6.32 \pm 0.03$  \\

Ar$^{3+}$ & $4.39^{+0.06} _{-0.05}$ & $4.67^{+0.04} _{-0.03}$ & $4.47^{+0.04} _{-0.03}$ &  $4.51 \pm 0.07 $  & $4.50 \pm 0.04$ & $4.61^{+0.04} _{-0.03}$ & $4.52^{+0.05} _{-0.04}$  \\

$^{*}$Fe$^{2+}$ & $5.77\pm 0.02$ & $5.94 \pm 0.05$ & $5.82\pm 0.02$ &  $5.75 \pm 0.05$  & $5.78 \pm 0.02$ & $5.76 \pm 0.03$ & $5.80 \pm 0.02$\\

$^{**}$Fe$^{2+}$ &$5.52\pm 0.03$ &$5.62 \pm 0.07$ & $5.57\pm 0.02$ & $5.40 \pm 0.06$  &$5.52 \pm 0.04$ & $5.48 \pm 0.03$ & $5.53 \pm 0.04$ \\

Fe$^{3+}$ & $5.68^{+0.13} _{-0.11}$ & $6.25^{+0.10} _{-0.09}$  & $5.70^{+0.09} _{-0.08}$ & <6.58& $5.73^{+0.11} _{-0.10}$ & $5.73^{+0.13} _{-0.12}$ & $5.75^{+0.11} _{-0.10}$  \\

Ni$^{2+}$ & $4.37 \pm 0.14$ & $4.50\pm 0.08$ & $4.33 \pm 0.17$ & $4.28^{+0.15} _{-0.11}$    & $4.32 \pm 0.16$& $4.36 \pm 0.12$& $4.38 \pm 0.10$\\

\hline
\end{tabular}
\end{adjustbox}
\begin{description}
\item $^*$ indicates that $T_{\rm e} (\text{high})$ was used. \\
\item $^{**}$ indicates that $T_{\rm e} (\text{low})$ was used. \\
\end{description}
\end{table*}

\subsection{Ionic abundances from RLs}
\label{subsec:ionic_from_RLs}

\subsubsection{He$^{+}$ abundance}
\label{subsubsec:ionic_from_He1r}

To estimate the He$^{+}$ abundance, we use the flux of some of the most intense He\thinspace I lines: $\lambda \lambda$3188, 3614, 3889, 3965, 4026, 4388, 4438, 4471, 4713, 4922, 5016, 5876, 6678, 7065, 7281. He\thinspace I $\lambda \lambda 4121,5048$ lines were discarded because they are contaminated by ghost lines (see Section~\ref{sec:siiv_coment}). The 15 selected lines correspond to both singlet and triplet configurations, as it is shown in Fig.~\ref{fig:grotrian_he}. The fluxes of triplet lines are affected by the metastability of the 2$^{3}$S level. The comparatively much longer lifetime of 2$^{3}$S means that transitions to this level can become optically thick, altering the flux ratios predicted by recombination theory for some He\thinspace I lines. For example, self-absorption of He\thinspace I $\lambda 3188$ photons can increase the flux of He\thinspace I $\lambda \lambda $3889, 5876 and 7065 lines at the expense of He\thinspace I $\lambda 3188$, which flux decreases accordingly. On the other hand, self-absorption of the He\thinspace I $\lambda 3889$ line is also important and increase the flux of He\thinspace I $\lambda 7065$ at the expense of He\thinspace I $\lambda 3889$. However, the sum of the fluxes of He\thinspace I $\lambda \lambda $3188, 3889, 4713, 5876, and 7065 
lines should remain independent of the optical depth \citep[parameterized by $\tau _{3889}$ or $\tau_{3188}$,][]{Porter07}.

In Table~\ref{tab:helium_2}, we show the He$^{+}$ abundances determined using the fluxes of He\thinspace I $\lambda \lambda$3188, 3889, 4713, 5876, and 7065 lines and the values of $n_{\rm e}$ and $T_{\rm e}$(He\thinspace I) corresponding to each component of each cut. In the same table, we also include the He$^{+}$ abundance obtained from the sum of the fluxes of all the individual lines of the table and re-distributing them assuming $\tau_{3188} = \tau_{3889}= 0$. 
In Table~\ref{tab:helium_1} we show the He$^{+}$ abundances determined from singlet lines and those triplet ones that are expected to be less affected by self-absorption \citep[see Table 2 from][]{Benjamin02}. Tables \ref{tab:helium_2} and \ref{tab:helium_1} show a good agreement between the average values of He$^{+}$/H$^+$ ratios included in Table~\ref{tab:helium_1} (the last row) and those obtained summing the fluxes of the lines included in Table~\ref{tab:helium_2}. This last table also shows that the self-absorption effects are less important in the high-velocity components than in the nebular one. This is noticeable in the lower dispersion of the abundances obtained with individual lines in the high-velocity components. As discussed in \citet[][see their figure 4.5]{osterbrock06} if the nebula has ionized zones at different velocities, the self-absorption effects can be  reduced due to the Doppler shift between the emitting and absorbing zones. For example, the effect of self-absorption in the intense He\thinspace I  $\lambda 5876$ line is notable in the nebular component,  giving He$^{+}$ abundances about 0.05 dex higher than the sum value. In this sense, the common procedure of using a flux-weighted average of He\thinspace I $\lambda 5876$ and other bright optical He\thinspace I lines (as $\lambda$4471 and $\lambda$6678) for obtaining the mean He$^{+}$ abundance would provide rather an upper limit of it.

Another interesting fact that can be noted in Table~\ref{tab:helium_1} is that the He$^+$ abundance determined from the He\thinspace I $\lambda 5016$ line is lower than the values  obtained from other lines in the high-velocity components. An abnormally low flux of this line was noted by \citet{Esteban04}, and this was attributed to self-absorption effects in the singlet configuration of He\thinspace I. \citet{Porter07} discussed this, proposing that the most likely explanation is a deviation from case B of the He\thinspace I $\lambda \lambda$ 537.0 and 522.0 lines, that go to the ground level, partially escaping before being reabsorbed. This is probably the case in the high-velocity components where any kind of self-absorption of photons emitted by the ``static'' nebular gas should be reduced. The adopted He$^{+}$/H$^{+}$ values are presented in Table~\ref{tab:rls_abundances}.

\subsubsection{O$^{2+}$ abundance}
\label{subsubsec:ionic_from_O2r}

In Table~\ref{tab:OII_abundances}, we present the O$^{2+}$ abundance obtained from RLs of \mbox{O}\thinspace \mbox{II}. We use $T_{\rm e}$(high) and the values of $n_{\rm e}$ obtained from \mbox{O}\thinspace \mbox{II} (see section~\ref{subsubsec:oii_pc}) and [Fe\thinspace III] lines for the nebular and high-velocity components, respectively. We used the recombination coefficients calculated by \citet{Storey17} that consider the distribution of population among the O$^{2+}$ levels with some improvements over similar estimates from \citet{bastin06}. Previous references \citep[as][]{Storey94} assumed that the O$^{2+}$ levels are populated according to their statistical weight, which is not suitable for densities below the critical one. 

In Table~\ref{tab:OII_abundances}, we also present the weighted average abundance for each multiplet. In the last row of Table~\ref{tab:OII_abundances} we give the  O$^{2+}$ abundance obtained averaging the values obtained for multiplets 1, 2, 10, 20 and $3d-4f$ transitions. These multiplets and transitions give consistent values and were also considered by \citet{Esteban04} for determining their mean values. However, we decided to consider only the abundance obtained from multiplet 1 as representative of the O$^{2+}$ abundance, as we show in Table~\ref{tab:rls_abundances}. This is because, although it gives values consistent with the average of the other aforementioned multiplets and  transitions, the inclusion of multiplets with fainter lines increases the formal uncertainties of the final mean O$^{2+}$ abundance. 

\subsubsection{Determination of the abundance of other heavy elements based on RLs.}
\label{subsubsec:other_rls_ionic_abundances}

Due to the high quality of our deep spectra, we were able to determine abundances of other heavy element ions such as O$^{+}$, C$^{2+}$ and Ne$^{2+}$ based on the fluxes of RLs and the recombination coefficients presented in Table~\ref{tab:rec_atomic_data}.

\begin{figure}
\includegraphics[width=\columnwidth]{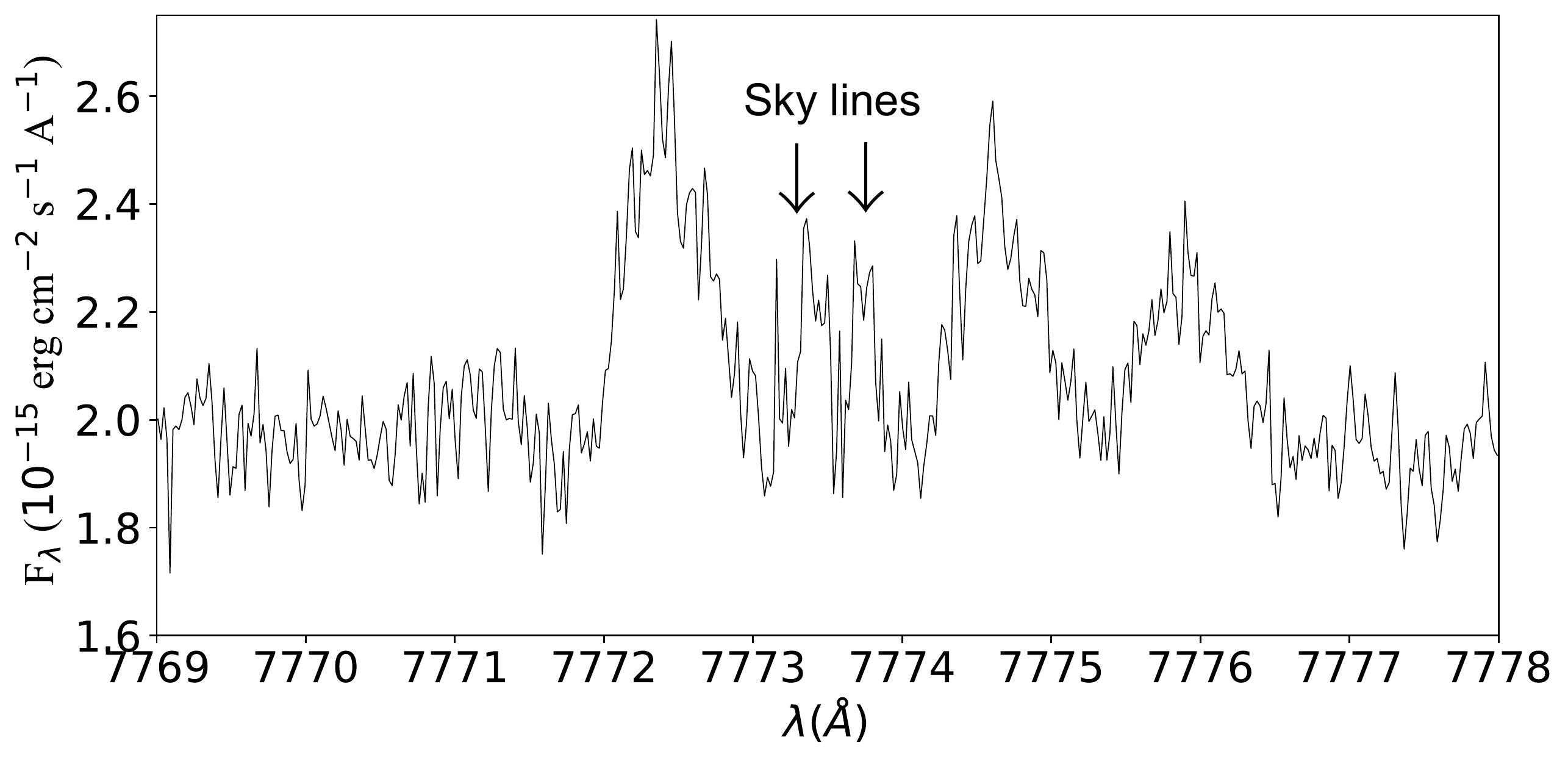}
\caption{Lines of multiplet 1 of O\thinspace I (3s$^{5}$S$^{0}$-3p$^{5}$P) in the spatial cut 2. No emission from HH~529~II is observed, only the nebular component is noticeable. These lines are produced from transitions of quintet levels \citep{grandi75a} and arise  purely by recombination. Due to high spectral resolution and the earth motion during the observations, these lines are free of blending with sky lines. }
\label{fig:oi_lines}
\end{figure}

O$^{+}$ abundances were obtained from the lines of multiplet 1 of O\thinspace I $\lambda \lambda$ 7771.94, 7774.17 and 7775.39 together with the adopted density and temperature of the low ionization zone for each component of each cut. Due to the high spectral resolution of our data, these O\thinspace I lines are not blended with telluric emission features, as is shown in Fig~\ref{fig:oi_lines}. We do not detect the lines of multiplet 1 of \mbox{O}\thinspace \mbox{I} in the high-velocity components. In these cases, we have estimated upper limits of their intensity and corresponding abundances considering an hypothetical line with a flux of 3 $\sigma$ of the rms of the adjacent continuum.  The resulting O$^+$ abundances and the estimated upper limits for the high-velocity components are shown in Table~\ref{tab:rls_abundances}.

\begin{figure}
\includegraphics[width=\columnwidth]{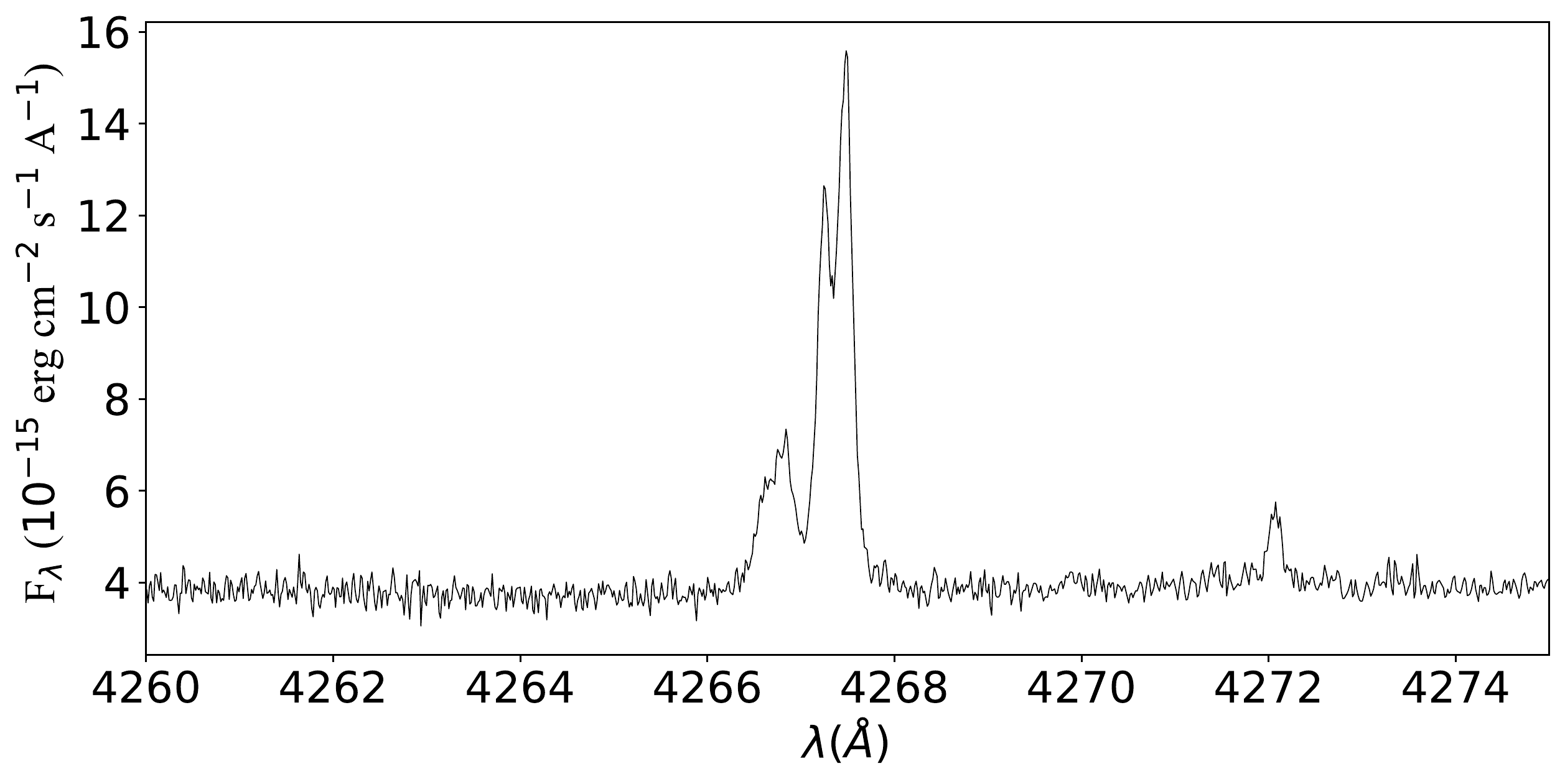}
\caption{Lines of multiplet 6 of C\thinspace II (3d$^{2}$D-4f$^{2}$F$^{0}$) in the spatial cut 2. Due to our high spectral resolution, we can partially separate $\lambda 4267.00$ from $\lambda 4267.18+\lambda4267.26 $ both in the component corresponding to HH~529~II and to the nebular one. }
\label{fig:cii_lines}
\end{figure}

\begin{figure}
\includegraphics[width=\columnwidth]{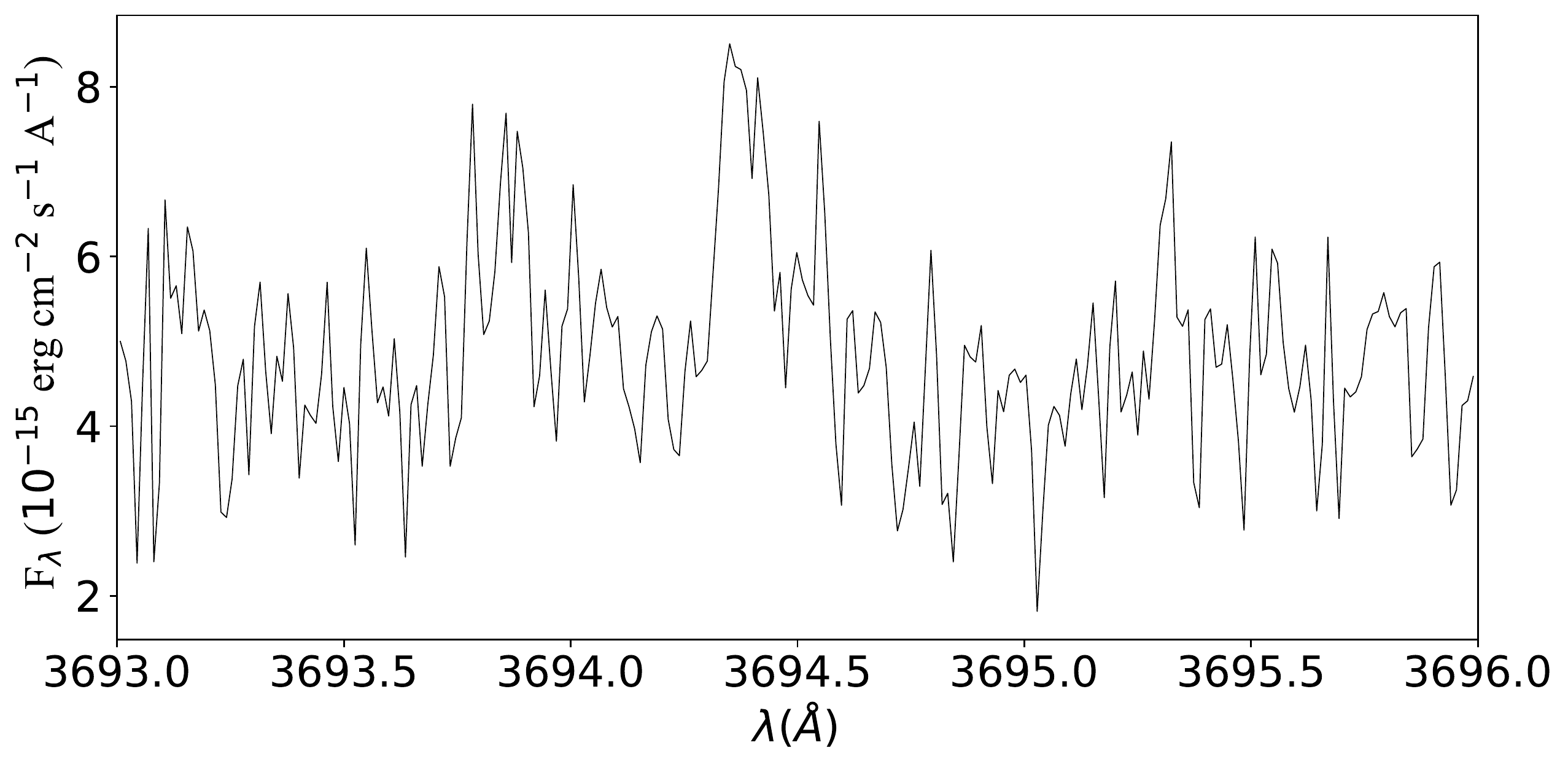}
\caption{Lines of multiplet 1 of Ne\thinspace II (3s$^{4}$P-3p$^{4}$P$^{0}$) in the spatial cut 2.}
\label{fig:neii_lines}
\end{figure}

For C$^{2+}$ and Ne$^{2+}$, we adopt the temperature of the high ionization zone for each component of each cut. C\thinspace II RLs from different transitions were considered to derive C$^{2+}$ abundances, as is shown in Table~\ref{tab:other_rls_abundances}. Multiplet 6 of C\thinspace II present two lines at 4267.00 and 4267.18+4267.26 \AA\, resolved at our spectral resolution, as shown in Fig.~\ref{fig:cii_lines}. In general, lines from all multiplets of C\thinspace II considered give consistent values of C$^{2+}$ abundances. RLs from multiplet 1 of Ne\thinspace II were used to calculate the Ne$^{2+}$ abundance. Although they are rather faint lines (see Fig.~\ref{fig:neii_lines}), the Ne$^{2+}$ abundances derived from Ne\thinspace II  $\lambda\lambda$ 3694.21 and 3766.26 lines for each component of cut 2 are consistent with each other. In addition, the Ne$^{2+}$ abundance we derive for the nebular component in cuts 2 and 3 is in good agreement with that obtained by \citet[][see their Table~11]{Esteban04}.

\begin{table*}
\centering
\caption{Chemical abundances based on RL's.  The units are logarithmic with $n(\text{H})=12$.}
\label{tab:rls_abundances}
\begin{adjustbox}{width=\textwidth}
\begin{tabular}{ccccccccccccc}
\hline
 & \multicolumn{1}{c}{Cut 1} & \multicolumn{2}{c}{Cut 2} & \multicolumn{2}{c}{Cut 3} & \multicolumn{1}{c}{Cut 4} \\
Ion &  Nebula & HH~529~II &  Nebula & HH~529~III &  Nebula &  Nebula & Combined cuts\\
\hline

He$^{+}$ & $10.89 \pm 0.02$ & $10.95 \pm 0.03$ & $10.90 \pm 0.01$ & $10.95 \pm 0.03$ & $10.90 \pm 0.02$ & $10.91 \pm 0.02$ & $10.91 \pm 0.02$\\

O$^{+}$ & $8.34 \pm 0.10$ & < 7.91 & $8.25 \pm 0.06$ & <7.95 & $8.28 \pm 0.07$ & $8.27 \pm 0.07$ & $8.19 \pm 0.07$\\

O$^{2+}$ & $8.47 \pm 0.04$ & $8.83 \pm 0.07$ & $8.52 \pm 0.02$ & $8.84 \pm 0.09$ & $8.52 \pm 0.03$ & $8.53 \pm 0.03$ & $8.58 \pm 0.02$\\

C$^{2+}$ & $8.34 \pm 0.03$ & $8.46 \pm 0.02$ & $8.35 \pm 0.03$ & $8.56 \pm 0.03$ & $8.34 \pm 0.03$ & $8.33 \pm 0.02$ & $8.37 \pm 0.03$ \\

Ne$^{2+}$ & - & $8.60 \pm 0.06$ & $8.07 \pm 0.03$ & - & $8.04 \pm 0.15$&- &-\\

\hline
\end{tabular}
\end{adjustbox}
\end{table*}

\section{Temperature fluctuations}
\label{sec:temp_fluc}

We followed the $t^2$-paradigm postulated by \citet{Peimbert67}, by using Eq.~(10) from \citet{Peimbert69} and Eq.~(10) from \citet{Peimbert04}, together with the measured $T_{\rm e}$(\mbox{H}\thinspace \mbox{I}) and $T_{\rm e}$(\mbox{[O}\thinspace \mbox{III]}) in order to estimate $t^2$  \citep{Peimbert03,Esteban04,garciarojas04,garciarojas05,garciarojas07-2}.  Implicitly, this approach assumes that $t^{2}(\text{H}^{+})\approx t^{2}(\text{O}^{2+})$ and $T_0(\text{H}^{+})\approx T_0(\text{O}^{2+})$. The same procedure has been followed with eqs.~(13) and (14) from \citet{Peimbert69}, together with the measured $T_{\rm e}$(\mbox{[S}\thinspace \mbox{III]}) and $T_{\rm e}$(\mbox{[N}\thinspace \mbox{II]}), respectively, in order to estimate representative values of $t^{2}$ for different ionization zones \citep{Peimbert69,Esteban98}. The analogous procedure has been used with Eq.~(11) from \citet{Zhang05}, to use the $t^2$ dependence of the measured $T_{\rm e}$(\mbox{He}\thinspace \mbox{I}).  In Table~\eqref{tab:t2_for_all_comb}, we show the  $t^{2}$ and $T_{\text{0}}$ values obtained for each combination. We have to emphasize the excellent agreement between using $T_{\rm e}$(\mbox{H}\thinspace \mbox{I}) and  $T_{\rm e}$(\mbox{He}\thinspace \mbox{I}) together with the $T_{\rm e}$ diagnostics based on CEL ratios. 

\begin{table*}
\centering
\caption{ $t^2$ and $T_0$ derived from the combinations of different temperature estimates within the formalism of \citet{Peimbert67} for the spectrum of all cuts combined.}
\label{tab:t2_for_all_comb}
\begin{adjustbox}{width=\textwidth}
\begin{tabular}{ccccccc}
\hline
&\multicolumn{2}{c}{$T_{\rm e}$(\mbox{[O}\thinspace \mbox{III]})}&\multicolumn{2}{c}{$T_{\rm e}$(\mbox{[S}\thinspace \mbox{III]})}&\multicolumn{2}{c}{$T_{\rm e}$(\mbox{[N}\thinspace \mbox{II]})}\\
&$t^2$&$T_0$&$t^2$&$T_0$&$t^2$&$T_0$\\
\hline
$T_{\rm e}(\text{H}\thinspace \text{I})_{\text{BJ}}$&$0.020 \pm 0.017$&$7770 \pm 850$&$0.051 \pm 0.030$&$8240 \pm 980$&$0.068 \pm 0.023$&$8510 \pm 960$\\
$T_{\rm e}(\text{H}\thinspace \text{I})_{\text{PJ}}$& $0.019 \pm 0.025$& $7800 \pm 1260$& $0.050 \pm 0.042$ & $ 8250 \pm 1420$ & $0.068 \pm 0.033$ & $8530 \pm 1430$  \\
$T_{\rm e}$(\mbox{He}\thinspace \mbox{I})&$0.018 \pm 0.012$&$7840 \pm 520$&$0.054 \pm 0.024$&$8160 \pm 570$&$0.075 \pm 0.018$&$8360 \pm 570$\\
\hline
\end{tabular}
\end{adjustbox}
\end{table*}

However, the above procedure may not be entirely accurate. From the definition of $T_0$ and $t^2$ -- eqs.~(9) and (12) from \citet{Peimbert67} -- it is clear that these quantities depend on the integrated volume of gas. Thus, since each ion X$^{i+}$ will have its own Str\"omgren sphere, each one will have a representative $t^2(\text{X}^{i+})$ and $T_0 (\text{X}^{i+})$. Considering another ion, Y$^{i+}$, the assumption $t^2(\text{X}^{i+}) \approx t^2(\text{Y}^{i+})$ will be only correct if X$^{i+}$ and Y$^{i+}$ occupy the same nebular volume. Based on a set of Cloudy photoionization models with different input parameters, \citet{kingdon95} derived $t^2$ in two manners: as $t^2_{\text{str}}$ from the formal definition and the $t^2_{\text{obs}}$ obtained from the comparison of $T_{\rm e}$(\mbox{H}\thinspace \mbox{I}) and $T_{\rm e}$(\mbox{[O}\thinspace \mbox{III]}). They found that generally $t^2_{\text{str}} \neq t^2_{\text{obs}}$, with the difference increasing with the $T_{\text{eff}}$ of the ionizing sources. However, for the $T_{\text{eff}}$ typical of the ionizing stars of H\thinspace II regions (between 30,000 and 50,000 K), the approximation $t^2_{\text{str}} \approx t^2_{\text{obs}}$ seems to be valid. The main drawback one faces in determining $t^2_{\text{obs}}$ is its high intrinsic uncertainty.

Assuming the two ionization zones scheme for H\thinspace II regions, a better approximation to $t^2$ can be obtained using  eqs.~(7) and (10) from \citet{peimbert02}. Then we need to estimate the fraction of O$^+$ in the total O abundance. For the spectrum of the combined cuts, this value varies from 0.36 to 0.23 according to whether the abundances are determined from RLs or CELs, respectively. A reasonable approximation is to take the average value $\sim 0.3$. On the other hand, He$^+$ should be present in both, the O$^+$ and O$^{2+}$ zones. Although there may be coexistence of He$^0$ and H$^+$, the volume that He$^0$ occupies should be small at the ionization conditions of the observed area of the Orion Nebula and it can be assumed that the volume containing H$^+$ and He$^+$ should be approximately the same. This assumption is reinforced by the fact that the parameter $\eta=(\text{O}^{+}/\text{O}^{2+})(\text{S}^{2+}/\text{S}^{+})$ \citep{Vilchez88}, which is a measure of the radiation hardness and is  anticorrelated with the $T_{\text{eff}}$ of the ionizing source, has a value of log$\left(\eta\right)=0.74$ for the ``combined cuts'' spectrum. \citet{pagel92} showed that for log$\left(\eta\right)<0.9$, the amount of He$^{0}$ is negligible for a large
variety of photoionization models. Therefore, we can assume $T_0(\text{H}^{+})\approx T_0(\text{He}^{+})$ and $t^2 (\text{H}^{+})\approx t^2(\text{He}^{+})$. 
Based on the previous discussion, we use the $T_0-t^2$ dependence of the measured $T_{\rm e}$(\mbox{H}\thinspace \mbox{I}) and $T_{\rm e}$(\mbox{He}\thinspace \mbox{I}), for the ``combined cuts'' spectrum, obtaining $t^2(\text{H}^{+})\approx 0.036$ and $T_0(\text{H}^{+}) \approx 8000 \text{ K}$. Using these values in Eq.~(10) from \citet{peimbert02} and assuming that the volume occupied by O$^{+}$ and N$^{+}$ is the same, and that to first order, $ \frac{ T_{\rm e}(\text{[O\thinspace III]})}{ T_{\rm e}(\text{[N\thinspace II]})} \approx \frac{ T_0(\text{O}^{2+})}{T_0(\text{N}^{+})}
\approx 0.85$, we obtain $T_0(\text{O}^{2+})\approx 7580$ and  $T_0(\text{N}^{+})\approx 8950$. Then, we estimate $t^2_{\text{high}}=t^{2}(\text{O}^{2+})\approx0.025$ and $t^2_{\text{low}}=t^{2}(\text{N}^{+})\approx0.050$. 

The remarkably good agreement between these values and those presented in Table~\ref{tab:t2_for_all_comb} reinforces the suitability of the temperature fluctuations paradigm to describe the results in the ``combined cuts'' spectrum. Considering the numerical values obtained, we adopt the average values $t^2_{\text{high}}=0.021 \pm 0.003 $, $t^2_{\text{inter}}=0.051 \pm 0.009 $ and $t^2_{\text{low}}= 0.064 \pm 0.011$, where the uncertainties correspond to the standard deviation of the average. Unfortunately, $T_{\rm e}$(H\thinspace I)  based on the Balmer and Paschen discontinuities can not be calculated for the individual components of the different cuts, and the estimations of $t^2$ must rely exclusively on the calculated $T_{\rm e}$(He\thinspace I) . However, calculations similar to those used to obtain the values presented in Table~\ref{tab:t2_for_all_comb} for the individual components of each cut show similar results. These values are presented in Table~\ref{tab:t2_per_comp}. Considering the higher uncertainty of the estimated $t^2$ based on $T_{\rm e}$(He\thinspace I) without using $T_{\rm e}$(H\thinspace I), we adopt the values of the ``combined cuts'' spectrum as representative for the other components of each cut. 

Following the same scheme described in Section~\ref{subsec:ionic_from_CELS}, we recalculate the ionic abundances assuming temperature fluctuations and the results are shown in Table~\ref{tab:cels_abundances_witht2}.

\begin{table*}
\centering
\caption{Chemical abundances based on CELs derived within the paradigm of temperature inhomogeneities ($t^2>0$).The units are logarithmic with $n(\text{H})=12$.}
\label{tab:cels_abundances_witht2}
\begin{adjustbox}{width=\textwidth}
\begin{tabular}{ccccccccccccc}
\hline
 & \multicolumn{1}{c}{Cut 1} & \multicolumn{2}{c}{Cut 2} & \multicolumn{2}{c}{Cut 3} & \multicolumn{1}{c}{Cut 4} \\
Ion &  Nebula & HH~529~II &  Nebula & HH~529~III &  Nebula &  Nebula & Combined cuts\\
\hline

O$^{+}$ & $8.18^{+0.10} _{-0.08}$  & $7.65^{+0.20} _{-0.12}$ & $8.14^{+0.12} _{-0.09}$ & $7.74^{+0.34} _{-0.17}$  & $8.05^{+0.13} _{-0.09}$  & $8.12^{+0.13} _{-0.09}$ & $8.05^{+0.11} _{-0.08}$  \\

O$^{2+}$ & $8.48^{+0.06} _{-0.05}$ & $8.72^{+0.04} _{-0.03}$ & $8.50^{+0.05} _{-0.04}$ &  $8.64^{+0.04} _{-0.03}$ & $8.51 \pm 0.04$ & $8.52^{+0.05} _{-0.04}$ & $8.53^{+0.05} _{-0.04}$  \\

N$^{+}$  & $7.18^{+0.05} _{-0.04}$ & $6.36^{+0.11} _{-0.08}$& $7.17^{+0.07} _{-0.05}$ &  $6.59^{+0.18} _{-0.11}$ & $7.09^{+0.07} _{-0.05}$ & $7.16^{+0.07} _{-0.06}$ & $7.07^{+0.06} _{-0.05}$  \\

Ne$^{2+}$ & $7.86^{+0.07} _{-0.06}$ & $8.12 \pm 0.04 $ & $7.89^{+0.06} _{-0.05}$  &  $7.99^{+0.05} _{-0.04}$ & $7.91^{+0.05} _{-0.04}$ & $7.94^{+0.06} _{-0.05}$ & $7.93^{+0.06} _{-0.05}$  \\

S$^{+}$ & $5.75^{+0.07} _{-0.06}$ & $5.04^{+0.12} _{-0.08}$  & $5.75^{+0.07} _{-0.06}$ &$5.30^{+0.19} _{-0.13}$   & $5.70^{+0.08} _{-0.07}$  & $5.77^{+0.08} _{-0.06}$ &  $5.66^{+0.07} _{-0.06}$  \\

S$^{2+}$& $6.87^{+0.05} _{-0.04}$& $7.01^{+0.06} _{-0.05}$  & $6.94^{+0.06} _{-0.05}$  & $6.97^{+0.07} _{-0.06}$  & $6.95^{+0.06} _{-0.05}$ & $6.91 \pm 0.05$ & $6.94 \pm 0.05$  \\

Cl$^{+}$ & $3.87 \pm 0.05$ & $3.09^{+0.16} _{-0.14}$ & $3.86^{+0.06} _{-0.05}$ & <3.49  & $3.81^{+0.06} _{-0.05}$ & $3.90^{+0.07} _{-0.06}$ & $3.76^{+0.06} _{-0.05}$   \\

Cl$^{2+}$ & $5.00^{+0.07} _{-0.06}$ & $5.15^{+0.09} _{-0.07}$ &$5.06^{+0.08} _{-0.06}$& $5.15^{+0.11} _{-0.08}$  & $5.09^{+0.08} _{-0.06}$ & $5.08^{+0.08} _{-0.06}$ & $5.07^{+0.07} _{-0.06}$  \\

Cl$^{3+}$ & $3.30 \pm 0.06$ & $3.71 \pm 0.05 $ & $3.38 \pm 0.04$ & $3.50^{+0.11} _{-0.10}$  & $3.43 \pm 0.04$ & $3.55^{+0.05} _{-0.04}$ & $3.48^{+0.06} _{-0.05}$    \\

Ar$^{2+}$ & $6.42 \pm 0.04$ & $6.51 \pm 0.03 $ & $6.42^{+0.04} _{-0.03}$ & $6.46 \pm 0.03 $   & $6.43 \pm 0.03$ & $6.40^{+0.04} _{-0.03}$ & $6.43^{+0.04} _{-0.03}$  \\

Ar$^{3+}$ & $4.55^{+0.07} _{-0.06}$ & $4.84^{+0.05} _{-0.04}$ & $4.63^{+0.06} _{-0.05}$ & $4.66^{+0.08} _{-0.07}$  & $4.66^{+0.05} _{-0.04}$ & $4.79^{+0.06} _{-0.05}$ & $4.69^{+0.06} _{-0.05}$  \\

Fe$^{{2+}^{*}}$ & $5.93 \pm 0.02$& $6.11 \pm 0.05$ &$5.97 \pm 0.02$& $5.91 \pm 0.05$ &$5.94 \pm 0.02$&$5.92 \pm 0.03$&$5.96 \pm 0.01$\\

Fe$^{{2+}^{**}}$ & $5.75 \pm 0.02$& $5.84 \pm 0.06$ &$5.81 \pm 0.02$&  $5.57 \pm 0.06$  &$5.75 \pm 0.02$&$5.72 \pm 0.03$&$5.76 \pm 0.02$\\

Fe$^{3+}$ & $5.99^{+0.16} _{-0.12}$ & $6.59^{+0.12} _{-0.10}$ & $6.00^{+0.12} _{-0.09}$ & <6.68 & $6.23^{+0.18} _{-0.16}$ & $6.05^{+0.16} _{-0.13}$ & $6.16^{+0.18} _{-0.15}$  \\

Ni$^{2+}$ & $4.52 \pm 0.19$& $4.68 \pm 0.09$ & $4.51 \pm 0.18$&  $4.42^{+0.20} _{-0.13}$ &$4.48 \pm 0.16$&$4.51 \pm 0.13$&$4.54 \pm 0.11$\\

\hline
\end{tabular}
\end{adjustbox}
\begin{description}
\item $^*$ indicates that $T_{\rm e} (\text{high})$ was used. \\
\item $^{**}$ indicates that $T_{\rm e} (\text{low})$ was used. \\
\end{description}
\end{table*}

\section{The Abundance Discrepancy Factor}
\label{sec:ADF}

A major problem in the analysis of photoionized regions is the discrepancy between the chemical abundances derived from RLs and CELs, known as the abundance discrepancy (AD) problem. The relatively weak RLs, give systematically higher abundances than CELs. This difference is commonly quantified through the abundance discrepancy factor \citep[ADF, ][]{liu00}, defined here as:

\begin{equation}
    \label{eq:ADF}
    \text{ADF}\left(\text{X}^{i}\right)=\text{log}\left(\frac{n\left(\text{X}^{i}\right)_{\text{RLs}}}{n\left(\text{X}^{i}\right)_{\text{CELs}}}\right).
\end{equation}

There is an extensive collection of works dedicated to this problem in the literature \citep[see][and references therein]{torrespeimbert80,liu01,stasinska07,garciarojas07,tsamis11,nicholls12,gomezllanos2020}. Although there is no definitive solution, there are several hypotheses to explain the AD. For example, temperature fluctuations (see Section~\ref{sec:temp_fluc}), which would primarily affect abundances based on CELs, underestimating the real values; semi-ionized gas clumps, overestimating abundances based on RLs and underestimating those of CELs; chemical inhomogeneities with different physical conditions, affecting both estimates depending on each specific case and so on. It is even possible that the AD is the result of the sum of various phenomena affecting each nebula in a different degree. Using a set of deep spectra of Galactic  H\thinspace II regions,  \citet{garciarojas07} found that the ADF is fairly constant around a factor 2, showing no trend with ionization degree, $T_{\rm e}$ or the effective temperature of the ionizing stars. They found that temperature fluctuations is the most likely explanation for the AD in H\thinspace II regions. 

In Table~\ref{tab:adf}, we present the ADF obtained from O$^{+}$, O$^{2+}$, Ne$^{2+}$ and C$^{2+}$ abundances determined from RLs and CELs for each component. The abundances based on CELs do not consider temperature fluctuations. In the case of C$^{2+}$, the value of the abundance from CELs have been taken from the UV observations reported by \citet{walter92}. We have considered the mean value of their positions number 5 and 7, which are the nearest to our slit and give 12+log(C$
^{2+}$/H$^+$) = 7.835. We do not estimate the ADF(C$^{2+}$) for the high-velocity component since the UV CELs values can  only be compared with the nebular component. We emphasize that the estimated value of $t^2$ comes from  the comparison of different temperature diagnostics and the formalism described in Section~\ref{sec:temp_fluc}. Therefore $t^2>0$, does not necessarily mean ADF $>0$, unless the measured value of $t^2$ is compatible with this.

From Table~\ref{tab:adf}, it is remarkable that the ADF is slightly different for each ion and higher in the high-velocity components. Comparing the values included in Table~\ref{tab:OII_abundances} and Table~\ref{tab:cels_abundances_witht2}, we can see that using the value of $t^2$ adopted for each ionization zone of the nebular components, the O$^{2+}$ abundances based on CELs become fairly consistent with those determined from RLs. In the case of the O$^+$ abundances, although the CELs abundances obtained with $t^2>0$ do not agree completely with those obtained from RLs, they become clearly more similar. Definitively, this is not the case for the Ne$^{2+}$ abundances, in which values determined from CELs and RLs still do not agree even considering  $t^2>0$. The results obtained for O$^{2+}$ and O$^{+}$ suggest that the temperature fluctuation paradigm may be capable of explaining the ADF, at least for these ions, the ones with the best abundance  determinations based on RLs. Among different scenarios, the existence of  H-deficient clumps has been advocated as a possible cause of the very high ADF values found in some PNe \citep[e.g][]{pequignot02}. Since the heating of ionized gas is mainly due by photoionization of H and He and the cooling by the emission of CELs of metallic ions, this scenario implies significant lower temperatures in the clumps \citep{pequignot02}. As we mentioned in Section~\ref{subsubsec:oii_pc}, the $T_{\rm e}$(O\thinspace II) determined for the nebular component of cut 2 (which must be representative of the other nebular components) is consistent with $T_{\rm e}$([O\thinspace III]) within the uncertainties, which rules out the aforementioned scenario in the nebular components analysed in this work.  
The situation seems to be different for the high-velocity components. Assuming $t^2>0$, the ionic abundances obtained from CELs do not increase enough to match the values obtained from RLs. For example, in the case of HH~529~II, considering the adopted value $t^2=0.021$, the ADF(O$^{2+}$) is reduced from 0.29 to 0.11 but is not zero. Even if we consider the value of $t^2=0.025$ from Table~\ref{tab:t2_per_comp}, the ADF(O$^{2+}$) would be 0.08. A similar situation is found in HH~529~III, where for $t^2=0.021$ the ADF(O$^{2+}$) is 0.20 while considering $t^2=0.030$ the ADF(O$^{2+}$) would be 0.12. Since we do not find evidence of higher temperature fluctuations than those previously commented, these results suggest the presence of another physical process apart (or in addition) to the classic description of temperature inhomogeneities to explain the ADF. A similar result was found by \citet{mesadelgado09} in the case of HH~202~S (see their Sec~5.5). For the high-velocity components, the presence of a H-deficient material can not be discarded as we will discuss in Section~\ref{subsec:overmetal}.

\begin{table*}
\centering\caption{Abundance discrepancy factor (ADF), defined in Eq.\ref{eq:ADF}, for different ions in the components of each cut.}
\label{tab:adf}
\begin{tabular}{lcccccccccc}
\hline
Cut & Component &\multicolumn{1}{c}{ADF(O$^{+}$)}&\multicolumn{1}{c}{ADF(O$^{2+}$)} & \multicolumn{1}{c}{ADF(Ne$^{2+}$)} & \multicolumn{1}{c}{ADF(C$^{2+}$)$^*$} \\

\hline
1& Nebular&$0.46 \pm 0.14$ &$0.15 \pm 0.07$&-&$0.51\pm 0.03$\\

2& HH~529~II&<0.55 &$0.29\pm 0.10 $&$0.79\pm 0.09$&-&\\

2& Nebular&$0.42\pm 0.12$&$0.17\pm0.05$&$0.37\pm0.04$&$0.52\pm 0.03$\\

3& HH~529~III&<0.44 & $ 0.36 \pm 0.11$&-&-\\

3& Nebular &$0.52\pm0.14$&$0.17\pm0.06$&$0.31\pm0.15$&$0.51\pm 0.03$\\

4& Nebular&$0.46 \pm 0.14$&$0.17\pm0.06$&-&$0.50\pm 0.02$\\

\multicolumn{2}{c}{Combined cuts}&$0.44\pm 0.12$&$0.21\pm0.05$&-&$0.54\pm 0.03$\\
\hline
\end{tabular}
\begin{description}
\item $^*$ We adopt 12+log(C$^{2+}$/H$^+$) = 7.835 from UV CELs considering the slit positions 5 and 7 of \citet{walter92}.\\
\end{description}
\end{table*}

\section{Total abundances}
\label{sec:total_abun}

\begin{table}
\centering
\caption{ICFs used to estimate the abundance of unseen ions.}
\label{tab:ICFs_used}
\begin{tabular}{llccccccccc}
\hline
Element & ICF Reference  \\
\hline
He&\citet{kunthsargent83}\\
C&\citet{berg19}\\
N&\citet{Peimbert69}\\
Ne&\citet{Peimbert69}\\
S& \citet{stasinska78}\\
Ar&\citet{izotov06}\\
Fe&\citet{rodriguez05}\\
Ni&\citet{delgadoinglada16}\\
\hline
\end{tabular}
\end{table}

We have to use ionization correction factors (ICFs) to estimate the contribution of unseen ions to the total abundance of some elements. Following the detailed analysis of \citet{arellanocorodova20}, we have used the ICF schemes for C, N, Ne and Ar adopted by those authors, which are shown in Table~\ref{tab:ICFs_used}. In the case of S, He, Fe and Ni, we use the ICFs from \citet{stasinska78}, \citet{kunthsargent83}, \citet{rodriguez05} and \citet{delgadoinglada16}, respectively. Results of total abundances based on CELs are presented in Table~\ref{tab:total_abundances_cels} and in Table~\ref{tab:total_abundances_cels_t2}, for the cases of $t^2=0$ and $t^2>0$, respectively. Total abundances based on RLs are presented in Table~\ref{tab:total_abundances_rls}. In this case, we do not expect significant changes in the total abundances within the temperature fluctuation paradigm due to the low dependence of RLs on temperature. The ICFs are generally based on the degree of ionization indicated by the abundance ratio of O ions. For consistency, in the case of abundances based on CELs, we use the degree of ionization determined also with CELs. An analogous procedure is applied for abundances determined from RLs.

\subsection{Total abundances with CELs}
\label{subsec:total_abun_CELS}

\subsubsection{Oxygen, Chlorine and Argon}
\label{subsec:O_Cl_Ar_CELs}

 The total abundances of O, Cl and Ar were obtained by adding the ionic abundances of the observed ions. Although in HH~529~III we could not estimate the Cl$^{+}$ abundance, its calculated upper limit shows that its contribution is negligible. It should be noted that, in the case of Ar, the ICF model of \citet{izotov06} indicates that the contribution of Ar$^+$/H$^+$ to the total Ar abundance is also negligible in all the analysed components. The Cl/O and Ar/O ratios are consistent with the solar values recommended by \citet{lodders19} within the uncertainties, whether we use abundances determined from CELs considering $t^2=0$ or $t^2>0$. In addition, there are no appreciable differences between the Cl/O and Ar/O ratios determined in the nebular and the high-velocity components.

\subsubsection{Nitrogen, Neon and Sulfur}
\label{subsec:N_Ne_S_CELs}


The total abundances of N, Ne and S  depend  strongly on the adopted ICF values. The schemes used for these elements are indicated in Table~\ref{tab:ICFs_used}.  The estimated fraction N/N$^{+}$ can reach values between 4 and 16 for the nebular and the high-velocity components, respectively. This indicates that the ICF values are rather uncertain at the high degree of ionization of the high-velocity components. However, in the nebular ones, the average value of $\text{log(N/O)}=-0.86 \pm 0.02$ is in very good agreement with the suggested solar value of $-0.88\pm 0.14$ \citep{lodders19}, while in the case of $t^2>0$, $\text{log(N/O)}=-0.98 \pm 0.02$ is still consistent within the relatively large uncertainties of the solar abundance ratio.


\citet{rubin2011} determined the Ne/H ratio of the Orion Nebula from FIR spectra taken with the {\it Spitzer Space Telescope}, that permitted to detect fine-structure [Ne\thinspace II] and [Ne\thinspace III] lines, avoiding the use of ICFs. They obtain 12+log(Ne/H) = 8.01$\pm$0.01, which is consistent with the Ne/H values we obtained for the nebular component assuming $t^2>0$. It is important to remark that the intensity of FIR CELs has a very small dependence on $T_{\rm e}$. Therefore, the agreement between the Ne/H ratios obtained from FIR CELs and optical ones assuming $t^2>0$ supports the temperature fluctuations paradigm for describing the spectral properties of the nebula.

The Ne/O and S/O ratios are rather similar in the nebular and high-velocity components. The average values of log(Ne/O) are $-0.64 \pm 0.02$ and $-0.61 \pm 0.02$ for $t^2=0$ and $t^2>0$, respectively, which are consistent with the solar value of $-0.58\pm 0.12$ \citep{lodders19} within the uncertainties. In the case of S/O, the average values of log(S/O) for $t^2=0$ and $t^2>0$ are $-1.50 \pm 0.05$ and $-1.63 \pm 0.05$, respectively, while the solar value is $-1.58\pm 0.08$ \citep{lodders19}.

\subsubsection{Nickel and Iron}
\label{subsubsec:total_abun_fe}

Ni/H abundances are estimated using the ICF scheme derived by \citet{delgadoinglada16} and are presented in Table~\ref{tab:total_abundances_cels} and Table~\ref{tab:total_abundances_cels_t2} for $t ^ 2 = 0$ and $t ^ 2> 0$, respectively. The estimation of this abundance is rather uncertain as discussed in Section~\ref{sec:ni2_ab_comment}.

In the case of Fe, considering the absence of He\thinspace II lines in our spectra, we do not expect to have Fe$^{4+}$ in the nebula and therefore Fe/H = Fe$^{+}$/H$^{+}$+Fe$^{2+}$/H$^{+}$+Fe$^{3+}$/H$^{+}$. We have determined the abundance of Fe$^{2+}$ and Fe$^{3+}$ in all the components of each cut except in  HH~529~III, where we could only estimate an upper limit to Fe$^{3+}$/H$^{+}$. In the high-velocity components, the absence of usually relatively intense [Fe\thinspace II] lines as $\lambda \lambda$4287,~5158~and~5262, together with the high ionization degree of the gas, indicates a negligible contribution of Fe$^{+}$ to the total abundance. Thus, in these cases Fe/H=Fe$^{2+}$/H$^{+}$+Fe$^{3+}$/H$^{+}$. In the nebular components, although a large number of [Fe\thinspace II] lines have been detected, their emission is mainly produced by fluorescence \citep{rodriguez99, verner00} and most of the observed lines will not provide reliable estimates of Fe$^{+}$ abundance. Unfortunately,  [Fe\thinspace II] $\lambda 8617$, a line almost insensitive to fluorescence  \citep{Lucy95,Baldwin96} can not be observed due to the physical gap of the CCDs in the Red Arm of UVES. However, previous studies with direct estimations of Fe$^{+}$ in the Orion Nebula as \citet{rodriguez02} or \citet{mesadelgado09}, obtain Fe$^{+}$/Fe$^{+2}$ ratios between 0.05 and 0.27. Considering the approximation Fe/H = Fe$^{2+}$/H$^{+}$+Fe$^{3+}$/H$^{+}$, the neglected Fe$^{+}$/H$^{+}$ ratio would contribute to Fe/H up to 0.06 dex in the worst case (calculating Fe$^{+2}$/H$^{+}$ with $T_{\rm e}(\text{high})$ and assuming Fe$^{+}$/Fe$^{+2}$ = 0.27). This maximum contribution is within the range of uncertainties associated with the sum of Fe$^{2+}$ and Fe$^{3+}$ abundances and therefore, it seems reasonable to consider Fe/H $\approx$ Fe$^{2+}$/H$^{+}$+Fe$^{3+}$/H$^{+}$ for the nebular component as well.

\citet{rodriguez05} proposed two ICFs for Fe, one derived from photoionization models and other based on observations with detection of [Fe\thinspace III] and [Fe\thinspace IV] lines. The values of Fe/H obtained using both ICFs are discrepant, perhaps due to errors in the atomic data of the ions involved. The true total Fe abundance is expected to be in between the values obtained from both ICFs \citep{rodriguez05,delgadoingladaetal14}. We use the aforementioned ICFs only for HH~529~III and we give its Fe/H ratio as the interval of values obtained from both ICFs, as it is shown in Table~\ref{tab:total_abundances_cels} and Table~\ref{tab:total_abundances_cels_t2}.

In HH~529~II, the abundances of Fe/H and Fe/O are higher than in the nebular components independently of whether the temperature $T_{\rm e}$(low) or $T_{\rm e}$(high) is considered to derive Fe$^{2+}$/H$^{+}$. The same behavior is observed in HH~529~III for $t^2 = 0$, although the uncertainty in Fe/H do not allow us to be conclusive in the case of $t^2>0 $. However, as is discussed in Section~\ref{subsec:iron_conditions}, the representative temperature to derive Fe$^{2+}$/H$^{+}$ in HH~529~II and HH~529~III is likely to be $T_{\rm e}$(high) while in the nebular components is $T_{\rm e}$(low). 

Considering the discussion above, the average log(Fe/O) value in the nebular components is $-2.53 \pm 0.02$ while for HH~529~II it is $-2.14 \pm 0.08$, both values computed assuming $t^2 = 0$. This represents an increase of the gaseous Fe abundance by a factor of 2.45 in HH~529~II. The same increase is observed when considering $t^2> 0$. For HH~529~III the increase is  between 1.78 and 4.37. Taking the solar value of $\text{log(Fe/O)}=-1.28 \pm 0.08$ recommended by \citet{lodders19}, we find that only 6\% of the total Fe is in gaseous phase in the nebular component, while this fraction increases to 14\% in HH~529~II and between 10\% and 25\% in  HH~529~III. In the case of HH~202~S, \citet{mesadelgado09} found that the gaseous phase fraction is around 44\%. The evidence of dust destruction on HH shocks is also present in non-photoionized objects \citep[see][and references therein]{Hartigan20}. This is shown by the relative enhancement of the  Fe emission lines with respect to the emission of other non-depleted elements in areas where shock waves are present. These results are consistent with theoretical studies predicting that fast shocks are effective at destroying dust grains \citep[see][and references therein]{jones94,mouri00}. However, it is possible to have partial depletion of Fe in jets \citep{antonicci14}. An evidence of surviving dust is the detection of thermal emission of dust at 11.7 $\mu$m coincident with HH~529~II and III as well as HH~202~S \citep{smith05}. A key factor is to explore correlations between the Fe abundance and some properties of the HH objects, such as their velocity, density or distance to the ionizing source.


\begin{table*}
\centering
\caption{Total abundances based on CELs with $t^2=0$.  The units are logarithmic with $n(\text{H})=12$.}
\label{tab:total_abundances_cels}
\begin{adjustbox}{width=\textwidth}
\begin{tabular}{ccccccccccccc}
\hline
 & \multicolumn{1}{c}{Cut 1} & \multicolumn{2}{c}{Cut 2} & \multicolumn{2}{c}{Cut 3} & \multicolumn{1}{c}{Cut 4} \\
Element &  Nebula & HH~529~II &  Nebula & HH~529~III &  Nebula &  Nebula & Combined cuts\\
\hline

O & $8.45 \pm 0.02$ & $8.57 \pm 0.03$ & $8.46 \pm 0.03$ & $8.53 \pm 0.03$  & $8.45 \pm 0.03$&$8.47 \pm 0.03$&$8.46 \pm 0.03$\\

N & $7.57 \pm 0.04$ & $7.40 ^{+0.16} _{-0.10}$ & $7.62 ^{+0.07} _{-0.05}$ &  $7.45 ^{+0.37} _{-0.17}$  &$7.60 ^{+0.08} _{-0.07}$ & $7.64 ^{+0.08} _{-0.06}$&$7.60 ^{+0.06} _{-0.05}$\\

Ne & $7.81 \pm 0.04$ &$7.94 \pm 0.03$ & $7.82 \pm 0.04$ & $7.84 \pm 0.04$  &$7.83 \pm 0.03$ & $7.86 \pm 0.04$&$7.82 \pm 0.04$\\

S & $6.89 \pm 0.04$&$7.18 ^{+0.07} _{-0.06}$ & $6.96 ^{+0.08} _{-0.09}$ & $7.09 ^{+0.11} _{-0.07}$  & $6.98 \pm 0.05$ & $6.94 \pm 0.05$ &$6.98 \pm 0.04$ \\

Cl & $4.92 \pm 0.06$ & $5.03 \pm 0.05$ & $4.97 \pm 0.06$ & $5.05 \pm 0.08$  & $4.99\pm0.06$ &  $4.99 \pm 0.06$ & $4.97\pm0.05$ \\

Ar & $6.32 \pm 0.03$ & $6.40 \pm 0.02$ & $6.32 \pm 0.03$ & $6.37 \pm 0.03$  & $6.34 \pm 0.02$& $6.30 \pm 0.03$&$6.33 \pm 0.03$\\

Fe$^{*}$ & $6.03 \pm 0.06$ & $6.42 \pm 0.07$ & $6.07 \pm 0.04$ &  6.24--6.63 &$6.06 \pm 0.05$&$6.05 \pm 0.06$&$6.08 \pm 0.05$\\

Fe$^{**}$ & $5.91 \pm 0.07$ & $6.34 \pm 0.08$ & $5.94 \pm 0.05$ &  5.90--6.28 &$5.94 \pm 0.07$&$5.92 \pm 0.08$&$5.95 \pm 0.07$\\

Ni & $4.59 \pm 0.14$ & $5.12 ^{+0.15} _{-0.10}$ & $4.58 ^{+0.18} _{-0.17}$ &  $4.75 ^{+0.29} _{-0.17}$ & $4.60 ^{+0.17} _{-0.16}$&$4.62 ^{+0.13} _{-0.12}$&$4.67 ^{+0.11} _{-0.10}$\\

\hline
\end{tabular}
\end{adjustbox}
\begin{description}
\item $^*$ indicates that $T_{\rm e} (\text{high})$ was used to compute Fe$^{++}$/H$^+$. \\
\item $^{**}$ indicates that $T_{\rm e} (\text{low})$ was used to compute Fe$^{++}$/H$^+$. \\
\end{description}
\end{table*}

\begin{table*}
\centering
\caption{Total abundances based on CELs with $t^2>0$.  The units are logarithmic with $n(\text{H})=12$.}
\label{tab:total_abundances_cels_t2}
\begin{adjustbox}{width=\textwidth}
\begin{tabular}{ccccccccccccc}
\hline
 & \multicolumn{1}{c}{Cut 1} & \multicolumn{2}{c}{Cut 2} & \multicolumn{2}{c}{Cut 3} & \multicolumn{1}{c}{Cut 4} \\
Element &  Nebula & HH~529~II &  Nebula & HH~529~III &  Nebula &  Nebula & Combined cuts\\
\hline

O & $8.66 \pm 0.05$ &$8.76 \pm 0.04$ & $8.66 \pm 0.05$ & $8.70 \pm 0.05$  & $8.64 \pm 0.04$ & $8.67 \pm 0.05$ & $8.65 \pm 0.05$\\

N & $7.66 ^{+0.11} _{-0.08}$ & $7.45 ^{+0.35} _{-0.16}$ & $7.69 ^{+0.14} _{-0.10}$ & $7.46 ^{+0.63} _{-0.28}$  & $7.68 ^{+0.16} _{-0.10}$& $7.71 ^{+0.15} _{-0.10}$ & $7.68 ^{+0.13} _{-0.09}$\\

Ne & $8.04 \pm 0.08$ & $8.16 \pm 0.04$ & $8.05 \pm 0.07$& $8.04 \pm 0.06$  & $8.04 \pm 0.06$& $8.08 ^{+0.08} _{-0.07}$ & $8.05 ^{+0.07} _{-0.06}$\\

S& $6.95 \pm 0.05$ & $7.24 ^{+0.10} _{-0.07}$ & $7.03 \pm 0.06$ & $7.15 ^{+0.14} _{-0.09}$  & $7.05 ^{+0.07} _{-0.06}$ & $7.01 ^{+0.06} _{-0.05}$& $7.05 ^{+0.06} _{-0.05}$\\

Cl &  $5.04 \pm 0.06$ & $5.17 \pm 0.09$ & $5.10 \pm 0.07$ & $5.17 \pm 0.11$ & $5.12 \pm 0.07$ & $5.12 \pm 0.07$ & $5.10 \pm 0.07$\\

Ar & $6.44 \pm 0.04$ & $6.52 \pm 0.03$ & $6.43 \pm 0.04$ & $6.47 \pm 0.04$ & $6.44 \pm 0.03$ & $6.42 \pm 0.04$ & $6.44 \pm 0.04$\\

Fe$^{*}$ & $6.26 \pm 0.09$ & $6.71 \pm 0.09$ & $6.29 \pm 0.06$&  6.38--6.72 & $6.41 \pm 0.12$ & $6.29 \pm 0.09$& $6.37 \pm 0.11$\\

Fe$^{**}$ & $6.19 \pm 0.10$ & $6.66 \pm 0.10$ & $6.22 \pm 0.07$&   6.04--6.37  & $6.35 \pm 0.14$ & $6.22 \pm 0.11$& $6.31 \pm 0.13$\\

Ni & $4.70 \pm 0.19$ & $5.22 ^{+0.26} _{-0.13}$ & $4.71 ^{+0.19} _{-0.18}$ & $4.81 ^{+0.42} _{-0.28}$  & $4.71 ^{+0.18} _{-0.16}$&  $4.72 ^{+0.15} _{-0.13}$& $4.78 ^{+0.13} _{-0.12}$\\

\hline
\end{tabular}
\end{adjustbox}
\begin{description}
\item $^*$ indicates that $T_{\rm e} (\text{high})$ was used to compute Fe$^{++}$/H$^+$. \\
\item $^{**}$ indicates that $T_{\rm e} (\text{low})$ was used to compute Fe$^{++}$/H$^+$. \\
\end{description}
\end{table*}

\subsection{Total abundances with RLs}
\label{subsec:total_rls}

\subsubsection{Helium}
\label{subsubsec:total_abun_he}

Considering the absence of an ionization front in HH~529~II and HH~529~III because of the non-detection of emission lines of neutral elements in their spectra, it is likely that the  He$^{0}$/H$^{+}$ ratio should be  negligible in the high-velocity components and, therefore, we can assume  $\text{He/H}=\text{He}^{+}/\text{H}^{+}$. In the nebular components, we estimate the fraction of neutral helium within the ionized zone making use of the ICF scheme by  \citet{kunthsargent83}, obtaining that the He$^{0}$/He fraction is approximately 10\%. This value is consistent with the other ICF schemes tested by \citet{mendez20} for the Orion Nebula. In Table~\ref{tab:total_abundances_rls}, we can see that the He/H ratios obtained for all the cuts are in complete agreement. 

\subsubsection{Oxygen}
\label{subsubsec:total_abun_o_rls}

The total O abundances based on RLs are determined directly from $\text{O}/\text{H}=\text{O}^{+}/\text{H}^{+}+\text{O}^{2+}/\text{H}^{+}$. In the high-velocity components, the estimated upper limits to the O$^{+}$ abundances indicate that this ion can contribute up to 0.05 to the total O abundance. Thus, for these high-velocity components the O abundance is assumed to be equal to the ionic abundance of O$^{2+}$.


\subsubsection{Carbon and Neon}
\label{subsubsec:total_abun_C_Ne}

In the case of the high-velocity components, due to the high degree of ionization estimated from the ionic O abundances based on RLs, we expect to have small or negligible contributions of the ions once ionized from Ne and C to their total abundances. For the nebular components, we use the same ICFs schemes of  \citet{Peimbert69} and \citet{berg19} for Ne and C, respectively, using ionic abundances based on RLs exclusively. It is important to note that this last ICF has been optimized for low-metallicy objects (up to 12+log(O/H)=8.0). However, \citet{arellanocorodova20} have shown that its use for higher metallicity objects provides consistent results. For comparison, in Table
~\ref{tab:total_abundances_rls}, we present the C/H ratio obtained using the ICF proposed by Amayo et al. (in prep, private communication), whose scheme is optimized for a wider range of metallicities, including the solar one. 

The resulting log(Ne/O) values based on RLs are $-0.44 \pm 0.03$ and $-0.23 \pm 0.10$ for the nebular components and HH~529~II, respectively. This indicates an overestimation of the Ne abundance based on RLs in HH~529~II, since it is significantly larger than the solar one. In the case of C, we obtain $\text{log(C/O)}=-0.20 \pm 0.02$ using the ICF of \citet{berg19} and $-0.26 \pm 0.02$ using the scheme of Amayo et al. (in prep) in the nebular components. This last value is more consistent with the recommended solar value of $-0.26 \pm 0.09$ by \citet{lodders19}. The log(C/O) value for HH~529~II is $-0.37\pm0.08$ while for HH~529~III is $-0.28 \pm 0.12$.

\begin{table*}
\centering
\caption{Total abundances based on RLs. The units are logarithmic with $n(\text{H})=12$.}
\label{tab:total_abundances_rls}
\begin{adjustbox}{width=\textwidth}
\begin{tabular}{ccccccccccccc}
\hline
 & \multicolumn{1}{c}{Cut 1} & \multicolumn{2}{c}{Cut 2} & \multicolumn{2}{c}{Cut 3} & \multicolumn{1}{c}{Cut 4} \\
Element &  Nebula & HH~529~II &  Nebula & HH~529~III &  Nebula &  Nebula & Combined cuts\\
\hline

O & $8.71 \pm 0.03$ & $8.83 \pm 0.07$ & $8.70 \pm 0.03$ &$8.84 \pm 0.09$ & $8.71 \pm 0.03$ & $8.72 \pm 0.03$ & $8.73 \pm 0.03$\\

He & $10.94 \pm 0.02$ & $10.95\pm 0.03$ &$10.94 \pm 0.01$ & $10.95\pm 0.03$ & $10.94 \pm 0.02$& $10.95 \pm 0.02$ & $10.94 \pm 0.02$ \\

C$^{*}$ & $8.56 \pm 0.04$ & \multirow{ 2}{*}{$ 8.46 \pm 0.02$}&$8.52 \pm 0.03$&\multirow{ 2}{*}{$ 8.56 \pm 0.03$}& $8.52 \pm 0.04$& $8.51 ^{+0.04} _{-0.03}$& $8.51 ^{+0.04} _{-0.03}$\\

C$^{**}$ & $8.48 ^{+0.08} _{-0.07}$ & &$8.45 \pm 0.05$& &$8.45 ^{+0.07} _{-0.06}$&$8.44 \pm 0.06$ & $8.45 ^{+0.06} _{-0.05}$\\

Ne &-&$8.60\pm 0.06$&$8.26 \pm 0.04$&-&$8.23 \pm 0.15$&-&-\\

\hline
\end{tabular}
\end{adjustbox}
\begin{description}
\item $^*$ Total abundances of the nebular components derived with the ICF of \citet{berg19}.\\
\item $^{**}$ Total abundances of the nebular components derived with the ICF of Amayo et al. (in prep.).\\
\end{description}
\end{table*}

\subsection{A slight higher metallicity in the high-velocity components?}
\label{subsec:overmetal}

An interesting result of our analysis is that the metal abundances are higher in the high-velocity components, HH~529~II and HH~529~III, than in the nebular ones. In the case of the O abundance, that difference can reach up to 0.14 dex, regardless if abundances are calculated with CELs or RLs. \citet{mesadelgado09} estimated that $0.12 \pm 0.03$ dex of log(O/H) is depleted into dust grains in the Orion Nebula. In principle one may explain the 0.14 dex increase of O/H in the HH objects as produced by dust destruction, and that all the O locked in grains has been released to the gas phase.  Nevertheless, the Ar/O, Ne/O, S/O and Cl/O ratios remain almost the same in all components. Since Ar and Ne are noble gases, they can not be depleted into dust grains and, therefore, lower abundance ratios would be expected if dust destruction is increasing the gaseous O abundance.  
In addition to this, considering that the O trapped onto dust grains is in olivine $(\text{Mg},\text{Fe})_2\text{SiO}_4$, pyroxene $(\text{Mg},\text{Fe})\text{SiO}_3$ or oxides like $\text{Fe}_2\text{O}_3$, then the gaseous O must grow in proportion to the release of elements like Fe to the gas phase. Considering this, \citet{mesadelgado09} estimated that 0.06 dex of log(O/H) can be attributed to dust destruction in HH~202~S. As we mention in Section~\ref{subsubsec:total_abun_fe}, the proportion of gaseous Fe present in in HH~529~II and HH~529~III is lower than in HH~202~S, and therefore, the expected increase of log(O/H) in the two bow shocks of HH~529 should be consequently less than 0.06 dex.  

BMB06 also report a higher O abundance in HH~529 with respect to the nebular one. The difference they obtained was slightly larger than ours, of around 0.2 dex. This value is confirmed in the later reanalysis of BMB06 data carried out by \citet{simondiaz2011}. However, as we discussed in Section~\ref{sec:line_inten}, part of the larger difference found by BMB06 with respect to our O/H ratio may be due to their underestimation of $\text{I(H}\beta\text{)}$, as we illustrate in Table~\ref{tab:comparison_balmer}.

Assuming that the abundance difference between the kinematical components is real, one possible explanation is that the bulk of the material of the HH objects comes from H-deficient material expelled by the source of the gas flow. As we mentioned in Section~\ref{sec:ADF}, an H-deficient ionized gas should be colder than one with normal chemical composition but this in not observed (see Table~\ref{tab:pc}). This may be because the possible over-metallicity is actually small, which might not significantly alter the temperature. The origin of the H-deficent material may be in the evaporation of protoplanetary discs around newly formed stars \citep{yuan11}, a probable scenario for the origin of HH~529. In this sense, it is a well-known fact that HH~529 is a source of IR emission \citep{Robberto05,smith05}, emitting strongly at $10\mu \text{m}$ and $11.7\mu \text{m}$. \citet{smith05} show that the $11.7\mu \text{m}$ radiation arises from thermal dust emission and is visible both behind the leading bow shock and within the jet body of HH~529 (see their Fig.~7). After analysing different scenarios, \citet{smith05} conclude that the dust may be entrained from the origin of the jet, which implies that at least part of the ejected material comes from a radius larger than the sublimation radius in the accretion disc of the source. Some of this material may originally come from H-deficient solids. However, the mechanism of expulsion of this H-deficient material and its interaction with the ambient gas requires a deeper analysis of solid body destruction in new formed stars, an idea further explored in planetary nebulae \citep{henney10}. \citet{tsamis11} obtained deep optical integral field spectroscopy of the LV2 proplyd in The Orion Nebula, determining, for the first time, the chemical composition in this kind of objects. They find that the abundances of O, C and Ne in the ionized gas of the proplyd are between 0.11 and 0.52 dex higher than in the rest of the nebula. This result is somehow qualitatively consistent with the overmetallicity we find for the bow-shocks of HH~529, supporting the possibility that the entrained material of the HH objects may come from the source of the gas flow. However, the abundance pattern found for LV2 is not confirmed in the proplyd HST~10, where the chemical composition is not substantially different from the nebular gas \citep{tsamis13}. 

 Finally, we should keep in mind that the apparently larger metallicity of the HH objects may be simply produced by the atomic data used for the analysis. As \citet{JuandeDios17} have discussed, uncertainties in the atomic data may be more important for high-density objects (densities above $10^4\text{ cm}^{-3}$) because there is less possibility to check them observationally. Subsequent analysis of new photoionized HH objects, whose analysis we are carrying out, can shed further light on this issue.

\section{kinematical analysis from UVES data}
\label{sec:kin_analisys}

We calculate the radial velocity of each line in the heliocentric framework by comparing its observed wavelength (after applying the radial velocity correction) with its theoretical  wavelength in air. All the theoretical values have been taken from the Atomic Line List v2.05b21 \citep{vanhoof18}. Wavelengths from this compilation list are mainly calculated from the theoretical energy difference between the levels connected by the transition. The exception are the hydrogenic lines, which include a weighted average of all the fine structure components. 

We detect some evident inaccuracies in the theoretical wavelengths of \mbox{[Cl}\thinspace \mbox{III]}, \mbox{[Cl}\thinspace \mbox{IV]} and \mbox{[Ne}\thinspace \mbox{III]} in the Atomic Line List v2.05b21. This conclusion is based on the discrepant velocities that those lines show with respect to the rest of lines in the high-velocity components, that show fairly similar velocities independently of the ionization state of the ions and elements (see Section~\ref{subsec:vel_struc}). For example, in the case of \mbox{[Ne}\thinspace \mbox{III]} $\lambda \lambda$3869, 3967, the Atomic Line List v2.05b21 gives $\lambda \lambda$3869.07$\pm$0.09 and 3967.79$\pm$0.10 based on the works of \citet{Persson91} and \citet{Feuchtgruber97}. These wavelengths give velocities about $-$20 km s$^{-1}$ displaced with respect to the mean velocity obtained for the rest of the lines. In this case, we decided to adopt the wavelengths $\lambda \lambda$3868.75 and 3967.46 obtained by \citet{Bowen55} from high-resolution spectroscopy of nebulae. The \mbox{[Cl}\thinspace \mbox{III]} and \mbox{[Cl}\thinspace \mbox{IV]} lines show a similar problem;  in this case, we adopt the reference wavelengths used by \citet{Esteban04} that give consistent velocities. The wavelengths adopted for \mbox{[S}\thinspace \mbox{III]} lines deserve special attention. The values given by the Atomic Line List v2.05b21 are $\lambda \lambda$6312.1$\pm$0.36, 8829.4$\pm$0.49, 9068.6$\pm$0.52 and 9530.6$\pm$0.57, taken from the work by \citet{Kaufman93}. There is a small (but noticeable at our spectral resolution) discrepancy in the velocity obtained for [S\thinspace III] $\lambda$6312 and the rest of the lines of about 10 km s$^{-1}$. Assuming the velocities measured for the H\thinspace I lines of HH~529~II, our best estimation of the rest wavelengths of the observed \mbox{[S}\thinspace \mbox{III]} lines are: $\lambda\lambda$6312.07$\pm$0.01, 8829.70$\pm$0.01, 9068.93$\pm$0.04 and 9530.98$\pm$0.01. 

\subsection{Radial velocity structure}
\label{subsec:vel_struc}

\begin{figure*}
\centering
\includegraphics[width=\textwidth]{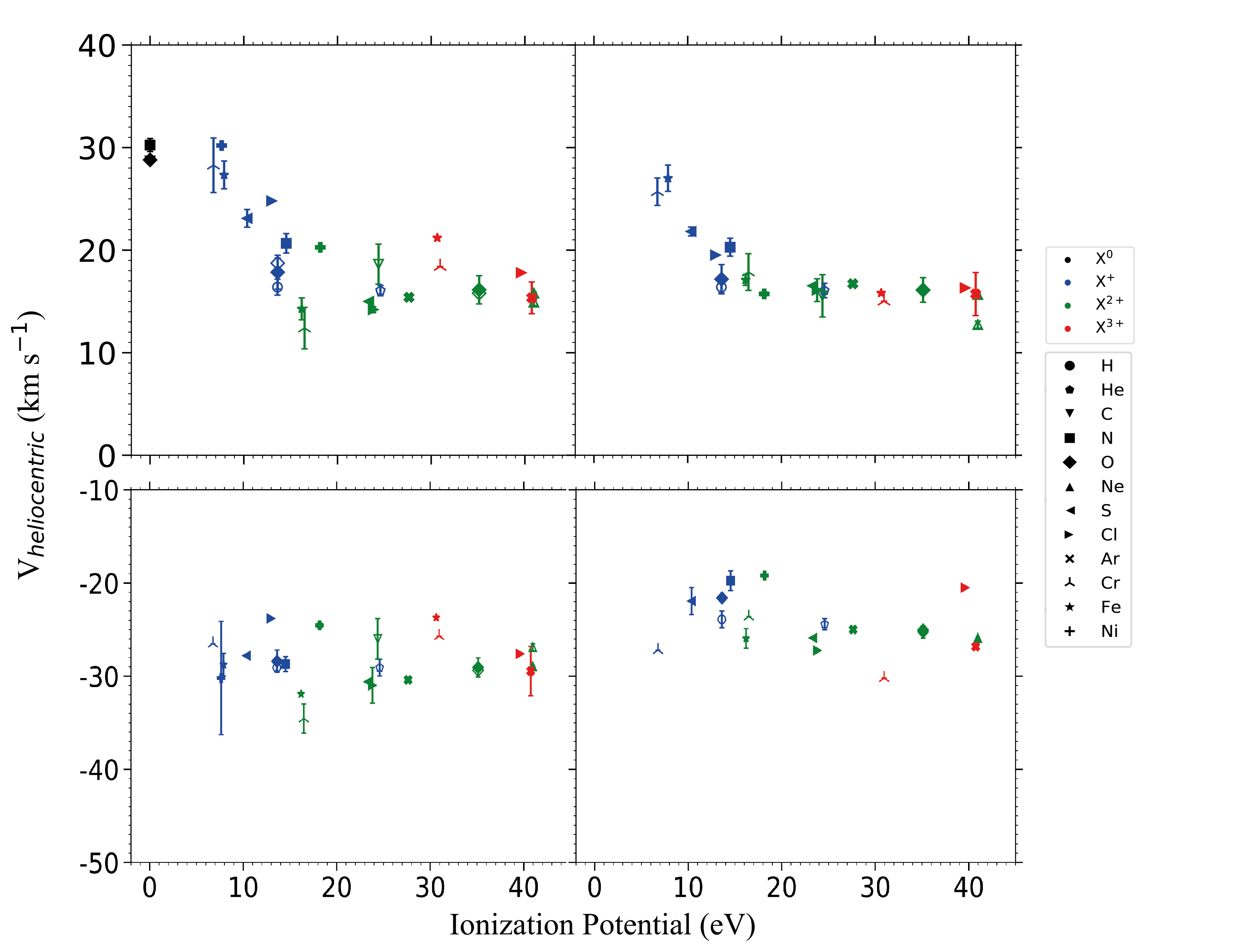}
\caption{Observed radial velocity of the ions as function of the ionization potential. The upper left, bottom left and bottom right panels correspond to the radial velocities derived by considering the rest-frame reference wavelength $\lambda_0$ and the observed one $\lambda$ in the heliocentric frame of reference for the nebular component of the cut 2, HH~529~II and HH~529~III, respectively. The upper right panel is defined with the difference of velocities between the nebular component and HH~529~II (subtraction of the upper and lower left panels) and rescaled using the velocity determined for H\thinspace I lines, whose rest-frame reference wavelengths $\lambda_0$ are the best determined ones.}
\label{fig:kin}
\end{figure*}

In Table~\ref{tab:kin_tab}, we present the average velocity and full width at half maximum (FWHM) of each ion observed in the nebular component of cut 2 and in HH~529~II and III. The behaviour of the nebular component of cut 2 is representative of what is observed in the nebular components of the other cuts. In each column, we include in parentheses the number of lines of each kind whose values have been averaged. In this analysis, we discard lines with known blends and those affected by ghosts or by telluric emissions/absorptions. For \mbox{O}\thinspace \mbox{I}, \mbox{O}\thinspace \mbox{II}, \mbox{C}\thinspace \mbox{II}  and \mbox{Ne}\thinspace \mbox{II} lines, we include only those used in Section~\ref{subsec:ionic_from_RLs} for abundance determinations, which are the lines that are assumed to be produced by pure recombination and are most probably not affected by fluorescence. In the special case of \mbox{[S}\thinspace \mbox{III]} lines we consider only the $\lambda 6312$ line, due to the aforementioned evident inaccuracies in the theoretical wavelengths of the rest of the \mbox{[S}\thinspace \mbox{III]} lines. Fig.~\ref{fig:kin} shows the heliocentric velocity as a function of ionization potential relation for the data collected in Table~\ref{tab:kin_tab}. 

From the upper left panel of Fig.~\ref{fig:kin}, it is clear that the nebular component presents a pattern consistent with the ``blister'' model for the Huygens Region of the Orion Nebula \citep[and references therein]{Odell01,Ferland01,Odell20}. The basic idea is that a layer of gas of the Orion Molecular Cloud (OMC) facing the direction towards the Sun is ionized by $\theta^{1}$ Ori C, which is located in the foreground of OMC. As the gas gets ionized, it is accelerated towards the observer. Velocities of \mbox{[O}\thinspace \mbox{I]}, \mbox{[C}\thinspace \mbox{I]}, and \mbox{[N}\thinspace \mbox{I]} are similar to the average velocity of the molecules in the OMC of $\sim 28$ km s$^{-1}$ \citep[][and references therein]{goudis82, odell18}, then a rapid drop in the observed velocity (which means an increase in velocity compared to the OMC's systemic velocity) of the ions whose ionization potential are between 6.77 and 13.6 eV is observed as well as a constant velocity after 13.6 eV. This behaviour has been observed in previous works  \citep{Kaler67, Fehrenbach77, odell92, Esteban99}. 

In the lower panels of Fig.~\ref{fig:kin}, we present the observed radial velocity of the ions as a function of their ionization potential for the high-velocity components: HH~529~II (lower left panel) and HH~529~III (lower right panel). Contrary to what the nebular components show, all the ions of the high-velocity components show a fairly constant radial velocity, independently of their ionization potential. Moreover, the high-velocity components do not show emission lines of neutral ions. These features are  consistent with the scenario of a fully ionized slab of gas moving at a different velocity with respect to the rest of the nebular gas.

In the upper right panel of Fig.~\ref{fig:kin}, we present the difference between the radial velocity pattern of the nebular component and HH~529~II (the subtraction of the upper and lower left panels of Fig.~\ref{fig:kin}) rescaled using the average radial velocity of 51 \mbox{H}\thinspace \mbox{I} lines in HH~529~II, whose rest-frame reference wavelengths $\lambda_0$ are the best determined among all the ions. Doing that  subtraction, we can see that the  dispersion of the data points represented initially in the upper left panel decreases substantially. This fact indicates that the dispersion is not due to errors associated with the determination of the wavelength of the lines or to a complex velocity structure, but to inaccuracies in the adopted rest-frame reference wavelengths or possibly in the wavelength calibration. By eliminating the aforementioned dispersion, we can demonstrate that the acceleration of the gas in the Orion Nebula is constant for  ionization potential between 6.77 and 13.6 eV, becoming zero for energies above, reaching a constant velocity of $16.4 \pm 0.8$ km s$^{-1}$.

There is a difference of $5.18 \pm 1.25$ km s$^{-1}$ between the radial velocities of HH~529~II and III. This is due to the presence of unresolved lower velocity components in HH~529~III, as it is shown in Fig.~\ref{fig:proper-motions}, which also contribute to increase the dispersion in the radial velocity structure of this component.

The radial velocities of the selected  \mbox{O}\thinspace \mbox{I}, \mbox{O}\thinspace \mbox{II} and \mbox{Ne}\thinspace \mbox{II} lines are practically the same as those of \mbox{[O}\thinspace \mbox{II]}, \mbox{[O}\thinspace \mbox{III]} and \mbox{[Ne}\thinspace \mbox{III]} CELs, respectively, in the nebular component. This reinforces the assumption that they are produced by pure recombination. For example, lets consider the \mbox{O}\thinspace \mbox{I} RLs from multiplet 1, which come from quintuplet levels. If these lines were  produced by fluorescence, they would be  emitted in neutral and partially ionized zones of the nebula and should show a radial velocity similar to the systematic one of the OMC. In fact, \mbox{O}\thinspace \mbox{I} lines from transitions between triplet levels (such as multiplet 4 $\lambda \lambda 8446.25,8446.36,8446.76 $) are displaced around $\sim10 \text{ km s}^{-1}$ with respect to the velocity of the aforementioned multiplet 1, clearly indicating the different nature of both multiplets, with the lines of multiplet 4 originating in starlight excitation \citep{grandi75b}.

\subsection{Electron temperature from thermal broadening of the line profiles}
\label{subsec:line_widths}

The observed line widths are the result of several physical processes. Apart from the instrumental width, $\sigma_{\text{ins}}$, the main contributors are the thermal width, $ \sigma_{\text{th}}$, the fine structure broadening, $ \sigma_{\text{fs}}$ and the non-thermal contribution, $\sigma_{\text{nt}}$, which includes effects such as turbulence and any other additional broadening process. Following \citet[][hereinafter GHLD08, see their equation 2]{garciadiaz08}, we use  Eq.~\eqref{eq:widths} to express the relationship commented above. 

\begin{equation}
    \label{eq:widths}
    \sigma_{\text{obs}}^{2}=\sigma_{\text{th}}^{2}+\sigma_{\text{fs}}^{2}+\sigma_{\text{ins}}^{2}+\sigma_{\text{nt}}^{2}.
\end{equation}{}

The thermal contribution of Eq.~\eqref{eq:widths} is the Doppler broadening and depends linearly on the temperature, $\sigma_{\text{th}}^{2}$ = 82.5 $T_4 /A$ (km s$^{-1}$)$^{2}$, where $A$ is the atomic weight of the emitting ion and $T_4 = T_{\text{e}}/10^{4}$ (GHLD08). 

In principle, using Eq.~\eqref{eq:widths} we can estimate $T_{\rm e}$ from the subtraction of the observed widths of \mbox{H}\thinspace \mbox{I} and \mbox{[O}\thinspace \mbox{III]} lines. The instrumental width affects the same for both kinds of lines and should be cancelled in the subtraction. GHLD08 estimated $\sigma_{\text{fs}}^{2}\left(\mbox{H}\thinspace \mbox{I}\right)$ = 10.233 (km s$^{-1}$)$^{2}$, finding that  $\sigma_{\text{fs}}^{2}\left(\mbox{[O}\thinspace \mbox{III]}\right)$ is negligible. On the other hand,  $\sigma_{\text{nt}}^{2}\left(\mbox{H}\thinspace \mbox{I}\right)$ is not strictly equal to $\sigma_{\text{nt}}^{2}\left(\mbox{[O}\thinspace \mbox{III]}\right)$, since the nebular volume occupied by both ions is different.  GHLD08 define $f$ as the fraction of the volume of H$^{+}$ occupied by O$^{2+}$ and (1-$f$) the fraction filled by O$^+$ and other ions with lower degree of ionization, as N$^{+}$. We assume the average value $\langle f \rangle=0.76$ estimated by GHLD08 for the Orion Nebula. Using equations 7 to 10 from GHLD08, we obtain  $T_{\text{e}} = 8340 \pm 410$ k for the nebular component. In the case of the high-velocity components, we assume $f=1.0$, obtaining $T_{\text{e}} = 8670 \pm 50$ K and $T_{\text{e}} = 10470 \pm 790$ K for HH~529~II and HH~529~III, respectively.

The resulting $T_{\rm e}$ values in the nebular component of cut 2 and HH~529~II are in remarkably good agreement with $T_{\rm e}$(\mbox{[O}\thinspace \mbox{III]}) from CEL ratios as shown in Table~\ref{tab:pc}. In the case of HH~529~III, the large difference between the values obtained from both methods may be due to the contamination by several unresolved velocity components, as it is shown in Fig.~\ref{fig:proper-motions} and discussed in Section~\ref{sec:proper-motions-hh}, that broadens the lines, providing overestimated temperatures.

\section{Proper motions of HH~529 II and III}
\label{sec:proper-motions-hh}

\begin{figure}
  \centering
  \includegraphics[width=\linewidth]{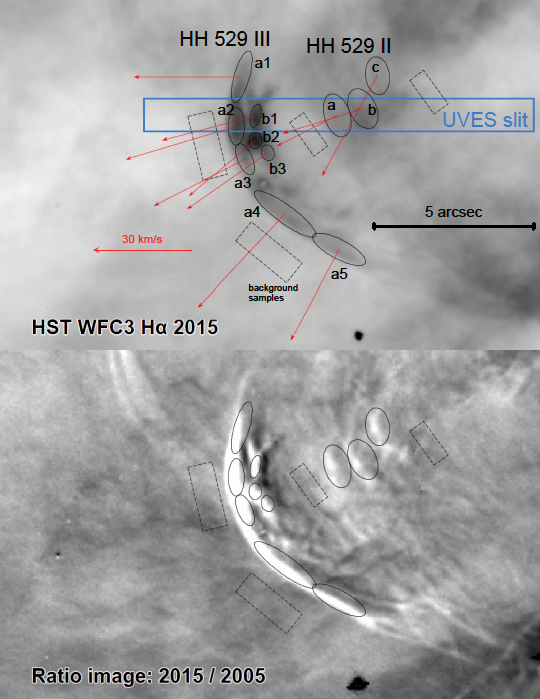}
  \caption{
    Tangential velocities of shock features in HH~529~II and III
    derived from 3 epochs of HST imaging.
    Upper panel shows various discrete features identified in the bow shocks
    (black ellipses with arrows indicating the average proper motion of each feature).
    Small dashed rectangles indicate regions where the nebular background brightness was measured
    and the large blue rectangle shows the position of the spectrograph slit.
    The backgound negative grayscale shows an \textit{HST} WFC3 image in the F656N filter from 2015.
    Lower panel shows the ratio between the 2015 image and an \textit{HST} ACS image
    in the F658N filter from 2005 (white means brighter in 2015).
    This highlights the changes in the nebula over that 10-year period,
    which are principally due to motions of the shocked gas.
  }
  \label{fig:proper-motions}
\end{figure}

The plane-of-sky motions of the bow shocks in HH~529 have been previously reported
in Table~3 of \citet{odellyhenney08} and in sec~3.3.1.3 of \citet{Odell15}.
However, the reported tangential velocities are very disparate,
so we have re-measured the proper motions, using HST imaging over 20~years as described in Section~\ref{sec:data}. The 1995 and 2015 images were aligned to the 2005 ACS image using Astrodrizzle%
\footnote{\url{https://drizzlepac.readthedocs.io}}
and rebinned to the ACS pixel scale of \SI{0.05}{arcsec}.
The 2005 image itself has been aligned to the absolute astrometric reference of 2MASS,
as painstakingly described in sec~3.3 of \citet{Robberto:2013a}. Proper motions are estimated for the two intervals, 1995--2005 and 2005--2015,
using the Fourier Local Correlation Tracking (FLCT) method
\citep{Welsch:2004a, Fisher:2008a}\footnote{
  We used version 1.07 of FLCT, obtained from \url{http://cgem.ssl.berkeley.edu/cgi-bin/cgem/FLCT/home},
  together with version 1.04 of the Python wrapper pyflct,
  obtained from \url{https://github.com/PyDL/pyflct}.}
with a kernel width of 10~pixels (\SI{0.5}{arcsec}).
For an assumed distance of \SI{417}{pc},
a shift of 1~pixel in 10~years corresponds to approximately 10 km s$^{-1}$.
A potential disadvantage of using the ACS data in this study
is that the F658N ACS filter is relatively broad and includes both
H\(\alpha\) \Wav{6563} and \nii{} \Wav{6583},
whereas the WFPC2 and WFC3 F656N filters are narrower and more effectively isolate \Wav{6563}.
Ionization gradients in the nebula can therefore contribute to differences in the images obtained,
which would obscure the signal due to the gas motions.
However, the degree of ionization in HH~529~III and II is so high that this turns out not to be an issue in this object. Results are presented in Table~\ref{tab:proper-motions} and Figure~\ref{fig:proper-motions}.

We find that HH~529~III consists of at least two distinct moving structures.
The large outer curved bow, which we call III~a, is relatively smooth,
spanning about \SI{7}{arcsec} in its brightest part,
but with fainter wings (best visible on the ratio image) that extend farther.
We cover the bright part of the bow with 5 sample ellipses: a1 to a5,
where a3 seems to be the apex of the bow but a2 is the one that falls in the UVES slit.
Roughly \SI{0.7}{arcsec} to the east of III~a is a smaller, knottier bow, which we call III~b and cover with 3 sample ellipses: b1, b2 and b3.
The brightest knot is b2, but it is the b1 sample that falls in the UVES slit.
HH~529~II is found to consist of three distinct bows with separations of order \SI{1}{arcsec}, which we call II~a, II~b, and II~c, with II~a and II~b falling in the UVES slit.

Considering the OMC reference frame \citep[$ 28 \pm 2$ km s$^{-1}$,][]{goudis82}, we obtain $v_{\text{tot}}$ values of $62.68 \pm 3.30$ km s$^{-1}$ and $67.01 \pm 3.09$ km s$^{-1}$ for HH~529~III and HH~529~II, respectively. To obtain those numbers we have used the weighted average tangential motion of $35 \pm 5 \text{ km s}^{-1}$, calculated with components a, b, a2 and b1 of HH~529 (see Fig.~\ref{fig:proper-motions}) and the radial velocities shown in Table~\ref{tab:kin_tab}). On the other hand, using the ratio between the average tangential and radial velocities and considering a systematic uncertainty of $\sim 2 \text{ kms}^{-1}$ between the OMC and the stellar source, we estimate a flow angle of $58 \pm 4^\circ$ for HH~529. This value is in agreement with the result of \citet{odellyhenney08} ($\sim$54$^\circ$) but discrepant with the value obtained by \citet{Odell15} ($\sim$83$^\circ$).

\begin{table}
  \caption{Proper motions of shock features}
  \label{tab:proper-motions}
  \begin{tabular}{lccccc}
    \toprule
     & UVES & {\(V_\text{t}\)} & {PA} & {Contrast}\\
    Feature & cut & {\si{km.s^{-1}}} & {deg} & {\(S(\mathrm{H\alpha}) / S(\mathrm{H\alpha,BG})\)} \\
    \midrule
    {(1)} & {(2)} & {(3)} & {(4)} & {(5)}  \\
    \addlinespace 
    HH~529~III a1 & & $33 \pm 3$ & $90  \pm  3$ & $0.46 \pm 0.14$\\
    HH~529~III a2 &3& $36 \pm 1$ & $107 \pm 7$ & $0.82 \pm 0.11$\\
    HH~529~III a3 & & $32 \pm 1$ & $117  \pm  3$ & $0.52 \pm 0.14$\\
    HH~529~III a4 & & $39 \pm 2$ & $137  \pm  5$ & $0.41 \pm 0.07$\\
    HH~529~III a5 & & $32 \pm 1$ & $152  \pm 10$ & $0.27 \pm 0.07$\\
    \addlinespace   
    HH~529~III b1 &3& $30 \pm 2$ & $105  \pm 3$ & $0.95 \pm 0.09$\\
    HH~529~III b2 & & $27 \pm 1$ & $130  \pm 1$ & $1.14 \pm 0.10$\\
    HH~529~III b3 & & $30 \pm 2$ & $125  \pm 4$ & $0.64 \pm 0.07$\\
    \addlinespace
    HH~529~II a &2& $21 \pm 9$ & $117  \pm 58$ & $0.25 \pm 0.03$\\
    HH~529~II b &2& $26 \pm 5$ & $107  \pm  4$ & $0.35 \pm 0.08$\\
    HH~529~II c & & $35 \pm 9$ & $151  \pm 87$ & $0.22 \pm 0.03$\\
    \bottomrule
    \addlinespace
    \multicolumn{5}{@{}p{\linewidth}@{}}{
    \textsc{Columns:}
    (1)~Name of shock feature (see Fig.~\ref{fig:proper-motions} for positions).
    (2)~Spatial cut of the UVES spectrum where this feature appears, if any.
    (3)~Mean tangential velocity for each feature,
    weighted by background-subtracted surface brightness,
    \(S(\mathrm{H\alpha})\), of each pixel.
    (4)~Mean position angle of proper motion,
    weighted in the same way.
    (5)~Mean relative H\(\alpha\) brightness with respect to nebular background (BG).
    For columns 3, 4, and 5, the \(\pm\) uncertainties correspond to the
    root-mean-square variation over each sample region and do not include systematic uncertainties,
    which are of order \SI{2}{km.s^{-1}}
    }
\end{tabular}
\end{table}

\section{Physical aspects of the high velocity components}
\label{sec:dym_impl}

Since the material in the HH outflows is moving highly supersonically with respect to the ionized sound speed in the nebula, it will give rise to shocks where the flow and nebula interact \citep{Hartigan:1987a}. Further internal shocks may form inside the outflow if its velocity varies with time \citep{Raga:1990a}. It is important to investigate the degree to which direct excitation by the shocks might be affecting our emission line analysis.

In this section, we first calculate the heating and compression expected behind a shock wave and then we use results of non-equilibrium Cloudy simulations to predict the relative contributions of the post-shock cooling zone and the equilibrium photoionized shell to the emission line spectrum of the knots.

\subsection{Shock compression and heating}
\label{subsec:shock-compr-heat}

A non-magnetised hydrodynamic shock is characterized by its Mach number \(\Mach = V\shock / \sound\), where \(V\shock\) is the shock velocity and \(\sound\) is the pre-shock adiabatic sound speed. On passing through the shock, the gas is heated \citep{ZelDovich:1967a} to a temperature \(T_1\), which is higher than the equilibrium photoionized temperature, \(T_0\):

\begin{equation}
  \label{eq:T1-T0}
  \frac{T_1}{T_0} = \frac{1}{16} \bigl( 5 \Mach^2 - 1 \bigr)
  \bigl( 1 + 3\Mach^{-2} \bigr),
\end{equation}

while at the same time it is compressed by a factor
\begin{equation}
  \label{eq:rho1-rho0}
  \frac{\rho_1}{\rho_0} = \frac{4 \Mach^2}{\Mach^2 + 3} .
\end{equation}
In both cases, a ratio of specific heats \(\gamma = 5/3\) is assumed, as is appropriate for ionized and atomic gas. 
The post-shock gas then cools in a radiative relaxation layer until it returns to the equilibrium temperature \(T_2 \approx T_0\), reaching a final density compression factor of :

\begin{equation}
  \label{eq:rho2-rho0}
  \frac{\rho_2}{\rho_0} = \frac53 \Mach^2 .
\end{equation}

The adiabatic sound speed in the equilibrium ionized gas is given by \(\sound = (\gamma k T_0 / \mu m_{\mathrm{H}})^{1/2}\),where \(k\) is the Boltzmann constant, \(T_0\) is the temperature, \(m_{\mathrm{H}}\) is the hydrogen mass and \(\mu\) is the mean atomic mass per particle. Assuming that all He is singly ionized with \(y = \mathrm{He/H} = 0.087\) (Table~\ref{tab:total_abundances_rls}) yields \(\mu \approx (1 + 4 y) / (2 + 2 y) \approx 0.62\), which combined with \(T_0 = \SI{8480}{K}\) (Table~\ref{tab:pc}) implies an adiabatic sound speed of \(\sim \SI{13.7}{km.s^{-1}}\). 

The case of a magnetized shock is considerably more complicated \citep{Bazer:1959a}, but the principle effect is that the component of the magnetic field, \(B\), parallel to the shock front provides extra pressure support (magnetic cushioning) in the post-shock gas \citep{Hartigan:1994a,Hartigan15}.
An approximate way to account for this is to replace the sound speed in the above equations by the fast magnetosonic speed: \(V_{\text{fast}} = (\sound^2 + V_{\text{A}}^2)^{1/2}\), where \(V_{\text{A}} = B / (4\pi \rho)^{1/2}\) is the Alfvén speed.
The ambient gas inside an \hii{} region is expected to have a low Alfvén speed of \(V_{\text{A}} \approx \SI{2}{km.s^{-1}} \ll \sound\) \citep{Arthur:2011a} so that the magnetic cushioning will be negligible
in shocks propagating in the ambient medium. On the other hand, the Alfvén speed in the jet itself \citep{Hansen:2015b, Pudritz:2019a} may be sufficiently high so as to limit the compression behind shocks driven into the jet.

Considering the \emph{isothermal} Mach number \(\Mach_{\text{s}} = \gamma^{1/2} V\shock /\sound\), we estimate $\Mach_{\text{s}}=5.31 \pm 0.28$ and $\Mach_{\text{s}}=5.68 \pm 0.27$ for HH~529~III and HH~529~II, respectively, on the assumption that both are external working surfaces. By using the estimated densities of the HH objects as post-shock densities, $n_2$, in Eq.~(\ref{eq:rho2-rho0}), we can estimate a value for the pre-shock densities, $n_0$. In the case of HH~529~III, we obtain $n_{0}=1070 \pm 310 \text{ cm}^{-3}$, a value that is consistent with photoionized areas of the Orion Nebula outside the Huygens region, where the density decreases radially \citep{mesadelgado08}. This would place HH~529~III in the foreground of the main ionization front. For HH~529~II, the pre-shock density, $n_{0}=370 \pm 70 \text{ cm}^{-3}$, would indicate that the impact with the ambient gas is taking place in a more external zone from where HH~529~III does, being located farther from the main ionization front (and closer to the observer). However, this scenario seems unlikely.

Considering the complex velocity structure of HH~529~II+III (see Fig.~\ref{fig:proper-motions}), that HH~529~III is almost a factor 3 denser than HH~529~II, that both objects present similar velocities (in the frame of reference of the OMC) and ionization degree, it is likely that  HH~529~II is rather an internal working surface of the jet beam~\citep{Masciadriyraga01}. In this case, the relevant $\Mach_{\text{s}}$ would be subsantially smaller than 5.68. This internal shock may be due to temporal velocity variations in the jet from the common source of HH~529 \citep[possibly located at the star COUP 666, ][]{Odell15}. 

The absence of [O\thinspace I] CELs (see Fig.~\ref{fig:cuts}) and the weakness of lines of low ionization potential ions (e.g. [O\thinspace II] or [N\thinspace II]) demonstrate  that HH~529~II and HH~529~III are fully photoionized objects and do not contain a trapped ionization front \citep{Masciadriyraga01}, contrary to what was found by \citet{mesadelgado09} in the case of HH~202~S. Therefore, the entire jet beam should be  observable in optical emission lines, contrary to partially ionized or neutral HH objects, where only limited regions can be observed in the optical \citep{raga00_proc, raga00_mnras}. Since HH~529 is flowing at an angle of $\sim 58^{\circ}$ with reference to the plane of the sky, the observed spectra of the HH objects should integrate both the compressed gas at the leading working surface and the gas of the jet beam behind. Which of these dominates the total emission depends on the relative densities of the jet and the ambient medium \citep{Hartigan89}. In the case of HH~529~III, it is possible that the two components, a and b, resolved by HST (see Fig.~\ref{fig:proper-motions}) represent respectively the bowshock (shocked ambient medium) and Mach disk (shocked jet). However, these are unresolved in our slit spectra.

\subsection{Shock emission versus shell emission}
\label{subsec:shock-emiss-vers}

In order to estimate the shock contribution to the line emission from the working surface, we are going to consider the high-temperature radiative relaxation layer (cooling zone) that lies immediately behind the shock. 

The post-shock gas in the working surface will be accelerated away from the jet axis by lateral pressure gradients, flowing sideways out of the working surface at the isothermal sound speed through a ``Mach ring'' \citep{Falle:1993a}, with radius approximately equal to that of the jet, \(r\jet\).  In a steady state, the outward mass flux through the Mach ring must be equal to the inward mass flux through the shock, while the isothermal Bernoulli equation shows that the density at the Mach ring is \(e^{-1/2} \rho\ws\), where \(\rho\ws\) is the density of the working surface. This yields the thickness of the working surface as \(H = \frac{1}{2} e^{1/2} r\jet / \Mach_\text{s}\). We assume that \(r\jet = \SI{2}{mpc}\), which is half the observed lateral extent of the HH~529~II knots%
\footnote{At the distance of the Orion Nebula, \(1'' \approx \SI{2}{mpc}\).}. The thickness of the cooling zone, \(d_{\text{cool}}\) is approximately the immediate post-shock velocity multiplied by the cooling time. This implies that, in our case,  \(d_{\text{cool}} / H \approx 0.1\), being thin compared with the total thickness of the working surface.

The contribution of this thin cooling zone should be equal to the kinetic energy flux through the shock:
\begin{equation}
  \label{eq:Fshock}
  F_1 = \frac{1}{2} \rho_0 V_\text{s}^3 \quad \si{erg.cm^{-2}.s^{-1}},
\end{equation}
whereas the radiative flux from the cooled equilibrium gas in the working surface is
\begin{equation}
  \label{eq:Fshell}
  F_2 = n_2^2 \Lambda_0 H \quad \si{erg.cm^{-2}.s^{-1}},
\end{equation}
where $\Lambda_0$ is the cooling coefficient of the equilibrium gas. Assuming \(\Lambda_0 = \SI{2.5e-24}{erg.cm^3.s^{-1}}\)\citep{osterbrock06}, the ratio of Eq.~\ref{eq:Fshock} and Eq.~\ref{eq:Fshell} is then:

\begin{equation}
  \label{eq:ratio-shock-shell}
\frac{F_1}{F_2}=195 \times \Mach_{\text{s} }^2  \times \left(\frac{n_2}{\text{cm}^{-3}}\right)^{-1} \times \left( \frac{r_\text{jet}}{ \text{mpc}} \right)^{-1}.
\end{equation}

We find that \(F_1 / F_2 \approx 0.1\) for HH~529~III, whereas for HH~529~II, assuming that HH~529~II moves at $20 \text{ km s}^{-1}$ with respect to the internal velocity of the jet beam, \(F_1 / F_2 \approx 0.03\). However, it should be noted that, the \(F_1 / F_2\) value found for HH~529~III is rather an upper limit to the real contribution of the cooling area since we have used the total velocity with respect to the OMC to define the Mach number, while a shock within the photoionized gas of the Orion Nebula must consider its internal velocity structure, which moves radially towards the observer (see Sec.~\ref{subsec:vel_struc}), as well as partially does HH~529. In addition to this, although HH~529~III is observed as a prominent arch, the possibility that it is preceded by other shocks cannot be ruled out. As recent analyzes have shown, there is at least one high ionization shock to the east of HH~529~III that may be related to the gas flow of HH~529 \citep[Labeled as ``East Shock'' in ][]{Odell15}, in which case it would imply a lower Mach number than the one used.

From an observational point of view, the spectrum of HH~529~III have logarithmic values of $I(\text{ [O\thinspace III] }\lambda 5007)/I(\text{H}\beta) = 0.69$ and $I(\text{ [N\thinspace II] }\lambda 6584)/I(\text{H}\alpha) = -1.34$, while HH~529~II have $I(\text{ [O\thinspace III] }\lambda 5007)/I(\text{H}\beta) = 0.69$ and $I(\text{ [N\thinspace II] }\lambda 6584)/I(\text{H}\alpha) = -1.60$. These values are in complete agreement with those expected in star-forming regions, as shown by the curve derived by \citet{kauffmann03} in their Eq.~(1), based in photoionization models. Furthermore, the values of $\text{log}(I(\text{ [S\thinspace II] }\lambda 6716+31)/I(\text{H}\alpha))$ in both components are $\sim -2.45$, also consistent with the typical values of ionized nebulae and very far from the range of values observed in objects ionized by shock energy (between -0.5 and 0.5), such as supernova remnants and non-photoionized HH-objects \citep{canto81,riera89}.  
 
Finally, a value of \(F_1 / F_2 \approx 0.1\) in HH~529~III (i.e. a Mach number $\sim 5$) would imply a post-shock temperature as high as \SI{70000}{K} (see Eq.~\ref{eq:T1-T0}) in the thin cooling zone. At such high-temperature, the gas emits predominantly at far-ultraviolet (FUV) wavelengths, via lines such as $\text{C\thinspace III] }\lambda 1909$ and $\text{C\thinspace IV }\lambda 1549$, whereas most of the optical lines studied in this paper are minor coolants in such conditions. Therefore, even in this extreme case, the optical spectra analyzed in this work would have a contribution probably much smaller than a factor of 0.1. 

We have calculated time-dependent photoionized shock models using the Cloudy plasma physics code \citep{Ferland:2013a, Ferland:2017a} in order to investigate this, taking the $\text{[O\thinspace III] }\lambda 5007$ line as a typical example. We find that for gas in photoionization equilibrium with the radiation field from the Trapezium stars, a fraction \(f_{5007} = 0.33\) of the total radiative cooling is due to this line. For the cooling zones behind low-velocity shocks, this fraction initially increases slightly with shock velocity, reaching a maximum value \(f_{5007} = 0.37\) for \(V_\text{s} \approx \SI{20}{km.s^{-1}}\), but subsequently declines, falling to \(f_{5007} = 0.11\) for \(V_\text{s} \approx \SI{70}{km.s^{-1}}\) as the FUV lines take over the majority of the cooling. The 5007-emission-weighted average temperature of the cooling zone saturates at \(T \approx \SI{16000}{K}\) for  \(V_\text{s} > \SI{50}{km.s^{-1}}\), indicating that the conditions in the post-shock [\ion{O}{3}] emission zone become insensitive to the shock velocity. For recombination lines such as H\(\alpha\), the emissivity declines with increasing temperature, so that the contribution of the post-shock zone to the total emission is negligible. On the other hand, for some weak lines we would expect a relatively larger shock-excited contribution. For instance, an auroral line such as $\text{[O\thinspace III] }\lambda 4363$ has an excitation temperature that is roughly double that of $\lambda 5007$, so will be relatively enhanced in the cooling zone. However, in both high-velocity components, $T_{\rm e}(\text{[O\thinspace III]})$ is consistent with the values of the Orion Nebula, which implies that there are no significant effects. Although in our case, all the indicators used show that the shock contribution in the observed optical spectra of HH~529~II and HH~529~III is negligible, the detailed effects of post-shock cooling on an optical spectrum clearly deserves a deeper study.

\section{Summary and Conclusions}
\label{sec:summary}

We have observed two of the bow shocks of the photoionized Herbig-Haro object HH~529: HH~529~II and HH~529~III, inside the Orion Nebula, under photometric conditions with the UVES echelle spectrograph at VLT. Our observations have an effective spectral resolution of $\sim$ 6.5 km s$^{-1}$ and cover a spectral range of 3100-10420\AA. We defined 4 spatial cuts in our 10 arcsecs-long slit to separate HH~529~II from HH~529~III. Our spatial and spectral resolution permitted us to resolve the blueshifted high-velocity components of HH~529~II and III from the nebular emission of the Orion Nebula. Thus, we analysed 6 1D spectra: 4 of the nebular emission and one of each bow shock. We measured up to 633 emission lines in the Orion Nebula and 376 and 245 in the cases of HH~529~II and III, respectively. We defined an additional spectrum labeled as ``combined cuts'' with the sum of all components to study the impact of the HH objects in a single low-spectral resolution longslit observation. We also take advantage of the 20 years of archival $HST$ imaging to analyse proper motions of HH~529 and their physical impact. 

Considering the absence of emission lines of neutral elements such as [O\thinspace I] and the rather faint emission from low ionization ions such as [O\thinspace II] in HH~529~II and III, we conclude that they are fully photoionized and present a high ionization degree. We analyze the influence of the shock energy in our spectra of HH~529~II and HH~529~III and all evidences indicate a certainly minor contribution, presenting values consistent with normal H\thinspace II regions. We analysed 5 density diagnostics based on CEL ratios. We found a mean value of $n_{\rm e}\approx 6000 \pm 1000 \text{ cm}^{-3}$ for the nebular components. We obtain $n_{\rm e}=11880 \pm 1860 \text{ cm}^{-3}$ and $n_{\rm e}=30200 \pm 8080 \text{ cm}^{-3}$ for HH~529~II and III, respectively, concluding that at so high values, the density diagnostics based on [Fe\thinspace III] lines are more suitable than the usual ones. We determined the density of each component using the RLs of multiplet 1 of O\thinspace II, finding that it does not differ from the one derived from CELs in the nebular components. However, the results based on O\thinspace II RLs are inconsistent with other diagnostics in the case of HH~529~II and III.

Using CEL ratios, we study 6 $T_{\rm e}$-diagnostics. We derive mean values of $T_{\rm e}\text{(low)}\approx 10000 \pm 200 \text{ K}$ and $T_{\rm e}\text{(high)}\approx 8500 \pm 150 \text{ K}$ for the temperature of the low and high ionization degree zones, respectively, in the nebular components. For HH~529~II we obtain $T_{\rm e}\text{(low)}= 10150^{+570} _{-510}\text{ K}$ and $T_{\rm e}\text{(high)}= 8270\pm 110\text{ K}$. For HH~529~III, we derive $T_{\rm e}\text{(low)}= 11040^{+920} _{-970}\text{ K}$ and $T_{\rm e}\text{(high)}= 8630 \pm 120\text{ K}$. These results indicate that the temperature are very similar in the nebular and the high-velocity components, although there is a slight increase in temperature in the shock front (HH~529~III). For the nebular component of cut 2, we were able to estimate $T_{\rm e}$ from O\thinspace II RLs, obtaining  $T_{\rm e}(\text{O}\thinspace \text{II})=9350 \pm 1090$. The good agreement between $T_{\rm e}(\text{O}\thinspace \text{II})$ and $n_{\rm e}(\text{O}\thinspace \text{II})$ with the physical conditions obtained with CELs in the nebular components demonstrate that the emission of CELs and RLs of O$^{2+}$ come basically from the same gas, ruling out the possibility of cold clumps dominating the emission in RLs. For all components, we derive $T_{\rm e}$(He\thinspace I) from He\thinspace I RL ratios while in the ``combined cuts'' spectra we were also able to derive $T_{\rm e}$(H\thinspace I) using both the Balmer and the Paschen discontinuities of the nebular continuum. In all the derived physical conditions, we found no significant deviation between the results of the ``combined cuts'' spectrum and the individual nebular ones. We conclude that the emission of HH~529~II and HH~529~III do not alter the physical conditions and abundances derived from low spectral resolution spectra in areas of size of the order or larger than 10 arcsecs. 

Based on the different temperature determinations available  and following the Peimbert's $t^2$-formalism, we estimate $t^2_\text{high}=0.021 \pm 0.003$, $t^2_\text{inter}=0.051 \pm 0.009$ and $t^2_\text{low}=0.064 \pm 0.011$ for the high, intermediate, and low ionization zones both for the nebular and the high-velocity components. We derive ionic abundances of O$^{+}$, N$^{+}$, S$^{+}$, Cl$^{+}$, Ni$^{2+}$, Fe$^{2+}$, S$^{2+}$, Cl$^{2+}$, O$^{2+}$, Ne$^{2+}$, Ar$^{2+}$, Cl$^{3+}$, Ar$^{3+}$ and Fe$^{3+}$ based on CELs both in the case of $t^2=0$ and $t^2>0$. We estimated ionic abundances of He$^{+}$, O$^{+}$, O$^{2+}$, C$^{2+}$ and Ne$^{2+}$ based on RLs. The mean ADF values for the nebular components are $\text{ADF}(\text{O}^{+})\approx 0.50 \pm 0.13 \text{ dex}$, $\text{ADF}(\text{O}^{2+})\approx 0.20 \pm 0.05 \text{ dex}$, $\text{ADF}(\text{C}^{2+})\approx 0.50 \pm 0.03 \text{ dex}$ and $\text{ADF}(\text{Ne}^{2+})\approx 0.35 \pm 0.10 \text{ dex}$. For HH~529~II, we obtained $\text{ADF}(\text{O}^{2+})= 0.29 \pm 0.10 \text{ dex}$ and $\text{ADF}(\text{Ne}^{2+})= 0.79 \pm 0.09 \text{ dex}$ while for HH~529~III we obtained $\text{ADF}(\text{O}^{2+})= 0.36 \pm 0.11 \text{ dex}$. The $t^2$ values we obtained are capable of account for the ADF(O$^{2+}$) in the nebular components but not in the ones corresponding to HH~529~II and III, where larger $t^2$ values would be needed to reproduce their ADF(O$^{2+}$). 

We estimate the total abundances of O, Cl, Ar and Fe without ICFs in the nebular components. 
In the cases of HH~529~II and III, in addition to the aforementioned elements (except Fe in HH~529~III), we were also able to estimate the total abundances of He and C without using an ICF. By using the Solar value of Fe/O as reference, we estimate that 6\% of the total Fe is in gaseous phase in the nebular components while in HH~529~II this fraction reaches 14\% and between 10\% and 25\% in HH~529~III. This increase should be due to destruction of dust grains in the shock fronts. We found a slight overabundance of heavy elements (around 0.12 dex) in the high-velocity components that can not be entirely due to dust destruction processes since it affects also the noble gases. We speculate that its possible origin may lie in the inclusion of H-deficient gas entrained after the evaporation of material in the outer part of the protoplanetary disc of the source of HH~529. 

We found a constant value of the radial velocity of the emission lines, irrespective of the ionization potential of the observed ions in HH~529~II ($v_r=-29.08 \pm 0.36 \text{ km s}^{-1}$) and HH~529~III ($v_r=-23.90 \pm 0.89 \text{ km s}^{-1}$). In the nebular emission of the Orion Nebula, we demonstrate that the velocity varies linearly with the ionization potential for ions with ionization potential between 6.77 and 13.6 eV, reaching a constant velocity of $16.4 \pm 0.8 \text{ km s}^{-1}$ for ionization potentials greater than 13.6. From the thermal broadening of the line profiles, we derived $T_{\rm e}=8340 \pm 410$, $T_{\rm e}=8670 \pm 50$ and $T_{\rm e}=10470 \pm 50$ for the nebular components, HH~529~II and III, respectively. The determination for HH~529~III is anomalously large due to the contamination by unresolved velocity components.

We determined the proper motions of HH~529~II and III by using HST imaging over 20 years. We found several discrete features identified in the bow shocks. We estimated an average tangential velocity of $35 \pm 5 \text{ km s}^{-1}$ for the HH~529~II-III system. We also estimate a flow angle with respect to the sky plane of $58 \pm 4^\circ$. Several indicators evidence that HH~529~II corresponds to an internal working surface of the jet beam.


\section*{DATA AVAILABILITY}
The lines measured in the spectra are entirely available in online tables annexed to this article. Table~\ref{tab:sample_spectra} is an example of the content found in the online tables. The rest of information is found in tables or references of this paper.

\section*{Acknowledgements}
This work is based on observations collected at the European Southern Observatory, Chile, proposal number ESO 092.C-0323(A). We are grateful to the anonymous referee for his/her helpful comments. We acknowledge support from the State Research Agency (AEI) of the Spanish Ministry of Science, Innovation and Universities (MCIU) and the European Regional Development Fund (FEDER) under grant with references AYA2015-65205-P and AYA2017-83383-P. JG-R acknowledges support from an Advanced Fellowship from the Severo Ochoa excellence program (SEV-2015-0548). The authors acknowledge support under grant P/308614 financed by funds transferred from the Spanish Ministry of Science, Innovation and Universities, charged to the General State Budgets and with funds transferred from the General Budgets of the Autonomous Community of the Canary Islands by the MCIU. KZA-C acknowledges support from Mexican CONACYT posdoctoral grant 364239. JEM-D acknowledges support of the Instituto de Astrof\'isica de Canarias under the Astrophysicist Resident Program and acknowledges support from the Mexican CONACyT (grant CVU 602402). AM-D acknowledges support from the FONDECYT project 3140383. WJH acknowledges support from DGAPA-UNAM PAPIIT IN107019.



\bibliographystyle{mnras}
\bibliography{Mendez}

\newpage


\appendix

\section{The alleged observation of Si~IV \texorpdfstring{\boldmath$ \lambda 4088.86$}.  line. }
\label{sec:siiv_coment}

Several authors used the $I$(\mbox{O}\thinspace \mbox{II} $\lambda$4649.13)/$I$(\mbox{O}\thinspace \mbox{II} $\lambda$4089.29) ratio to derive $T_{\rm e}$ based on its theoretical dependence on $T_{\rm e}$ and insensitivity to $n_{\rm e}$ \citep[see e.~g.][]{garciarojas07, fangyliu13, McNabb13, wesson18}. Nevertheless, \citet{peimbert13} discourages its use, due (among other reasons) to the possible contamination of \mbox{O}\thinspace \mbox{II} $\lambda$4089.29 by the  \mbox{Si}\thinspace \mbox{IV} $\lambda 4088.86$ line, which would lead to underestimating $T_{\rm e}$. The authors consider that \mbox{Si}\thinspace \mbox{IV} $\lambda 4088.86$ has been detected in 2 H\thinspace II regions observed with UVES echelle spectrograph at VLT telescope: the Orion Nebula \citep{Esteban04} and 30 Doradus \citep{Peimbert03}. Although it is a real possibility that a line like \mbox{Si}\thinspace \mbox{IV} $\lambda 4088.86$ may be detected in an H\thinspace II region, much of the flux attributed to this line is actually due to an observational artifact of UVES spectrum. 

Fig.~\ref{fig:echelle} shows the echelle orders extracted in an UVES blue arm spectrum using dichroic~\#2 ($\Delta \lambda=3750-4995$). Optical reflections produced by the dichroic~\#2 in the blue arm can be noted as vertical lines crossing the echelle orders. These artifacts are negligible with the exception of those produced by the most intense lines: \mbox{[O}\thinspace \mbox{III]} $\lambda 4959$, H$\beta$ and \mbox{[O}\thinspace \mbox{III]} $\lambda 5007$. The last of those lines does not enter in any complete echelle order in this arm, but it is partially observed at the edge of the CCD, together with its associated high-velocity component. As a consequence 4 main sources of ``ghost lines'' can be noticed. The third of them (from left to right), affects exactly the $\lambda 4089.07$ position in the echelle order number 11 (bottom up) in our observations. Approximately at this wavelength we expect to have the high velocity component of \mbox{O}\thinspace \mbox{II} $\lambda 4089.29$ in cuts 2 and 3, but it must be free of emission from HH~529~II and III in cut 4. Fig.~\ref{fig:oii_4089} shows the emission around $\lambda 4089.29$ in the spectra of cut 4, a pretty similar image than the Fig. 2 from \citet{peimbert13}.

In our spectra, an hypothetical \mbox{Si}\thinspace \mbox{IV} $\lambda 4088.86$ line should be observed at $\lambda 4089.08$, considering the kinematical structure of the nebular component and the high ionization potential of the line (see Section~\ref{subsec:vel_struc}). This means that in case of being detected, the \mbox{Si}\thinspace \mbox{IV} $\lambda 4088.86$ line would be indistinguishable from the ghost line at $\lambda 4089.07$. We have measured the intensity of ghost lines coming from the same source than $\lambda 4089.07$ along the echelle orders but excluding those ones which are blended with other nebular lines. Fig.~\ref{fig:ghost_fit} shows the decreasing trend of the intensity of ghost emission with respect to its source from higher to lower orders, as well as a least squares fit to predict ghost emission in order 11, where the emission feature at $\lambda 4089.07$ lies. The predicted ghost emission in $\lambda 4089.07$ is $\frac{F(\lambda)}{F(\text{H}\beta)} = 0.007$ while the rms of the noise associated with the continuum in cut 4 around $\lambda 4089.07$ represents a possible contribution of $\frac{F(\text{rms})}{F(\text{H}\beta)} = 0.004$. On the other hand, the measured flux of $\lambda 4089.07$ is $\frac{F(\lambda)}{F(\text{H}\beta)} = 0.012$. Thus, the emission observed at $\lambda 4089.07$ is consistent with purely ghost emission. This ghost emission affects in a similar way the spectra of the Orion Nebula and 30 Dor analysed by \citet{Esteban04} and \citet{Peimbert03}, respectively.

The main drawback of the $T_{\rm e}$ diagnostic based on the $I$(\mbox{O}\thinspace \mbox{II} $\lambda$4649.13)/$I$(\mbox{O}\thinspace \mbox{II} $\lambda$4089.29) ratio in H\thinspace II regions is that $\lambda 4089.29$ is rather weak, providing uncertain $T_{\rm e}$ values. Therefore, this diagnostic  will only be useful when the \mbox{O}\thinspace \mbox{II} lines are well measured, or when the object shows significant difference between the $T_{\rm e}$ of the zone where the \mbox{O}\thinspace \mbox{II} lines are formed and the rest of the nebula \citep{wesson18}. 
Finally, although \mbox{Si}\thinspace \mbox{IV} lines  are expected to be rather faint in normal H\thinspace II regions, this may not be the case in high-ionization planetary nebulae (PNe) as NGC 3918 \citep{garciarojas15}.

\section{Cl$^{2+}$ and Cl abundances}
\label{sec:cl_comment}

\citet{dominguezguzman19} have proposed the use of $T_{\rm e}$([N\thinspace II]) to determine the Cl$^{2+}$ abundance in a sample of 37  Galactic and extragalactic H\thinspace II regions, including the Orion Nebula \citep[using the data of][]{Esteban04}. They argue that $T_{\rm e}$([N\thinspace II]) reduces the dispersion of the Cl/O ratio and remove trends in the Cl/O {\it versus} O/H relation. However, in a detailed study of each specific object, the optimal temperature to adopt can differ from what statistically would be the best choice. 
Considering the IP of 23.8 eV of Cl$^{2+}$, in between of those of N$^+$ (14.5 eV) and O$^{2+}$ (35.1 eV), we expect that a $T_{\rm e}$ representative of an intermediate ionization zone as $T_{\rm e}$([S\thinspace III]) should be more appropriate for determining the Cl$^{2+}$ abundance. The relationship of $T_{\rm e}$([S\thinspace III]) with $T_{\rm e}$([N\thinspace II]) and $T_{\rm e}$([O\thinspace III]) may depend on the ionization degree of the gas \citep{Berg20}. In our spectra, we are able to calculate the total Cl abundance because we measure CELs of all the ionization species of Cl that are expected to be present in the Orion Nebula. We test the resulting Cl abundance considering three different temperatures: $T_{\rm e}$(low), $T_{\rm e}$([S\thinspace III]) and $T_{\rm e}$(high) for deriving the Cl$^{2+}$/H$^+$ ratio. In Table~\ref{tab:cl_comp}, we present the Cl$^{2+}$ and Cl abundances as well as their corresponding log(Cl/O) values using the three aforementioned temperatures.

Using $T_{\rm e}$(low), we obtain a log(Cl/O) value in the high-velocity components about 0.1 dex lower with respect to the value found in the nebular ones of the different cuts. This  suggests that $T_{\rm e}$(low) is slightly underestimating the Cl/H ratio, although within the uncertainties. Conversely, the Cl/O ratio becomes more consistent when using $T_{\rm e}$([S\thinspace III]) or $T_{\rm e}$(high) to estimate the Cl$^{2+}$ abundance. We obtain a mean log(Cl/O) value of $-3.63 \pm 0.04$, $-3.42 \pm 0.03$ and $-3.50 \pm 0.03$ when adopting $T_{\rm e}$(low), $T_{\rm e}$(high) or $T_{\rm e}$([S\thinspace III]) to calculate the Cl$^{2+}$ abundance, respectively. The log(Cl/O) value obtained using $T_{\rm e}$([S\thinspace III]) is the one closest to the solar value of $-3.50 \pm 0.09$ recommended by \citet{lodders19}. Therefore, we finally adopt $T_{\rm e}$([S\thinspace III]) to estimate the Cl$^{2+}$ abundance.

\section{Ni$^{2+}$ abundance}
\label{sec:ni2_ab_comment}

The first estimation of the Ni abundance in an H\thinspace II region was made by \citet{osterbrock92} for the Orion Nebula. They used estimates of the atomic data of Ni ions, considering [Ni\thinspace II] and [Ni\thinspace  III] lines. Since then, the number of Ni abundance determinations in ionized nebulae is still very limited, both for PNe \citep{Zhang06,garciarojas13,delgadoinglada16} and H\thinspace II regions \citep{mesadelgado09,delgadoinglada16}. There is a considerable amount of \mbox{[Ni}\thinspace \mbox{II]} and \mbox{[Ni}\thinspace \mbox{III]} lines in our spectra. However, \mbox{[Ni}\thinspace \mbox{II]} lines are affected by fluorescence \citep{Lucy95} and their use to calculate the abundance of Ni$^{+}$ is restricted to some particular cases as low-excitation nebulae \citep{Zhang06}. On the other hand,  \mbox{[Ni}\thinspace \mbox{III]} lines are, in principle, not affected by fluorescence effects and can be used to derive Ni$^{2+}$ abundances. From all the detected [Ni\thinspace III] lines, we chose [Ni\thinspace III] $\lambda \lambda$6000, 6534, 6682, 6797, 6946 and 7890 because they are not affected by blends or telluric absorption. 

As we commented in Section~\ref{sec:physical_conditions}, all tested diagnostics based on \mbox{[Ni}\thinspace \mbox{III]} lines fail to provide reliable values of physical conditions, indicating inaccuracies between the observed lines and the theoretical predictions. \citet{delgadoinglada16} studied the Fe/Ni ratio in eight PNe and three H\thinspace II regions, including the Orion Nebula. They used different datasets for these last objects, including the high spectral resolution ones from \citet{Esteban04} and  \citet{mesadelgado09} (which incluides HH~202~S), and some previously unpublished observations covering approximately the same area as the Position 1 of \citet{Esteban98} and the brightest part of the Orion Bar. We have compared our data with some observed flux ratios compiled by \citet{delgadoinglada16} looking for possible undetected line blends or observational inaccuracies. For convenience, we have compared the predicted and observed flux ratios of [Ni\thinspace III] $\lambda \lambda$6000, 6534 and 6946 lines that arise from the same atomic level and therefore we expect that their intensity ratios should be constant. In Table~\ref{tab:intensity_ni3}, we compare the observed flux ratios and the predicted ones using the atomic data of Ni$^{2+}$ by \citet{Bautista01}. As Table~\ref{tab:intensity_ni3} shows, the [Ni\thinspace III] $\lambda$6534/$\lambda$6000 and $\lambda$6946/$\lambda$6000 intensity ratios are not inconsistent with the predicted ones. The most discrepant case is HH~529~II, although it is the component with the greatest dispersion. The intensity ratios based on the data sets with smaller dispersion seems to indicate a systematic value of $\lambda$6534/$\lambda$6000$\sim 1.6$ and $\lambda$6534/$\lambda$6000$\sim 0.30$ which is compatible with an overestimation of the flux of the [Ni\thinspace III] $\lambda 6000$ line. However, we discard the presence of sky emission affecting this line. Beside to this, we do not find strong candidates of unusual lines in the literature or in the Atomic Line List v2.05b21 with a wavelength difference below to 0.15 \AA\ (a conservative limit for an appropriate deblending of Gaussian profiles at our spectral resolution). In addition, as can be seen in  Table~\ref{tab:Ni3_abundances}, we find an inconsistent pattern of Ni$^{2+}$ abundances in all cuts and components for the six selected lines, with differences up to 0.8 dex. This may be a problem of the currently available atomic data of Ni$^{2+}$, which could not be accurate enough for deriving confident values of its ionic abundances. Therefore, our Ni$^{2+}$/H$^+$ abundances must be interpreted with care, since they may include unknown uncertainties.

\section{Supporting material}

\begin{figure*}
  \begin{subfigure}{7.5cm}
    \centering\includegraphics[height=4cm,width=\columnwidth]{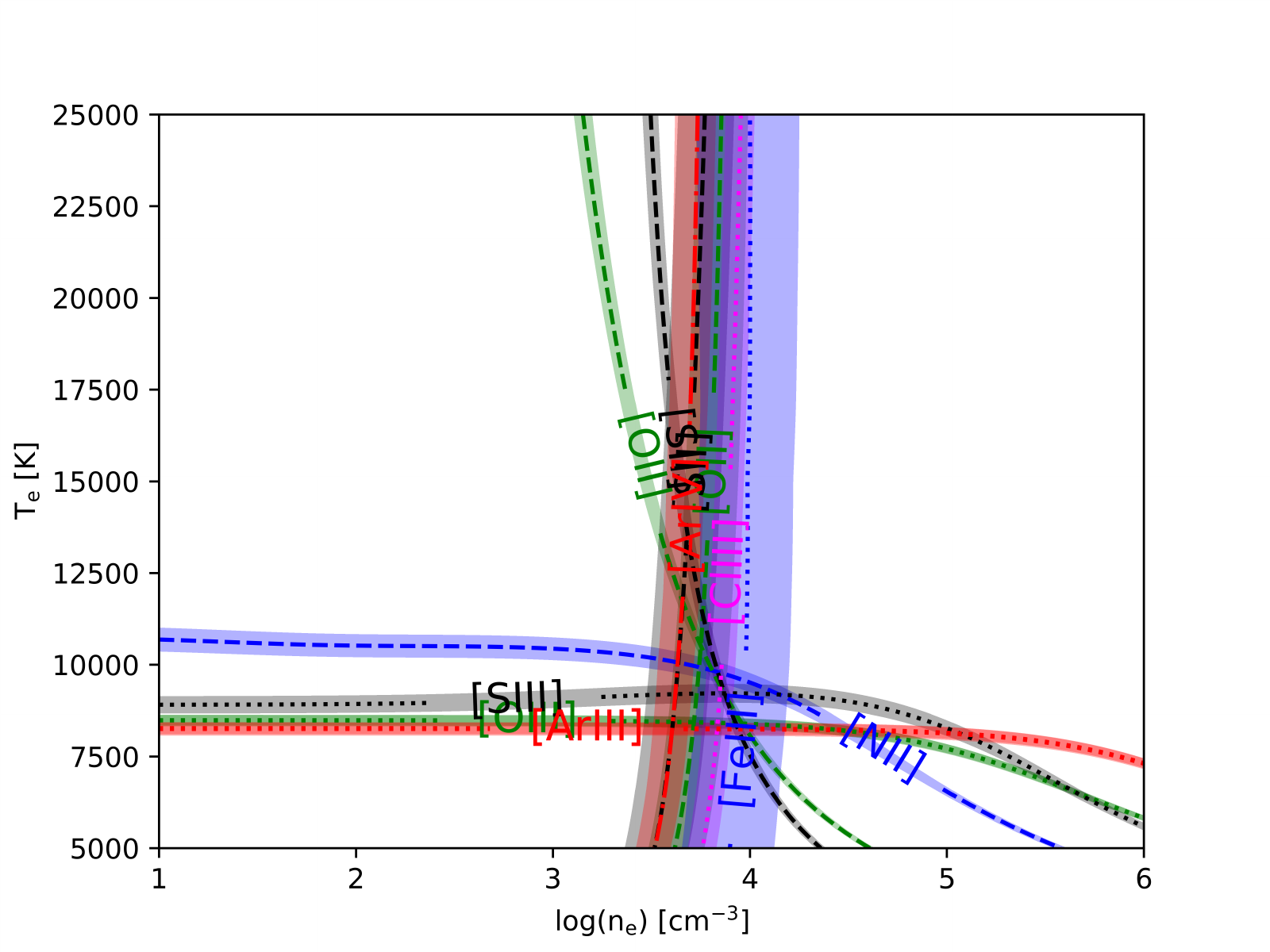}
    \caption{Cut 1, nebular component.}
  \end{subfigure}
  \begin{subfigure}{7.5cm}
    \centering\includegraphics[height=4cm,width=\columnwidth]{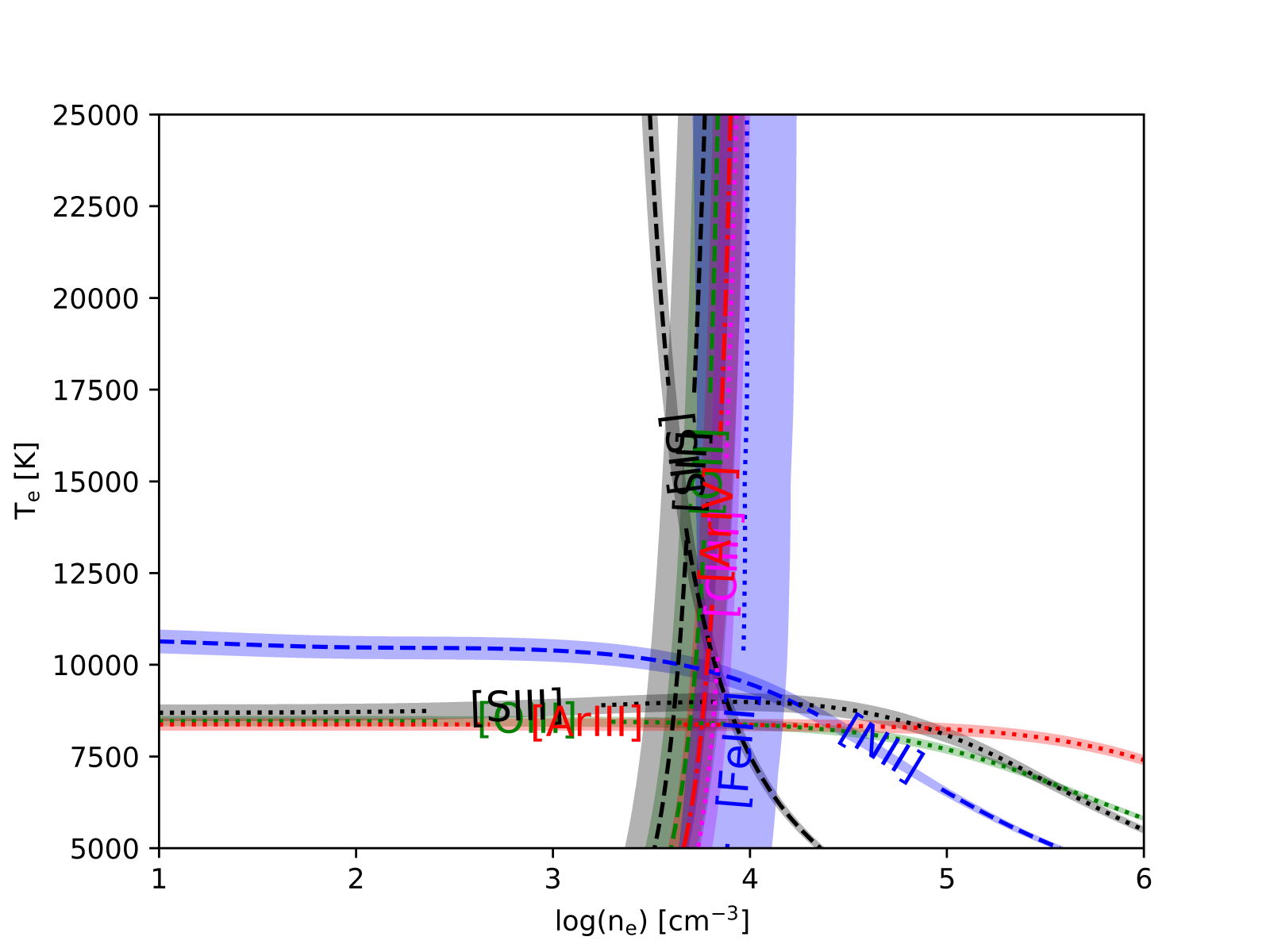}
    \caption{Cut 2, nebular component.}
  \end{subfigure}
 
  \begin{subfigure}{7.5cm}
    \centering\includegraphics[height=4cm,width=\columnwidth]{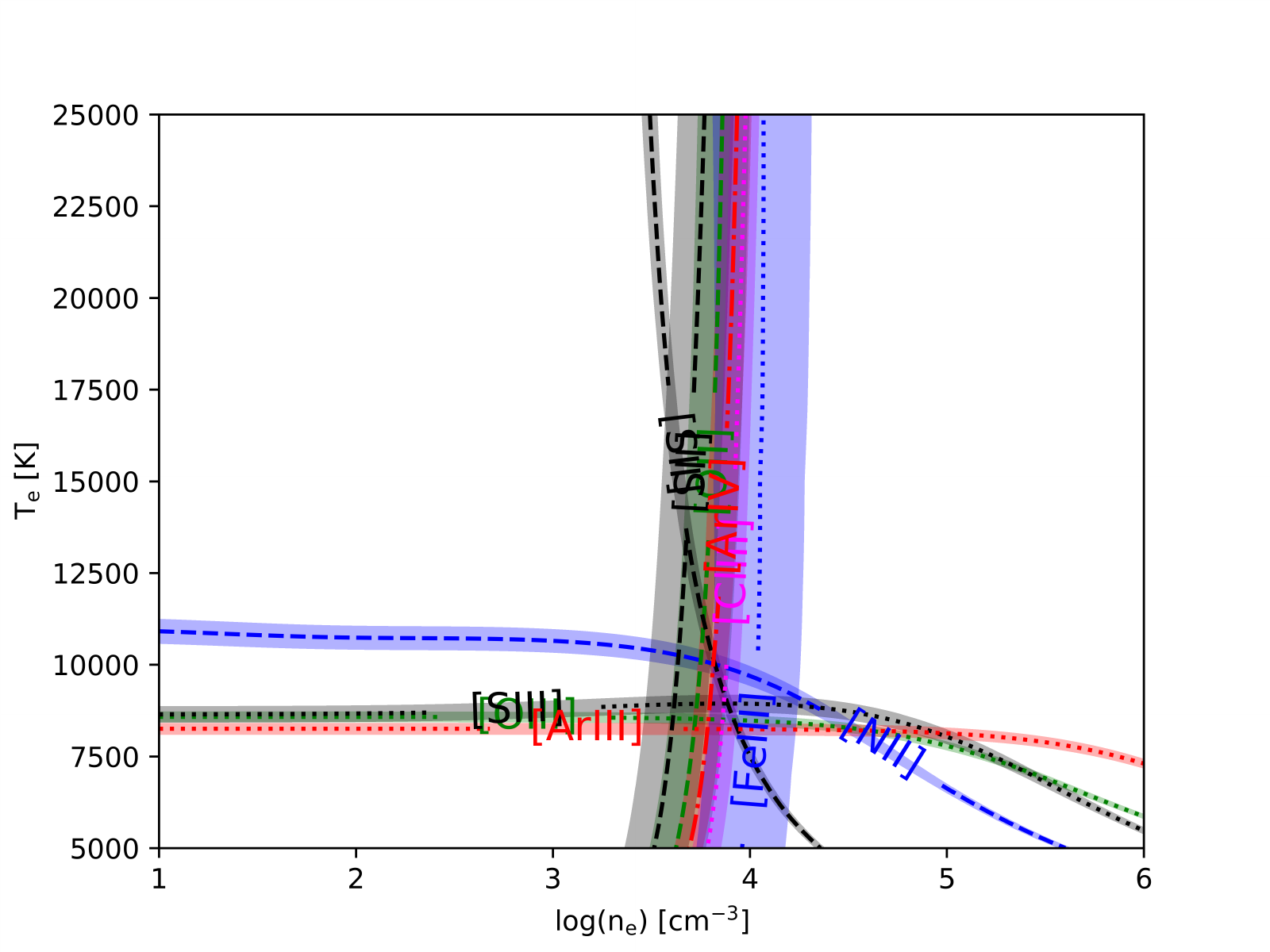}
    \caption{Cut 3, nebular component.}
  \end{subfigure}
  \begin{subfigure}{7.5cm}
    \centering\includegraphics[height=4cm,width=\columnwidth]{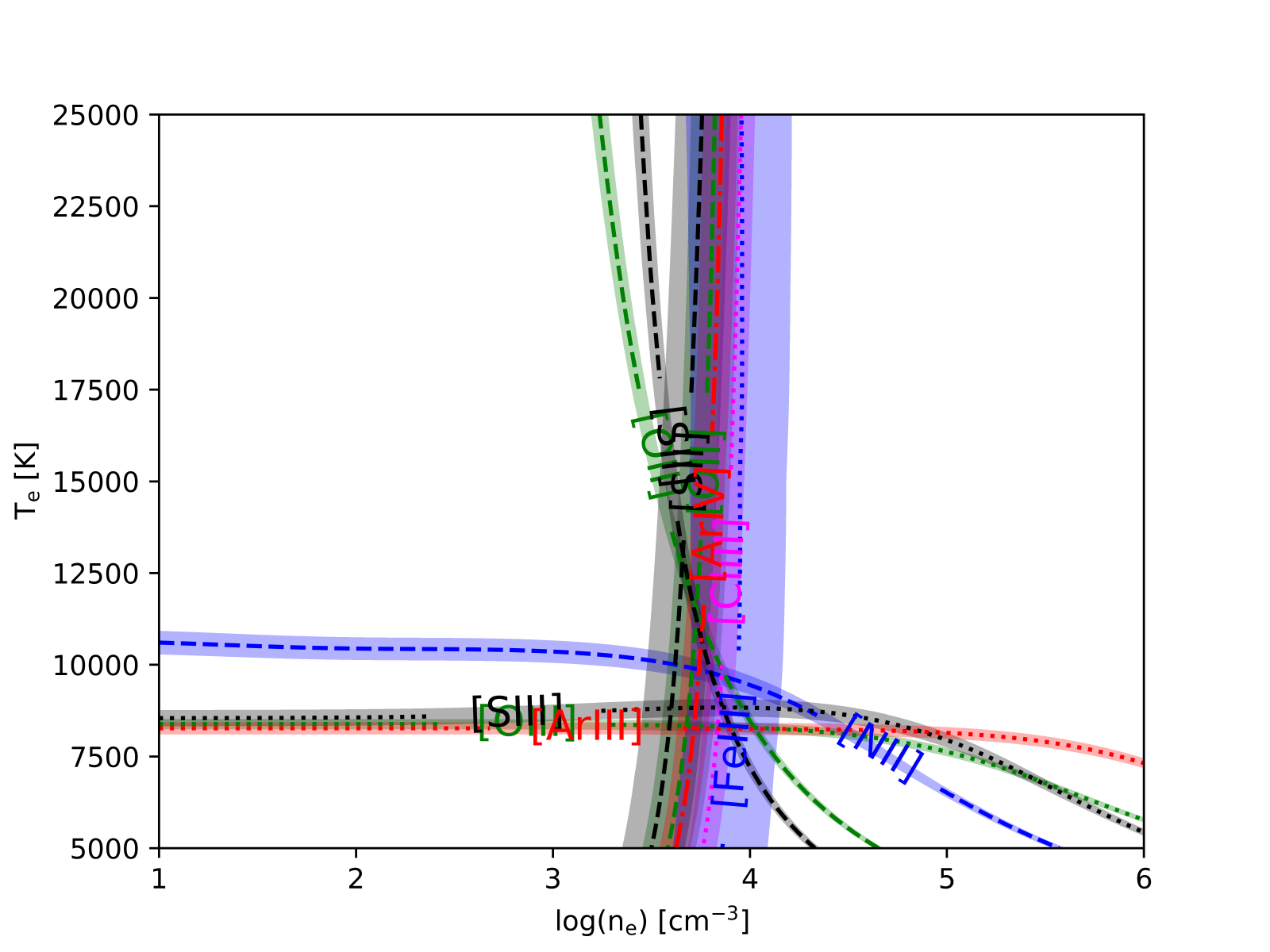}
    \caption{Cut 4, nebular component.}
  \end{subfigure}

  \begin{subfigure}{7.5cm}
    \centering\includegraphics[height=4cm,width=\columnwidth]{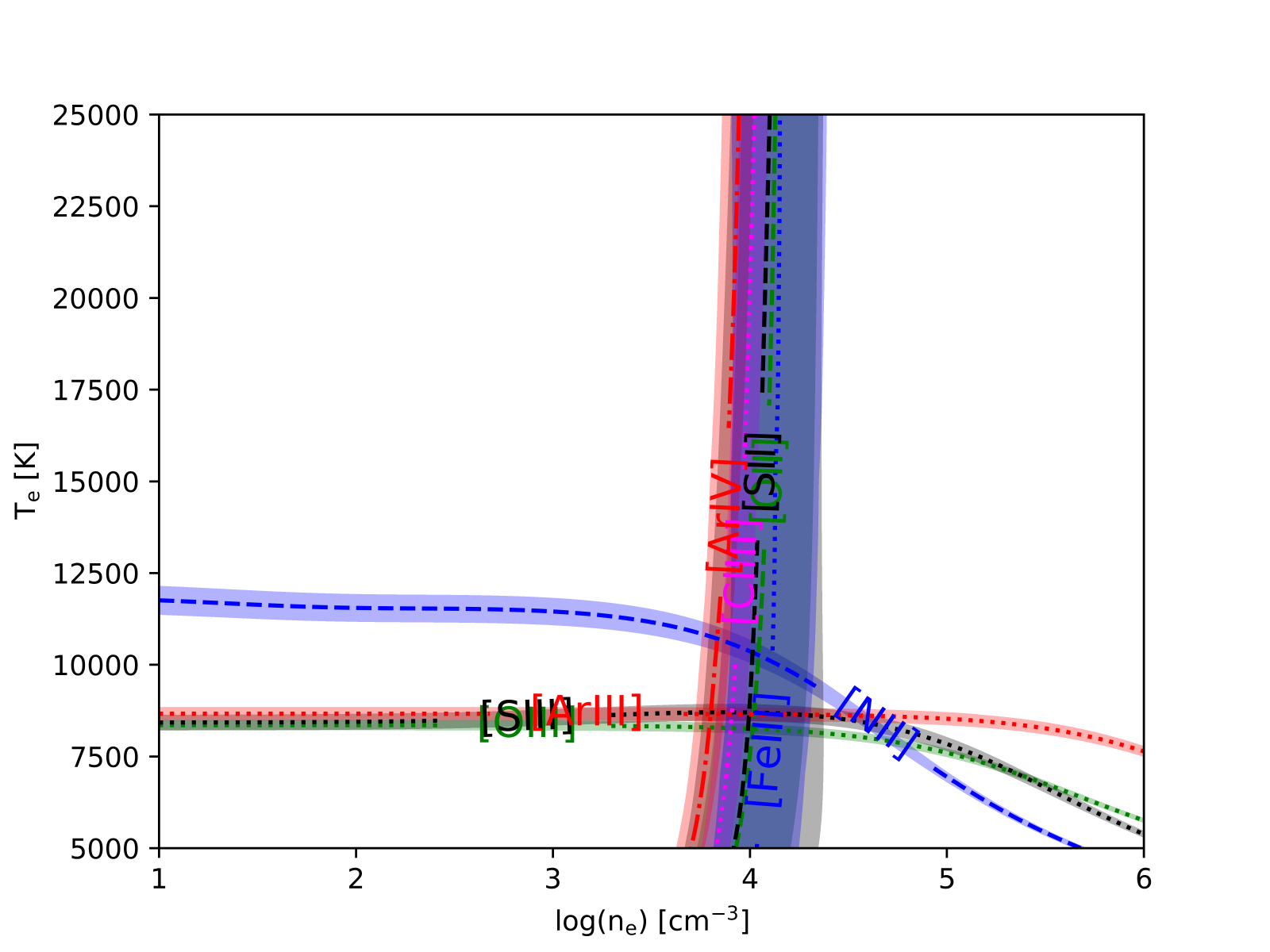}
    \caption{HH~529~II}
  \end{subfigure}
  \begin{subfigure}{7.5cm}
    \centering\includegraphics[height=4cm,width=\columnwidth]{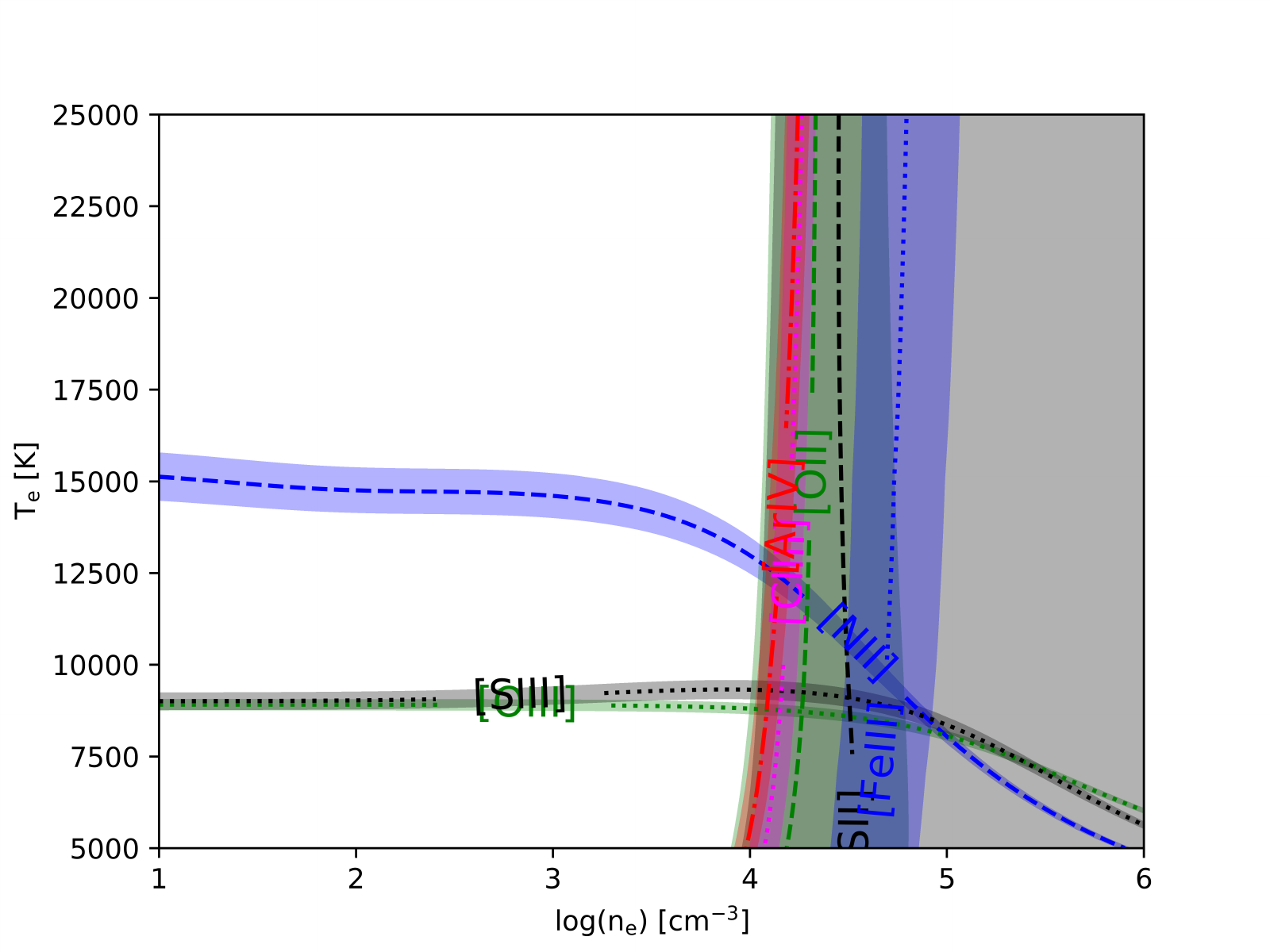}
    \caption{HH~529~III}
  \end{subfigure}

  \begin{subfigure}{12cm}
    \centering\includegraphics[height=6cm,width=\columnwidth]{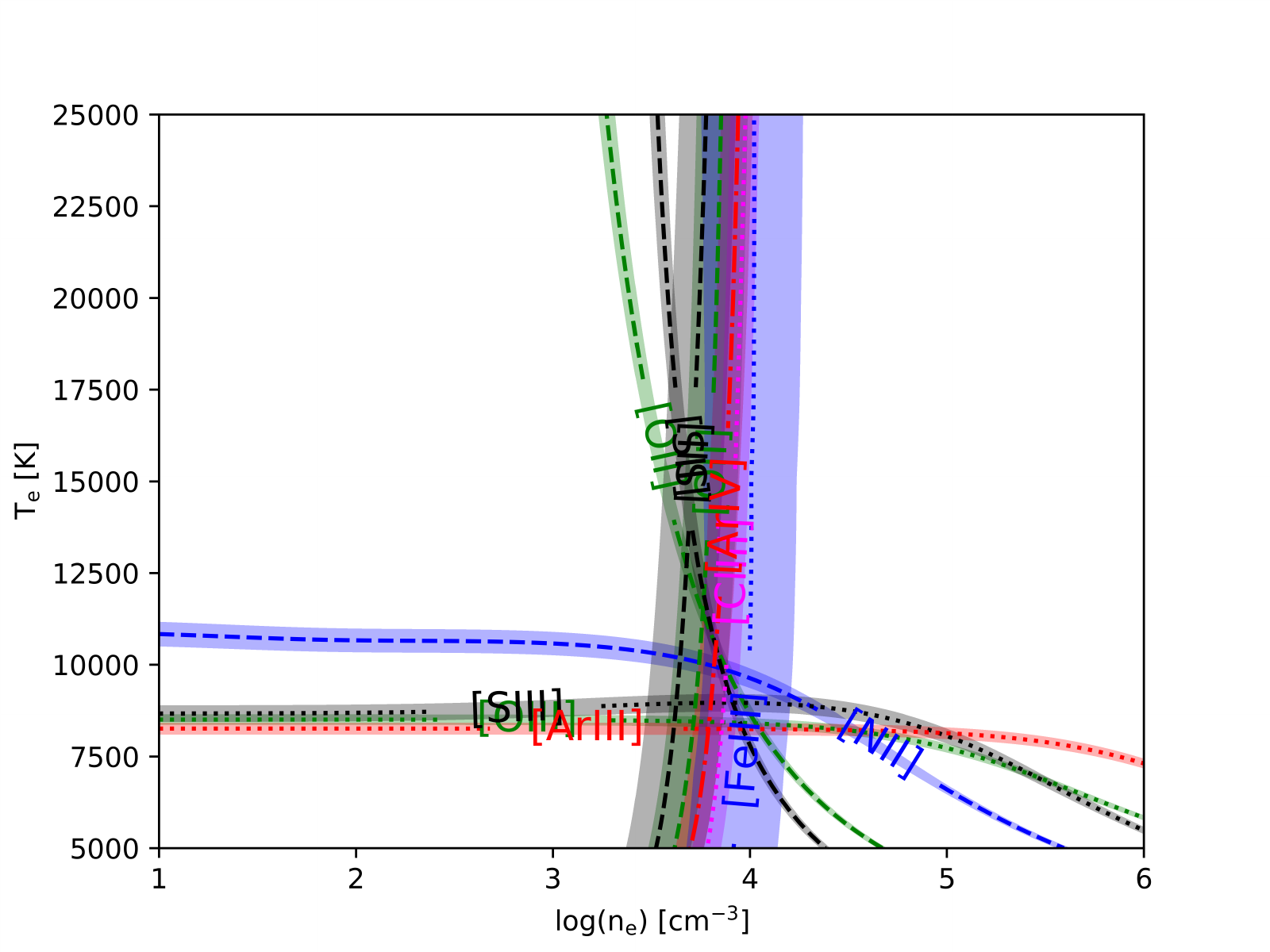}
    \caption{All cuts combined.}
  \end{subfigure}
  \caption{Plasma diagnostic plots for each of the 7 components analysed in this work. The labeled diagnostics correspond to those discussed in Section~\ref{subsec:physical_conditions_cels}, whose results are presented in Table~\ref{tab:pc}.}
\label{fig:plasma}
\end{figure*}

\begin{figure}
\includegraphics[width=\columnwidth]{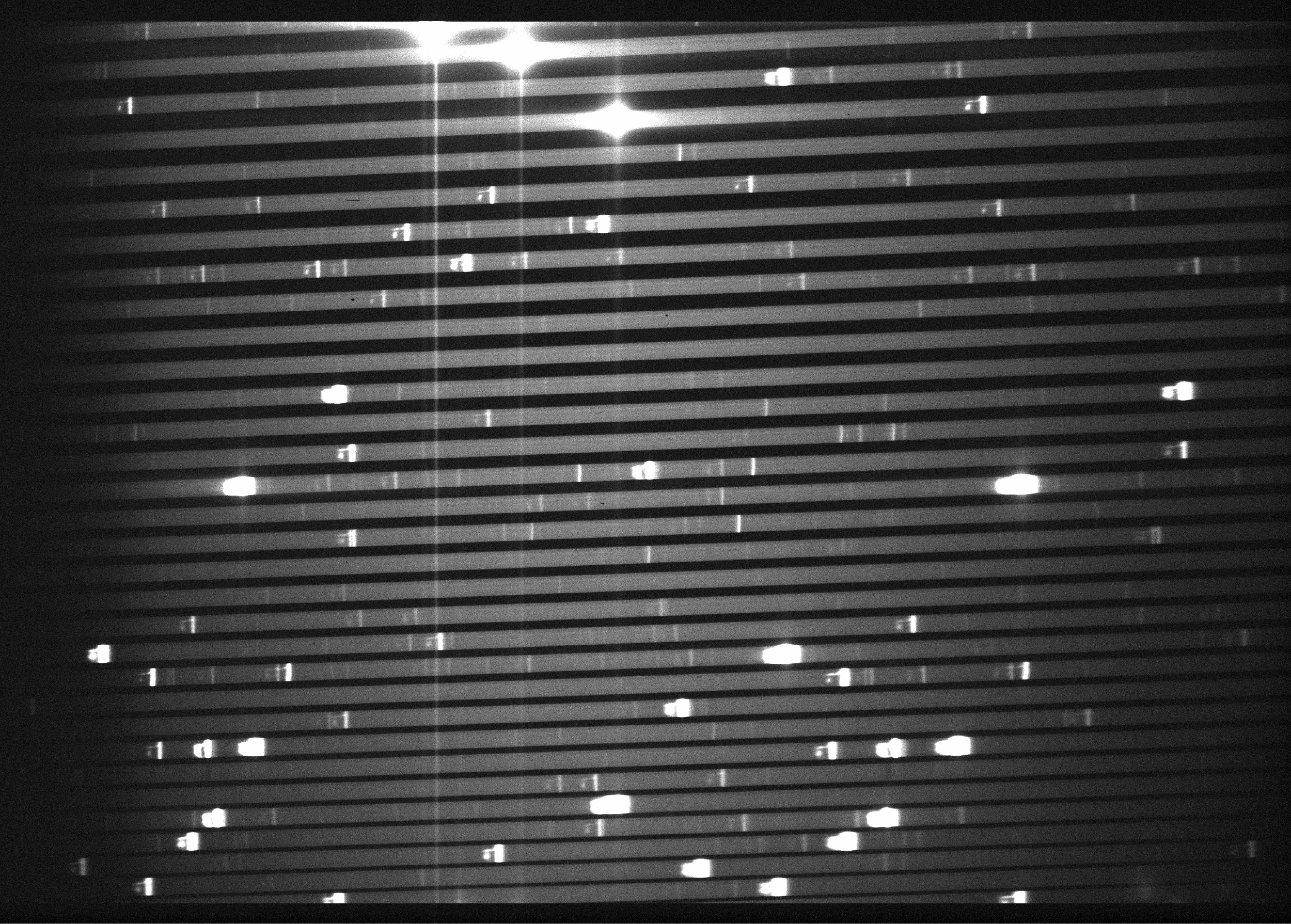}
\caption{Image of part of the echelle orders extracted in the UVES blue arm using dichroic~\#2 setting ($\Delta \lambda=3750-4995$ \AA). The contrast highlights reflections in the optical system of the spectrograph that can affect some lines. We have established that order 1 is the order at the bottom and 31 at the top.}
\label{fig:echelle}
\end{figure}

\begin{figure}
\includegraphics[width=\columnwidth]{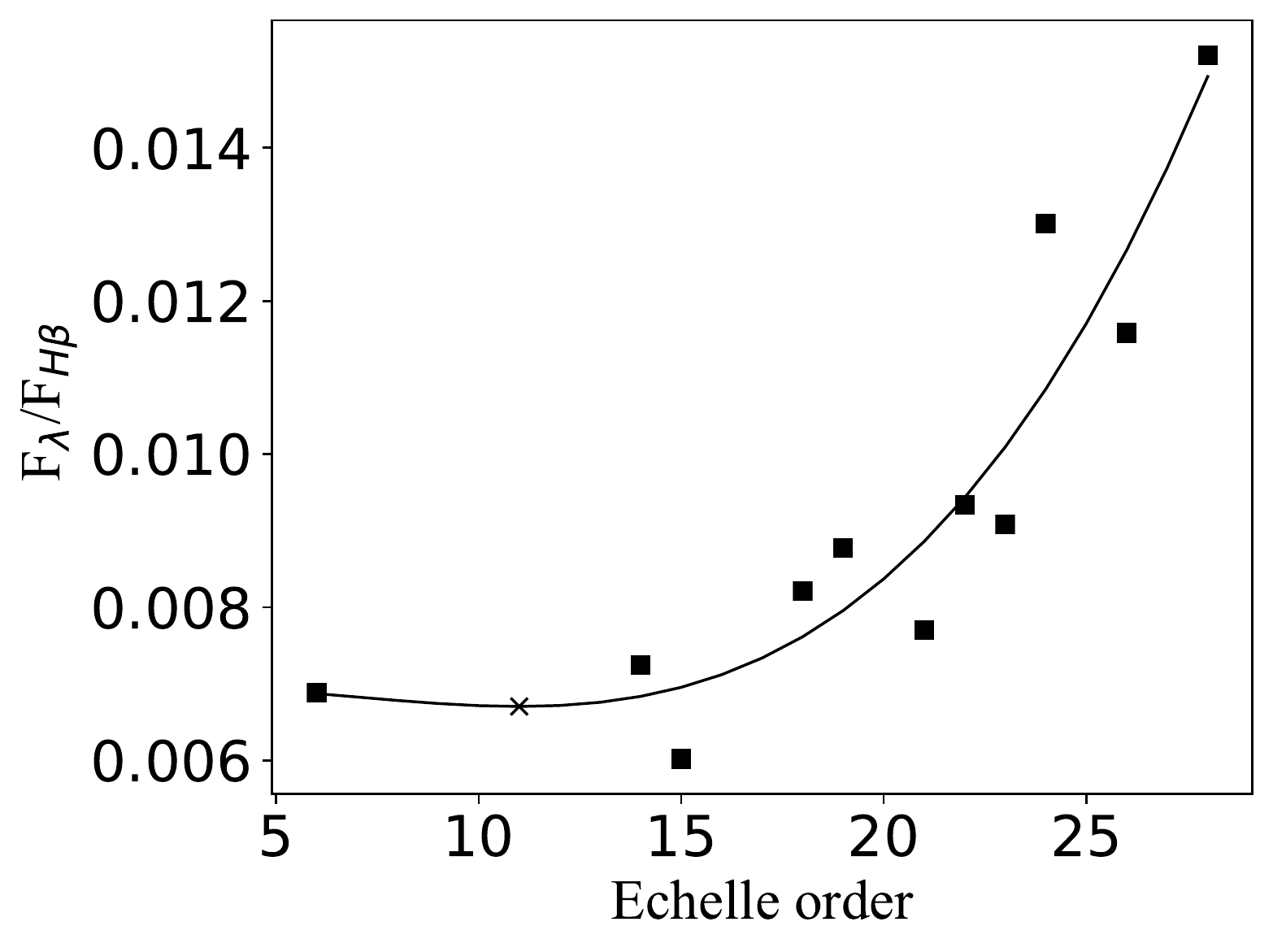}
\caption{Least Squares fit of the flux emission for the third source of ``ghost lines'' in the echelle orders. The prediction for the ghost line at $\lambda = 4089.07$ (in the order 11) is marked with a cross.}
\label{fig:ghost_fit}
\end{figure}

\begin{figure*}
\includegraphics[width=\textwidth]{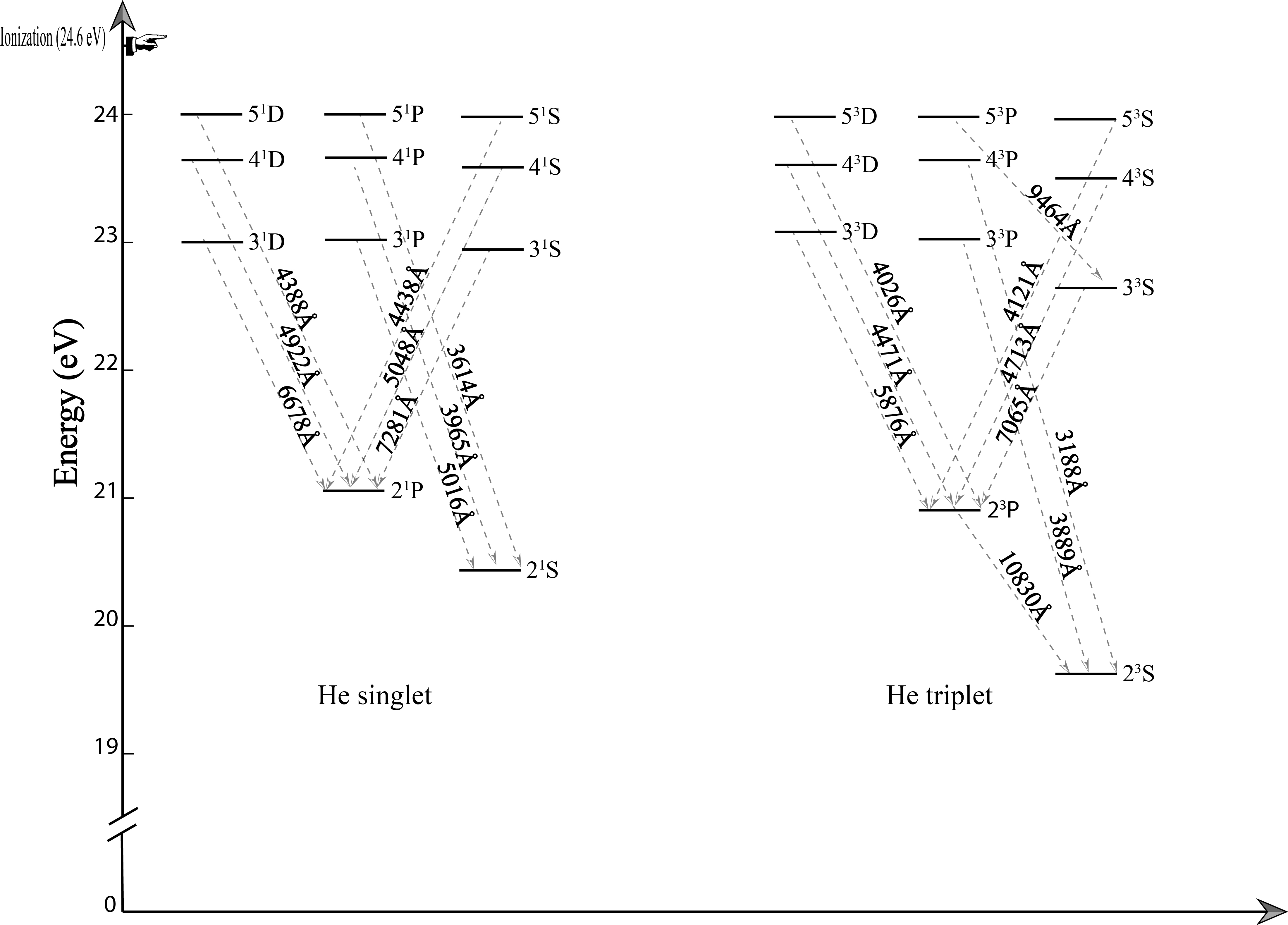}
\caption{Grotrian diagram of \mbox{He}\thinspace \mbox{I} for both configurations: triplet and singlet.}
\label{fig:grotrian_he}
\end{figure*}

\begin{figure*}

\begin{subfigure}{8.5cm}
    \centering\includegraphics[height=6cm,width=\columnwidth]{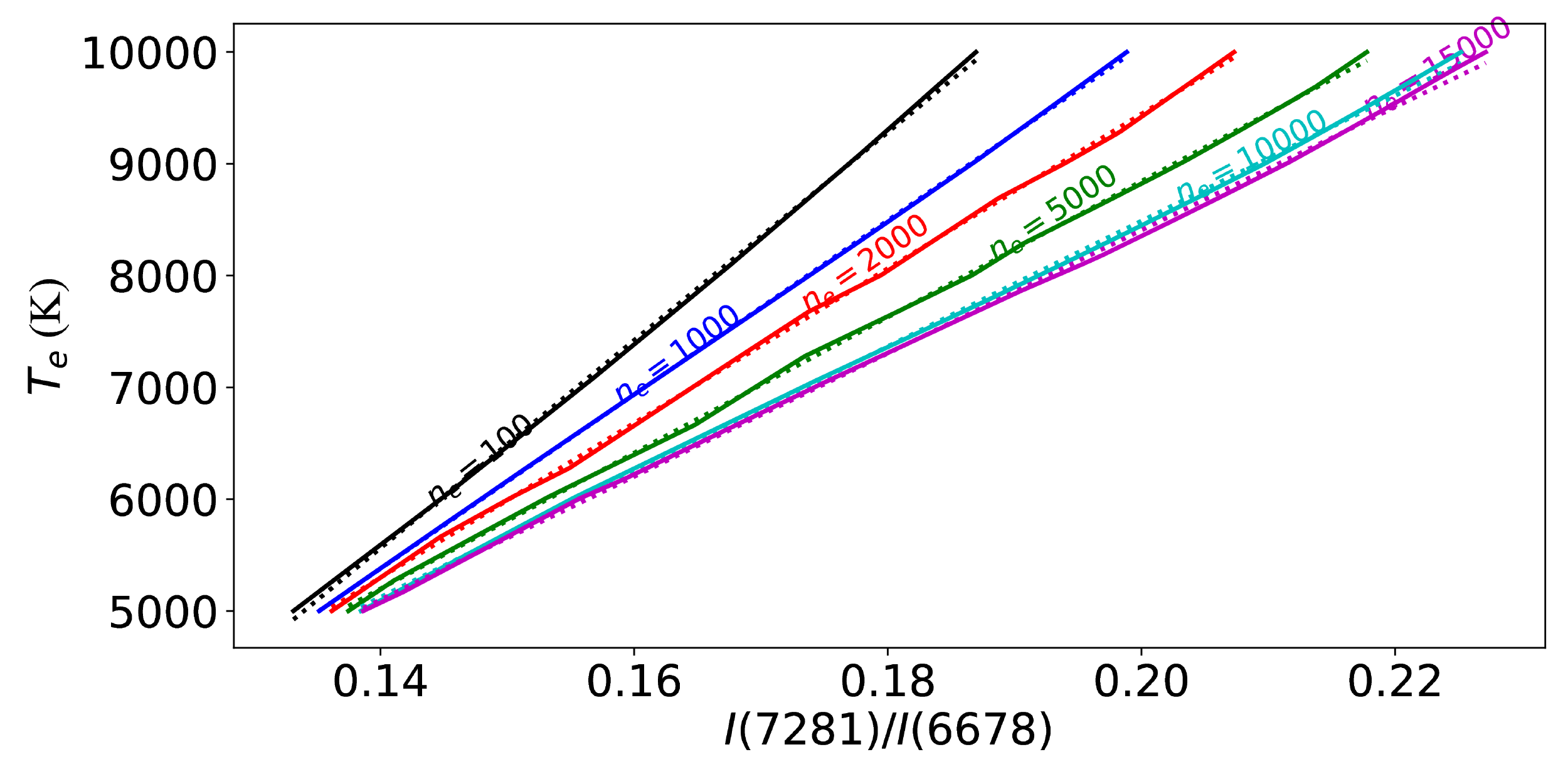}
    \caption{A linear fit is accurate for $T_{\rm e}\leq$ 10000 K.}
    \label{fig:7281_temperature_helium_dependence_a}
\end{subfigure}
\begin{subfigure}{8.5cm}
    \centering\includegraphics[height=6cm,width=\columnwidth]{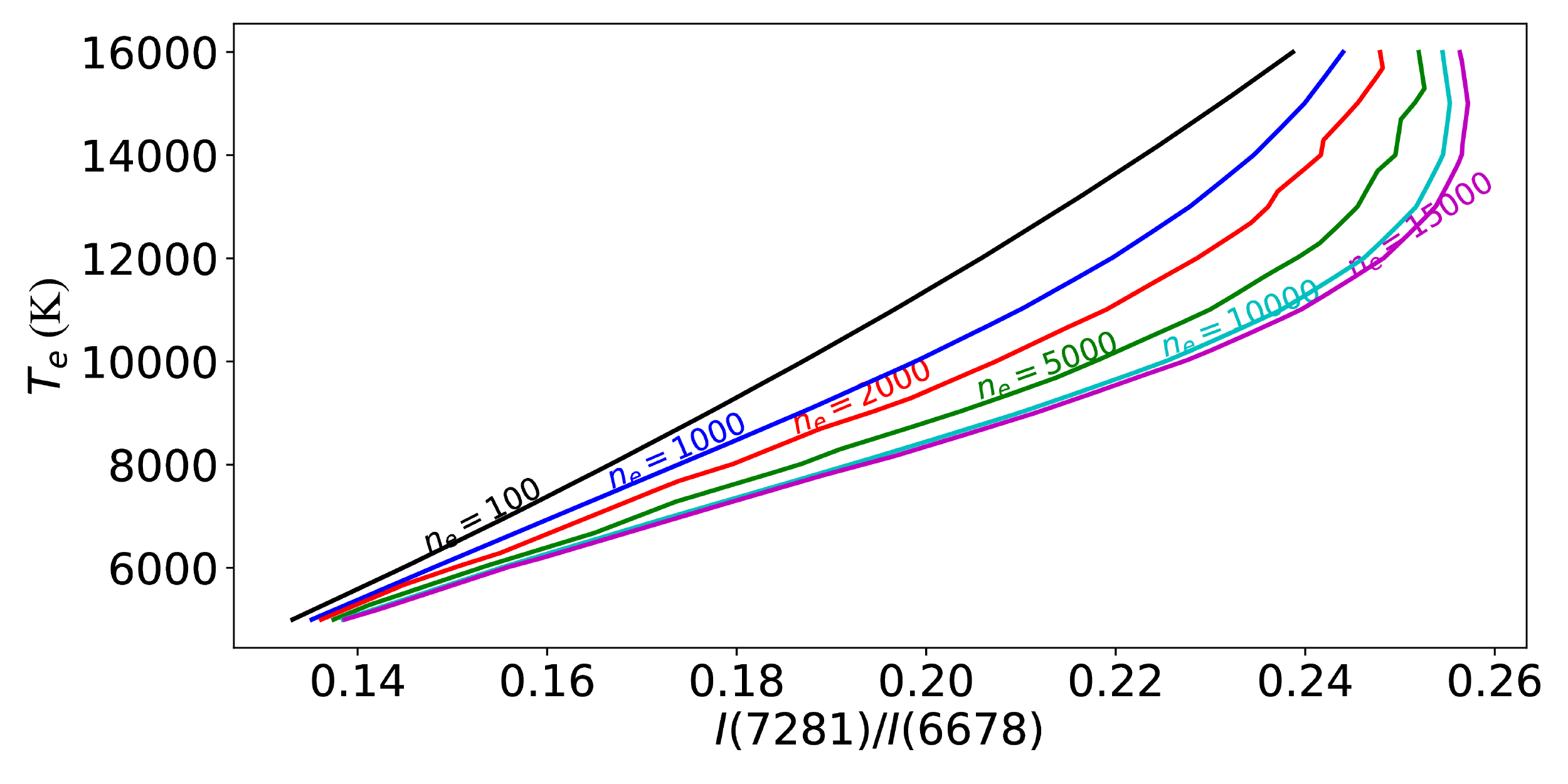}
    \caption{A linear fit is not accurate for $T_{\rm e} >$ 10000 K.}
    \label{fig:7281_temperature_helium_dependence_b}
\end{subfigure}
\caption{Dependence of $I$(\mbox{He}\thinspace \mbox{I} $\lambda$7281)/$I$(\mbox{He}\thinspace \mbox{I} $\lambda$6678) on the physical conditions.}
\label{fig:7281_temperature_helium_dependence}

\end{figure*}

\begin{figure}
\includegraphics[width=\columnwidth]{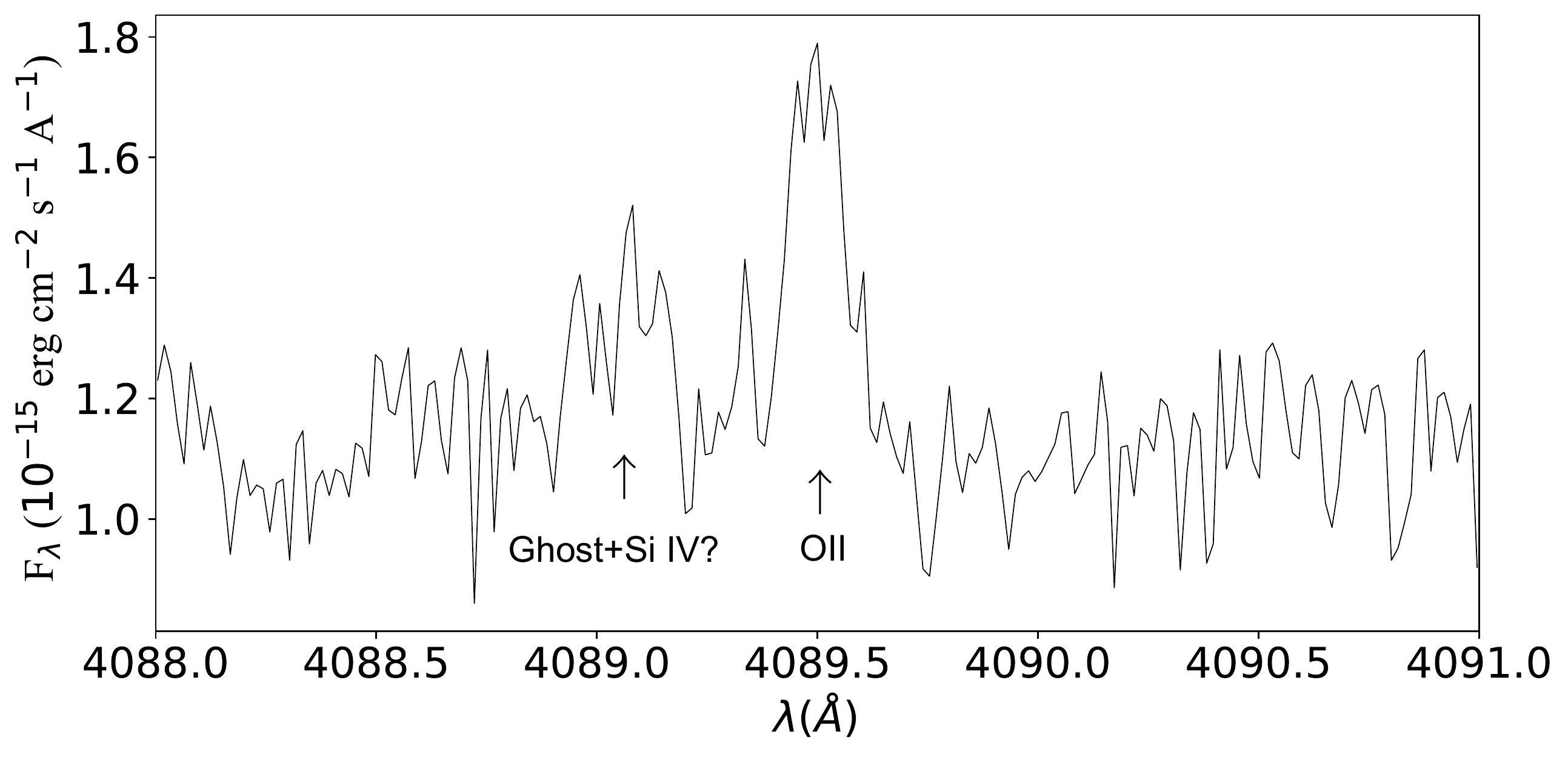}
\caption{Emission spectrum of the spatial cut 4 around $\lambda 4089.29$. There are no high-velocity components in this spatial cut.}
\label{fig:oii_4089}
\end{figure}

In this appendix we include the following material:
\begin{itemize}
    
    \item Table~\ref{tab:comparison_balmer}: Comparison of the Balmer line ratios between \citet{Blagrave06} and this work.

    \item Table~\ref{tab:sample_spectra}: Sample of 15 lines of the spectra of cut 2. The complete line list of all cuts is appended in the online supporting material.
    
    \item Table~\ref{tab:atomic_data} Atomic data set (Transition probabilities and collision strengths) used in this work for the analysis of CELs.
    
    \item Table~\ref{tab:rec_atomic_data} Effective recombination coefficients used in this work for recombination lines.
    
    \item Table~\ref{tab:critical_densities} Critical densities of some lines whose ratios are commonly used as density diagnostics. 
    
    \item Table~\ref{tab:fe3_ratios_theo} Comparison of some observed [Fe\thinspace III] intensity ratios with the theoretical predictions.

    \item Table~\ref{tab:slopes}: Slopes and intercepts calculated to estimate $T_{\rm e}$(He\thinspace I) using Eq.\eqref{eq:helium_fit} for a range of usual densities. 
    
    \item Table~\ref{tab:Fe3_abundances_low} and Table~\ref{tab:Fe3_abundances_high}: Ionic abundances of Fe$^{2+}$/H$^{+}$ derived in all studied components with 11 CELs by using $T_{\rm e} (\text{low})$ and $T_{\rm e} (\text{high})$, respectively.
    
    \item Table~\ref{tab:cl_comp}: Cl$^{2+}$/H$^{+}$, Cl/H and log(Cl/O) abundances derived with $T_{\rm e} (\text{low})$, $T_{\rm e} (\text{[S\thinspace III]})$ and $T_{\rm e} (\text{high})$, respectively, for all studied components.

    \item Table~\ref{tab:intensity_ni3}: Comparison between the predicted and the measured intensity ratios of Ni$^{2+}$ 2F multiplet ($\lambda \lambda 6000, 6534, 6946$).

    \item Table~\ref{tab:Ni3_abundances}: Ionic abundances of Ni$^{2+}$/H$^{+}$ derived per line in all studied components.

    \item Table~\ref{tab:helium_2}: He$^{+}$/H$^{+}$ abundances derived from He\thinspace I lines highly affected by self-absorption effects.

    \item Table~\ref{tab:helium_1}: He$^{+}$/H$^{+}$ abundances derived from He\thinspace I lines not affected by self-absorption effects.

    \item Table~\ref{tab:OII_abundances}: O$^{2+}$/H$^{+}$ abundances derived with RLs of several multiplets and transitions. 

    \item Table~\ref{tab:other_rls_abundances}: O$^{+}$/H$^{+}$, C$^{2+}$/H$^{+}$ and Ne$^{2+}$/H$^{+}$ abundances derived with RLs.

    \item Table~\ref{tab:t2_per_comp}: values of $t^2$ derived for each component, estimated by using $T_{\rm e}$(He\thinspace I), $T_{\rm e}$([O\thinspace III]), $T_{\rm e}$([S\thinspace III]) and $T_{\rm e}$([N\thinspace II]).
    
    \item Table~\ref{tab:kin_tab}: Average radial velocity $\langle V \rangle$ and Full Width at Half Maximum $\langle \text{FWHM} \rangle$ of the lines observed in the nebular component and in HH~529~II-III. 
    
    \item Fig.~\ref{fig:plasma}: Plasma diagnostics used in each analysed component to determine physical conditions based on CEL ratios. 
    
    \item Fig.~\ref{fig:echelle}: Echelle orders extracted using the UVES blue arm with the dichroic~\#2.
    
    \item Fig.~\ref{fig:ghost_fit}: Least squares fit of the ghost emission that affects the spectral position $\lambda = 4089.07$.
    
    \item Fig.~\ref{fig:grotrian_he}: Grotrian diagram of the levels of He\thinspace I. 
    
    \item Fig.~\ref{fig:7281_temperature_helium_dependence}: Dependence of $I$(\mbox{He}\thinspace \mbox{I} $\lambda$7281)/$I$(\mbox{He}\thinspace \mbox{I} $\lambda$6678) on the physical conditions.
    
    \item Fig.~\ref{fig:oii_4089}: Emission spectrum of cut 4 around $\sim \lambda 4089$.
    
 \end{itemize}

\begin{table*}
\centering
\caption{Reddening-corrected Balmer line ratios from \citet{Blagrave06} and this work.}
\label{tab:comparison_balmer}
\begin{tabular}{lcccccccccccc}
\hline
 & \multicolumn{2}{c}{BMB06} & \multicolumn{3}{c}{This Work} & \\
Ratio &  High-velocity &  Nebular &  HH~529~II & HH~529~III &  Nebula & Predicted Value\\
\hline
$\text{H}\alpha/\text{H}\beta$ & $3.65 \pm 0.18$ & $3.19\pm 0.03$ & $2.77 \pm 0.08$ & $2.87 \pm 0.11$ & $2.82 \pm 0.09$ & 2.85\\
$\text{H}\gamma/\text{H}\beta$ & $0.74 \pm 0.04$ & $0.69\pm 0.01$ & $0.46 \pm 0.01$ & $0.46 \pm 0.01$ & $0.46 \pm 0.01$ & 0.47\\
$\text{H}\delta/\text{H}\beta$ & $0.34 \pm 0.02$ & $0.40\pm 0.01$ & $0.26 \pm 0.01$ & $0.26 \pm 0.01$ & $0.26 \pm 0.01$ & 0.26\\
$\text{H}\varepsilon/\text{H}\beta$ & $0.27 \pm 0.02$ & $0.31\pm 0.01$ & $0.16 \pm 0.01$ & $0.15 \pm 0.01$ & $0.16 \pm 0.01$& 0.16 \\
$\text{H}\gamma/\text{H}\alpha$ & $0.20 \pm 0.01$ & $0.22\pm 0.01$ & $0.17 \pm 0.01$ & $0.16 \pm 0.01$ & $0.17 \pm 0.01$ & 0.17\\
$\text{H}\delta/\text{H}\alpha$ & $0.09\pm 0.01$ & $0.13 \pm 0.01$ & $0.09 \pm 0.01$ & $0.09 \pm 0.01$ & $0.09 \pm 0.01$ & 0.09\\
$\text{H}\varepsilon/\text{H}\alpha$ & $ 0.07 \pm 0.01$ & $0.10\pm 0.01$ & $0.06 \pm 0.01$ & $0.05 \pm 0.01$ & $0.06 \pm 0.01$ &0.06\\
\hline
\end{tabular}
\end{table*}

\begin{table*}
\caption{Sample of 15 lines of the cut 2 spectra. This cut includes emission from HH~529~II and the Orion Nebula. F$\left( \mbox{H}\beta \right)$=100. This is an example of the content found in the online tables for all cuts.}
\label{tab:sample_spectra}
\begin{adjustbox}{width=\textwidth}
\begin{tabular}{ccccccccccccccccccccc}
\hline
& &\multicolumn{6}{c}{HH~529~II}& \multicolumn{6}{c}{The Orion Nebula}&\\
$\lambda_0$(\AA) & Ion & $\lambda$(\AA) & $v_r\text{ km s}^{-1}$ & $\text{FWHM} \text{ km s}^{-1}$& F$\left( \lambda \right)$/F$\left( \mbox{H}\beta \right)$ & I$\left( \lambda \right)$/I$\left( \mbox{H}\beta \right)$ & Error (\%) & $\lambda$(\AA) & $v_r\text{ km s}^{-1}$ & $\text{FWHM} \text{ km s}^{-1}$&  F$\left( \lambda \right)$/F$\left( \mbox{H}\beta \right)$ & I$\left( \lambda \right)$/I$\left( \mbox{H}\beta \right)$ & Error (\%) & Notes \\
\hline
4638.86 & $\mbox{O}\thinspace \mbox{II}$ & 4638.40 & -29.44 & 20.68 $\pm$ 1.67 & 0.091 & 0.098 & 6 & 4639.08 & 14.51 & 14.86 $\pm$ 0.60 & 0.042 & 0.045 & 4 &  \\
4640.64 & $\mbox{N}\thinspace \mbox{III}$ & 4640.07 & -36.53 & 22.68 $\pm$ 3.59 & 0.037 & 0.040 & 10 & 4640.81 & 11.27 & 11.95 $\pm$ 0.98 & 0.013 & 0.014 & 6 &  \\
4641.81 & $\mbox{O}\thinspace \mbox{II}$ & 4641.35 & -29.42 & 18.60 $\pm$ 0.58 & 0.152 & 0.164 & 4 & 4642.04 & 15.15 & 13.76 $\pm$ 0.24 & 0.083 & 0.089 & 2 &  \\
4643.09 & $\mbox{N}\thinspace \mbox{II}$ & 4642.64 & -28.76 & 16.21 $\pm$ 5.62 & 0.026 & 0.028 & 21 & 4643.32 & 15.15 & 12.01 $\pm$ 1.52 & 0.015 & 0.016 & 9 &  \\
4649.13 & $\mbox{O}\thinspace \mbox{II}$ & 4648.67 & -29.36 & 18.90 $\pm$ 0.38 & 0.240 & 0.258 & 3 & 4649.37 & 15.78 & 12.51 $\pm$ 0.07 & 0.114 & 0.122 & 2 &  \\
4650.84 & $\mbox{O}\thinspace \mbox{II}$ & 4650.38 & -29.35 & 18.89 $\pm$ 1.45 & 0.088 & 0.095 & 6 & 4651.06 & 14.49 & 13.28 $\pm$ 0.52 & 0.040 & 0.043 & 4 &  \\
4658.17 & $\mbox{[Fe}\thinspace \mbox{III]}$ & 4657.67 & -31.86 & 13.52 $\pm$ 0.04 & 0.887 & 0.950 & 2 & 4658.38 & 13.83 & 13.32 $\pm$ 0.01 & 0.770 & 0.820 & 2 &  \\
4661.63 & $\mbox{O}\thinspace \mbox{II}$ & 4661.18 & -29.26 & 27.59 $\pm$ 1.91 & 0.124 & 0.133 & 6 & 4661.86 & 14.47 & 15.76 $\pm$ 0.60 & 0.048 & 0.051 & 4 &  ghost in neb comp \\
4667.11 & $\mbox{[Fe}\thinspace \mbox{III]}$ & 4666.57 & -35.00 & 19.53 $\pm$ 1.62 & 0.064 & 0.068 & 7 & 4667.28 & 10.61 & 18.95 $\pm$ 0.59 & 0.042 & 0.045 & 4 &  ghost \\
4673.73 & $\mbox{O}\thinspace \mbox{II}$ & 4673.30 & -27.88 & 29.00 $\pm$ 13.54 & 0.021 & 0.022 & 29 & 4674.00 & 17.02 & 11.80 $\pm$ 2.54 & 0.007 & 0.007 & 16 &  \\
4676.23 & $\mbox{O}\thinspace \mbox{II}$ & 4675.79 & -28.51 & 19.68 $\pm$ 2.89 & 0.046 & 0.049 & 11 & 4676.47 & 15.09 & 13.21 $\pm$ 0.69 & 0.026 & 0.028 & 5 &  \\
4699.22 & $\mbox{O}\thinspace \mbox{II}$ & * & * & * & * & * & * & 4699.42 & 12.50 & 24.18 $\pm$ 9.72 & 0.011 & 0.012 & 33 &  \\
4701.64 & $\mbox{[Fe}\thinspace \mbox{III]}$ & 4701.13 & -32.77 & 12.63 $\pm$ 0.25 & 0.299 & 0.315 & 3 & 4701.85 & 13.14 & 13.39 $\pm$ 0.05 & 0.247 & 0.260 & 2 &  \\
4704.55 & $\mbox{Ni}\thinspace \mbox{III?}$ & * & * & * & * & * & * & 4704.88 & 20.78 & 26.83 $\pm$ 3.26 & 0.014 & 0.015 & 9 &  \\
4705.35 & $\mbox{O}\thinspace \mbox{II}$ & * & * & * & * & * & * & 4705.60 & 15.68 & 14.40 $\pm$ 1.02 & 0.017 & 0.018 & 5 &  \\
\hline
\end{tabular}
\end{adjustbox}
\end{table*}

\begin{table*}
\caption{Atomic data set used for collisionally excited lines.}
\label{tab:atomic_data}
\begin{tabular}{lcc}
\hline
\multicolumn{1}{l}{Ion} & \multicolumn{1}{c}{Transition Probabilities} &
\multicolumn{1}{c}{Collision Strengths} \\
\hline

O$^{+}$   &  \citet{Fischer04} & \citet{Kisielius09}\\
O$^{2+}$  &  \citet{Wiese96}, \citet*{Storey00} & \citet{Storey14}\\
N$^{+}$   &  \citet{Fischer04} & \citet{Tayal11}\\
Ne$^{2+}$  &  \citet{McLaughlin11} & \citet{McLaughlin11}\\
S$^{+}$   &  \citet{Podobedova09} & \citet{Tayal10}\\
S$^{2+}$  &  \citet{Podobedova09} & \citet{Grieve14}\\
Cl$^{+}$ &  \citet{Mendoza83} & \citet{Tayal04}\\
Cl$^{2+}$ &  \citet{Fritzsche99} & \citet{Butler89}\\
Cl$^{3+}$ &  \citet{Kaufman86}, \citet{Mendoza82a}, \citet{Ellis84} & \citet{Galavis95}\\
Ar$^{2+}$ &   \citet{Mendoza83_2}, \citet{Kaufman86}  & \citet*{Galavis95}\\
Ar$^{3+}$ &   \citet{Mendoza82b}  & \citet{Ramsbottom97}\\
Fe$^{2+}$ & \citet{Quinet96} , \citet{Johansson00} & \citet{Zhang96}\\
Fe$^{3+}$ & \citet{Fischer08} & \citet{Zhang97}\\
Ni$^{2+}$ & \citet{Bautista01} & \citet{Bautista01}\\
\hline
\end{tabular}
\end{table*}

\begin{table}
\caption{Effective recombination coefficients used for recombination lines.}
\label{tab:rec_atomic_data}
\begin{tabular}{lc}
\hline
\multicolumn{1}{l}{Ion} & \multicolumn{1}{c}{Reference}  \\
\hline
H$^{+}$   & \citet{Storey95}\\
He$^{+}$   & \citet{Porter12,Porter13}\\
O$^{+}$   & \citet{Pequignot91}\\
O$^{2+}$   & \citet{Storey17}\\
C$^{2+}$   & \citet{Davey00}\\
Ne$^{2+}$   & \citet{Kisielius98}\\
\hline
\end{tabular}
\end{table}

\begin{table}
\caption{Critical densities of density diagnostics for $T_{\rm e}=10000 \text{ K}$.}
\label{tab:critical_densities}
\begin{tabular}{lcc}
\hline
Ion & $\lambda$(\AA) & $n_{\text{crit}}(\text{ cm}^{-3}$)\\
\hline
[O\thinspace II]   &  3729, 3726 & $1.30\times 10^{3}$, $4.06\times 10^{3}$ \\

[S\thinspace II]   &  6731, 6716 & $3.06\times 10^{3}$, $1.16\times 10^{3}$ \\

[Cl\thinspace III]  &  5538, 5518 & $3.57\times 10^{4}$, $7.23\times 10^{3}$ \\

[Fe\thinspace III]  &  4658, 4702 & $5.17\times 10^{6}$, $3.09\times 10^{6}$  \\

[Ar\thinspace IV]  &  4740, 4711 & $1.26\times 10^{5}$, $1.39\times 10^{4}$ \\
\hline
\end{tabular}
\end{table}

\begin{table*}
\centering
\caption{Comparison of the observed [Fe\thinspace III] intensity ratios and theoretical ones predicted by the transition probabilities adopted in Table~\ref{tab:atomic_data}}
\label{tab:fe3_ratios_theo}
\begin{tabular}{lcccccccccccc}
\hline
 & \multicolumn{1}{c}{Cut 1} & \multicolumn{2}{c}{Cut 2} & \multicolumn{2}{c}{Cut 3} & \multicolumn{1}{c}{Cut 4} \\
\multicolumn{1}{c}{Ratio} &  Nebula & HH~529~II &  Nebula & HH~529~III &  Nebula &  Nebula & Combined cuts& Prediction \\
\hline
4667/4734$^{*}$ & - & $0.52 \pm 0.04$ & $0.42 \pm 0.02$ & - & $0.45 \pm 0.03$ & $0.43 \pm 0.03$ & - & 0.28 \\
4778/4734 & $0.51 \pm 0.04$ & $0.70 \pm 0.06$ & $0.51 \pm 0.03$ & $0.48 \pm 0.12$ & $0.44 \pm 0.04$ & $0.43 \pm 0.03$ & $0.51 \pm 0.04$ & 0.48  \\
4778/4667$^{*}$ & - & $1.35 \pm 0.14$ & $1.22 \pm 0.06$ & - & $1.09 \pm 0.07$ & $1.00 \pm 0.09$ & - & 1.74 \\
4607/4702$^{**}$ & $0.24 \pm 0.01$ & $0.22 \pm 0.02$ & $0.24 \pm 0.01$ & $0.24 \pm 0.05$ & $0.23 \pm 0.02$ & $0.24 \pm 0.02$ & $0.23 \pm 0.02$ & 0.17\\
4607/4770$^{**}$ & $0.72 \pm 0.05$ & $0.60 \pm 0.06$ & $0.64 \pm 0.03$ & $0.52 \pm 0.11$ & $0.64 \pm 0.06$ & $0.68 \pm 0.05$ & $0.64 \pm 0.05$ & 0.51 \\
4702/4770 & $2.96 \pm 0.15$ & $2.68 \pm 0.16$ & $2.73 \pm 0.10$ & $2.22 \pm 0.27$ & $2.74 \pm 0.19$ & $2.87 \pm 0.14$ & $2.72 \pm 0.14$ & 2.93 \\
4658/4755 & $5.50 \pm 0.20$ & $5.28 \pm 0.24$ & $5.26 \pm 0.15$ & $4.70 \pm 0.51$ & $5.32 \pm 0.24$ & $5.30 \pm 0.19$ & $5.31 \pm 0.19$ & 5.49  \\
5011/5085 & $3.19 \pm 0.72$ & $2.51 \pm 0.93$ & $5.00 \pm 1.00$ & $2.96 \pm 1.22$ & $3.61 \pm 0.46$ & $3.84 \pm 0.94$ & $3.66 \pm 0.96$ & 5.94 \\
5271/5412 & $10.41 \pm 1.36$ & $9.39 \pm 1.49$ & $10.48 \pm 0.81$ & - & $10.37 \pm 0.85$ & $9.51 \pm 1.24$ & $10.84 \pm 1.06$ & 11.01\\
4881/4987$^{***}$ & $4.88 \pm 0.31$ & $4.90 \pm 0.38$ & $5.07 \pm 0.23$ & $2.41 \pm 0.46$ & $5.27 \pm 0.41$ & $5.98 \pm 0.35$ & $4.85 \pm 0.37$ & 5.76  \\
\hline
\end{tabular}
\begin{description}
\item $^*$ $\lambda$4667 affected by ghost. \\
\item $^{**}$ [Fe\thinspace III] $\lambda 4607.12$ blended with  N\thinspace II $\lambda 4607.15$. \\
\item $^{***}$ [Fe\thinspace III] $\lambda 4987.29$ blended with  N\thinspace II $\lambda 4987.38$. \\
\end{description}
\end{table*}

\begin{table}
\caption{Slope and intercept for Eq.~\ref{eq:helium_fit} for a range of densities.}
\label{tab:slopes}
\begin{tabular}{ccc}
\hline
$n_{\rm e}$ (cm$^{-3}$) & $\alpha$ & $\beta$\\
\hline
100 & 92984 & -7455 \\
500 & 81830 & -6031 \\
1000 & 77896 & -5527 \\
2000 & 69126 & -4378 \\
3000 & 65040 & -3851 \\
4000 & 62517 & -3529 \\
5000 & 60744 & -3305 \\
6000 & 59402 & -3137 \\
7000 & 58334 & -3004 \\
8000 & 57456 & -2895 \\
9000 & 56715 & -2804 \\
10000 & 56077 & -2726 \\
12000 & 55637 & -2676 \\
15000 & 55087 & -2611 \\
20000 & 54364 & -2523 \\
25000 & 53796 & -2452 \\
30000 & 53329 & -2392 \\
40000 & 52591 & -2297 \\
45000 & 52289 & -2257 \\
50000 & 52019 & -2222 \\
\hline
\end{tabular}
\end{table}

\begin{table*}
\centering
\caption{Fe$^{2+}$/H$^{+}$ abundances based on $T_{\rm e} (\text{low})$. The units are logarithmic with $n(\text{H})=12$.}
\label{tab:Fe3_abundances_low}
\begin{tabular}{ccccccccc}
\hline
 & \multicolumn{1}{c}{Cut 1} & \multicolumn{2}{c}{Cut 2} & \multicolumn{2}{c}{Cut 3} & \multicolumn{1}{c}{Cut 4} \\
$\lambda$ &  Nebula & HH~529~II &  Nebula & HH~529~III &  Nebula &  Nebula & Combined cuts\\
\hline

4658  & $5.51 \pm 0.02$ & $5.59^{+0.08} _{-0.06}$ & $5.56^{+0.04} _{-0.03}$ & $5.34^{+0.15} _{-0.09}$ & $5.51 \pm 0.04$ & $5.47^{+0.05} _{-0.04}$ & $5.52 \pm 0.03$ \\

4702 & $5.51 \pm 0.02$ & $5.59^{+0.08} _{-0.06}$ & $5.56^{+0.04} _{-0.03}$ &  $5.38^{+0.15} _{-0.10}$  & $5.52 \pm 0.04$ & $5.47^{+0.05} _{-0.04}$ & $5.52^{+0.04} _{-0.03}$ \\

4734 & $5.48 \pm 0.03$ & $5.58^{+0.09} _{-0.07}$ & $5.56 \pm 0.04$  &  $5.36^{+0.16} _{-0.10}$  & $5.50^{+0.05} _{-0.04}$ & $5.47^{+0.05} _{-0.04}$ & $5.52 \pm 0.04$ \\

4755 & $5.51^{+0.03} _{-0.02}$ & $5.61^{+0.09} _{-0.07}$ & $5.58^{+0.04} _{-0.03}$ &  $5.41^{+0.16} _{-0.10}$  & $5.52^{+0.05} _{-0.04}$ & $5.49^{+0.05} _{-0.04}$ & $5.54^{+0.04} _{-0.03}$ \\

4770 & $5.51 \pm 0.03$ &$5.63^{+0.09} _{-0.07}$ & $5.59 \pm 0.04$ &  $5.50^{+0.15} _{-0.10}$ & $5.55^{+0.05} _{-0.04}$ & $5.48^{+0.05} _{-0.04}$ &  $5.56^{+0.04} _{-0.03}$ \\

4778 & $5.51 \pm 0.04$ &$5.75^{+0.09} _{-0.07}$& $5.59 \pm 0.04$ &  $5.36^{+0.18} _{-0.12}$  & $5.51^{+0.05} _{-0.04}$ & $5.42^{+0.06} _{-0.05}$ & $5.55 \pm 0.04$ \\

4881 & $5.53 \pm 0.02$ &$5.61^{+0.09} _{-0.06}$& $5.58^{+0.04} _{-0.03}$ & $5.38^{+0.15} _{-0.10}$   & $5.53 \pm 0.04$ & $5.49^{+0.05} _{-0.04}$ & $5.54 \pm 0.03$ \\

5011 & $5.52 \pm 0.04$ &$5.78^{+0.08} _{-0.07}$& $5.56 \pm 0.04$  & $5.55^{+0.15} _{-0.11}$  & $5.51^{+0.05} _{-0.04}$ & $5.51^{+0.06} _{-0.05}$ &  $5.57 \pm 0.04$\\

5271 & $5.55 \pm 0.02$ &$5.61^{+0.08} _{-0.06}$& $5.58^{+0.04} _{-0.03}$ & $5.40^{+0.14} _{-0.09}$ & $5.55 \pm 0.04$ & $5.52 \pm 0.04$ & $5.55 \pm 0.03$ \\

5412 & $5.57^{+0.06} _{-0.05}$ & $5.68^{+0.10} _{-0.08}$ & $5.60^{+0.05} _{-0.04}$ &  - & $5.57^{+0.05} _{-0.04}$ & $5.58^{+0.07} _{-0.06}$ & $5.56 \pm 0.05$ \\

8838 & $5.37^{+0.11} _{-0.10}$ & $5.41^{+0.19} _{-0.15}$ & $5.48^{+0.08} _{-0.07}$ &-& $5.34^{+0.09} _{-0.08}$ & $5.47^{+0.11} _{-0.09}$ & $5.34^{+0.09} _{-0.08}$ \\

\textbf{Weighted Average} &\boldmath$5.52 \pm 0.03$& \boldmath$5.62 \pm 0.07 $&\boldmath$5.57 \pm 0.02$&\boldmath$5.40 \pm 0.06$&\boldmath$5.52 \pm 0.04$&\boldmath$5.48 \pm 0.03$&\boldmath$5.53 \pm 0.04$\\

\hline
\end{tabular}
\end{table*}

\begin{table*}
\centering
\caption{Fe$^{2+}$/H$^{+}$ abundances based on $T_{\rm e} (\text{high})$. The units are logarithmic with $n(\text{H})=12$.}
\label{tab:Fe3_abundances_high}
\begin{tabular}{ccccccccc}
\hline
 & \multicolumn{1}{c}{Cut 1} & \multicolumn{2}{c}{Cut 2} & \multicolumn{2}{c}{Cut 3} & \multicolumn{1}{c}{Cut 4} \\
$\lambda$ &  Nebula & HH~529~II &  Nebula & HH~529~III &  Nebula &  Nebula & Combined cuts\\
\hline

4658  & $5.77^{+0.05} _{-0.04}$ & $5.92^{+0.03} _{-0.02}$ & $5.81 \pm 0.03$ & $5.72 \pm 0.03$  & $5.78 \pm 0.03$ & $5.75 \pm 0.03$ & $5.79 \pm 0.03$   \\

4702 & $5.76^{+0.05} _{-0.04}$ & $5.92 \pm 0.03$  & $5.80 \pm 0.03$ & $5.75 \pm 0.04$ & $5.78 \pm 0.03$ & $5.74 \pm 0.03$ &  $5.79^{+0.04} _{-0.03}$ \\

4734 & $5.74^{+0.05} _{-0.04}$ & $5.91^{+0.04} _{-0.03}$ & $5.81^{+0.04} _{-0.03}$ &  $5.73 \pm 0.06$ & $5.77^{+0.04} _{-0.03}$ & $5.75^{+0.04} _{-0.03}$ &  $5.79 \pm 0.04$ \\

4755 & $5.76^{+0.05} _{-0.04}$ & $5.94 \pm 0.03$ & $5.83 \pm 0.03$&  $5.79 \pm 0.05$  & $5.79 \pm 0.03$ & $5.76 \pm 0.03$ &$5.81 \pm 0.03$  \\

4770 & $5.76^{+0.05} _{-0.04}$ & $5.96 \pm 0.03$ & $5.84 \pm 0.03$ & $5.87 \pm 0.05$ & $5.81^{+0.04} _{-0.03}$ & $5.75^{+0.04} _{-0.03}$ & $5.82^{+0.04} _{-0.03}$ \\

4778 & $5.76^{+0.06} _{-0.05}$ &  $6.08 \pm 0.04$ & $5.84^{+0.04} _{-0.03}$ &  $5.74 \pm 0.10$ & $5.78^{+0.04} _{-0.03}$ & $5.70^{+0.05} _{-0.04}$ & $5.82 \pm 0.04$  \\

4881 & $5.79 \pm 0.04$ & $5.94 \pm 0.03$  & $5.83 \pm 0.03$ &  $5.76 \pm 0.04$ & $5.79 \pm 0.03$ & $5.76 \pm 0.03$ & $5.80 \pm 0.03$ \\

5011 & $5.76^{+0.06} _{-0.05}$ & $6.09 \pm 0.04$ & $5.79 \pm 0.03$ &   $5.90^{+0.07} _{-0.06}$ & $5.76 \pm 0.03$ & $5.77^{+0.05} _{-0.04}$ & $5.82 \pm 0.04$ \\

5271 & $5.77 \pm 0.04$ & $5.91 \pm 0.03$  & $5.81 \pm 0.03$ &  $5.74 \pm 0.04$ & $5.79 \pm 0.03$ & $5.77 \pm 0.03$ & $5.79 \pm 0.03$ \\

5412 & $5.80^{+0.07} _{-0.06}$ & $5.97^{+0.07} _{-0.06}$ & $5.83 \pm 0.04$ &  - & $5.81 \pm 0.04$ & $5.83 \pm 0.06$ &$5.80 \pm 0.05$  \\

8838 & $5.73^{+0.12} _{-0.11}$ & $5.87 \pm 0.13$ & $5.83^{+0.07} _{-0.06}$ & - & $5.72^{+0.08} _{-0.07}$ & $5.86 \pm 0.09$ &  $5.72^{+0.09} _{-0.08}$ \\

\textbf{Weighted Average} &\boldmath$5.77 \pm 0.02$ &\boldmath$5.94 \pm 0.05$ &\boldmath$5.82 \pm 0.02$ & \boldmath$5.75 \pm 0.05$ &\boldmath$5.78 \pm 0.02$&\boldmath$5.76 \pm 0.03$&\boldmath$5.80 \pm 0.02$
\\

\hline
\end{tabular}
\end{table*}

\begin{table*}
\caption{Cl$^{2+}$/H$^{+}$, Cl/H and log(Cl/O) abundances using $T_{\rm e}$(low), $T_{\rm e}$([S\thinspace III]) and $T_{\rm e}$(high). The units are logarithmic with $n(\text{H})=12$.}
\label{tab:cl_comp}
\begin{adjustbox}{width=\textwidth}
\begin{tabular}{ccccccccccccc}
\hline
& & & \multicolumn{3}{c}{$T_{\rm e}$(low)} & \multicolumn{3}{c}{$T_{\rm e}$([S\thinspace III])} & \multicolumn{3}{c}{$T_{\rm e}$(high)} \\
Cut& Component & O$^{2+}$/O & Cl$^{2+}$/H$^+$ & Cl/H & log(Cl/O) & Cl$^{2+}$/H$^+$ & Cl/H & log(Cl/O) & Cl$^{2+}$/H$^+$ & Cl/H& log(Cl/O) \\
\hline
1 &  Nebular & $0.73 \pm 0.04$ & $4.78 \pm 0.02$ &$4.83 \pm 0.02$&$-3.63\pm 0.03$&$4.88^{+0.06} _{-0.05}$&$4.92 \pm 0.06$&$-3.53 \pm 0.07$&$5.01 \pm 0.03$&$5.04 \pm 0.03$&$-3.42 \pm 0.05$\\
2 &  HH~529~II &$0.94 \pm 0.04$ & $4.80^{+0.08} _{-0.06}$ & $4.83 \pm 0.07$&$-3.74 \pm 0.07$&$5.01^{+0.06} _{-0.05}$&$5.03 \pm 0.05$&$-3.54 \pm 0.05$&$5.08\pm 0.03$&$5.10 \pm 0.03$&$-3.47 \pm 0.04$\\
2 &  Nebular & $0.77 \pm 0.05$ & $4.81^{+0.04} _{-0.03}$ & $4.86 \pm 0.04$&$-3.61\pm 0.06$&$4.93^{+0.06} _{-0.05}$ & $4.97 \pm 0.06$ &$-3.49\pm 0.07$& $5.03 \pm 0.03$ & $5.06 \pm 0.03$&$-3.41\pm 0.05$\\
3 &  HH~529~III & $0.90 \pm 0.05$ & $4.77^{+0.13} _{-0.09}$ & $4.80 \pm 0.12$ & $-3.72 \pm 0.13$ & $5.03^{+0.08} _{-0.07}$ & $5.05 \pm 0.08$ & $-3.48 \pm 0.09$ & $5.09^{+0.05} _{-0.04}$ & $5.11 \pm 0.05$ &$-3.42 \pm 0.06$ \\
3 &  Nebular & $0.79 \pm 0.05$ & $4.80^{+0.04} _{-0.03}$ &$4.85 \pm 0.04$&$-3.60 \pm 0.06$& $4.96^{+0.06} _{-0.05}$& $4.99\pm0.06$ &$-3.46 \pm 0.07$&$5.03^{+0.03} _{-0.02}$ &$5.06 \pm 0.03$& $-3.39 \pm 0.05$ \\  
4 &  Nebular & $0.78 \pm 0.05$ & $4.79 \pm 0.04$ & $4.85 \pm 0.04$ &$-3.62\pm 0.06$ & $4.95^{+0.06} _{-0.05}$ & $4.99 \pm 0.06$ &$-3.48 \pm 0.07$& $5.03 \pm 0.03$&$5.06\pm 0.03$&$-3.41\pm 0.05$ \\  
- & Combined cuts & $0.81 \pm 0.05$ & $4.79 \pm 0.03$&$4.84 \pm 0.03$&$-3.62\pm 0.05$& $4.94^{+0.05} _{-0.04}$&$4.97\pm0.05$&$-3.50 \pm 0.06$&$5.02\pm0.03$&$5.05\pm0.03$&$-3.41 \pm 0.05$ \\  
\hline
\end{tabular}
\end{adjustbox}
\end{table*}

\begin{table*}
\centering
\caption{Comparison of the observed [Ni\thinspace III] intensity ratios and theoretical ones  predicted by the transition probabilities adopted in Table~\ref{tab:atomic_data}}
\label{tab:intensity_ni3}
\begin{tabular}{cccccccccccc}
\hline
& & \citet{mesadelgado09} & \citet{delgadoinglada16} & \multicolumn{2}{c}{This work} &  \\
Ratio & \citet{Esteban04} & HH~202~S &  Orion Bar &  Nebular &  HH~529~II & Prediction\\
\hline
6534/6000& $2.09\pm0.94$  & $1.58 \pm 0.38$ & $1.46\pm 0.40$ & $1.54\pm 0.39$ & $3.35\pm1.40$ & 2.19\\
6946/6000& -&$0.28 \pm 0.09$&-& $0.31\pm 0.11$&$0.82\pm 0.52$&0.39\\
\hline
\end{tabular}
\end{table*}

\begin{table*}
\centering
\caption{Ni$^{2+}$/H$^{+}$ abundances per line. The units are logarithmic with $n(\text{H})=12$.}
\label{tab:Ni3_abundances}
\begin{tabular}{cccccccc}
\hline
 & \multicolumn{1}{c}{Cut 1} & \multicolumn{2}{c}{Cut 2} & \multicolumn{2}{c}{Cut 3} & \multicolumn{1}{c}{Cut 4} \\
$\lambda$ &  Nebula & HH~529~II &  Nebula & HH~529~III &  Nebula &  Nebula & Combined cuts\\
\hline

6000 & $4.52 \pm 0.08$ & $4.42^{+0.13} _{-0.11}$  & $4.34^{+0.07} _{-0.06}$ &  - & $4.40 \pm 0.07$ & $4.42^{+0.10} _{-0.09}$ & $4.35 \pm 0.08$ \\
6534 & $4.22^{+0.10} _{-0.11}$ &4.61:  & $4.19 \pm 0.06$ &4.32:   & $4.16 \pm 0.07$ & $4.18 \pm 0.11$ & $4.26 \pm 0.07$ \\
6682 & $4.76^{+0.14} _{-0.13}$ & - & $4.70^{+0.14} _{-0.13}$ & - & $4.44^{+0.14} _{-0.13}$ & -& -\\
6797 & $4.50^{+0.23} _{-0.22}$ & 4.94:  & $4.79 \pm 0.08$ & - & $4.77 \pm 0.10$ & 4.81: & $4.77 \pm 0.14$ \\
6946 & $3.94^{+0.26} _{-0.23}$ & $4.75^{+0.19} _{-0.17}$ & $4.23 \pm 0.11$& 4.37:  & $4.21 \pm 0.11$ & $4.23^{+0.15} _{-0.14}$ & 4.31:\\
7890 & $4.42 \pm 0.04$ & $4.52^{+0.08} _{-0.06}$ & $4.49 \pm 0.04$ &  $4.28^{+0.15} _{-0.11}$ & $4.45^{+0.05} _{-0.04}$  & $4.45^{+0.05} _{-0.04}$ & $4.46 \pm 0.04$ \\

\textbf{Weighted Average} & \boldmath$4.37 \pm 0.14$ &\boldmath$4.50\pm 0.08$&\boldmath$4.33 \pm 0.17$ &\boldmath$4.28^{+0.15} _{-0.11}$ & \boldmath$4.32 \pm 0.16$&\boldmath$4.36 \pm 0.12$&\boldmath$4.38 \pm 0.10$\\

\hline
\end{tabular}
\end{table*}

\begin{table*}
\centering
\caption{He$^+$/H$^+$ abundances determined using He\thinspace I triplet lines highly affected by self-absorption. The row of ``sum'' is the result of adding the measured intensity of the triplets presented and redistributing it assuming negligible self-absorption effects ($\tau=0$). The units are logarithmic with $n(\text{H})=12$.}
\label{tab:helium_2}
\begin{tabular}{cccccccc}
\hline 
$\lambda_0$ & \multicolumn{1}{c}{Cut 1} & \multicolumn{2}{c}{Cut 2} & \multicolumn{2}{c}{Cut 3} & \multicolumn{1}{c}{Cut 4} & \\
(\AA) &  Nebula & HH~529~II & Nebula & HH~529~III & Nebula & Nebula & Combined Cuts  \\
\hline 
3188 & $10.67 \pm 0.02$ & $10.92 \pm 0.02$ &  $10.63 \pm 0.02$ & $10.97 \pm 0.05$ & $10.62 \pm 0.02$ & $10.68 \pm 0.02$ & $10.71 \pm 0.02$  \\

3889 &$10.60 \pm 0.02 $& $10.93\pm 0.02$ &$10.52 \pm 0.02$& $10.75 \pm 0.02$ & $10.42 \pm 0.02$&$10.55\pm 0.02$&$10.61\pm 0.02$\\

4713 & $11.01 \pm 0.03$ & $11.10 \pm 0.03$  & $11.04 \pm 0.03$ & $11.02^{+0.03} _{-0.04}$  & $11.07 \pm 0.02$ & $11.11 \pm 0.02$ & $11.07 \pm 0.02$  \\

5876 & $10.96 \pm 0.01 $ & $10.95 \pm 0.01$ & $10.97 \pm 0.01 $ & $10.92 \pm 0.01$ & $10.98 \pm 0.01 $ & $10.96 \pm 0.01 $  & $10.97 \pm 0.01 $ \\

7065 &$11.34 \pm 0.04$& $11.22^{+0.05} _{-0.04}$ &$11.35\pm 0.04$& $11.18 \pm 0.06$  &$11.34 \pm 0.04$&$11.37 \pm 0.04$& $11.34 \pm 0.04$\\

\textbf{Sum} &\boldmath$10.91 \pm 0.02 $&\boldmath$10.97\pm 0.02 $&\boldmath$10.90 \pm 0.02 $ &\boldmath$10.93\pm 0.03  $ & \boldmath$10.89 \pm 0.02 $ &\boldmath$10.90 \pm 0.02$&\boldmath$10.91 \pm 0.02$\\

\hline
\end{tabular}
\end{table*}

\begin{table*}
\centering
\caption{He$^+$/H$^+$ abundances determined with He\thinspace I singlet lines and triplet lines less affected by  self-absorption effects. The units are logarithmic with $n(\text{H})=12$.}
\label{tab:helium_1}
\begin{tabular}{cccccccc}
\hline 
$\lambda_0$ & \multicolumn{1}{c}{Cut 1} & \multicolumn{2}{c}{Cut 2} & \multicolumn{2}{c}{Cut 3} & \multicolumn{1}{c}{Cut 4} & \\
(\AA) &  Nebula & HH~529~II & Nebula & HH~529~III & Nebula & Nebula & Combined Cuts  \\
\hline 

3614 & $10.93 \pm 0.02$ & $10.85 \pm 0.04$ & $10.89 \pm 0.02$ &  $11.09 \pm 0.05$  & $10.86 \pm 0.02$ & $10.85 \pm 0.02$ & $10.89 \pm 0.02$  \\

3965 & $10.88 \pm 0.01$ & $10.86 \pm 0.02$ & $10.89 \pm 0.01$ &  $10.93 \pm 0.02$ & $10.87 \pm 0.01$ & $10.88 \pm 0.01$ & $10.89 \pm 0.01$  \\

4026 & $10.89 \pm 0.01$ & $10.97 \pm 0.01$ & $10.90 \pm 0.01$ &  $11.00 \pm 0.01$  & $10.93\pm 0.01$ & $10.93 \pm 0.01$ & $10.93 \pm 0.01$  \\

4388 & $10.90 \pm 0.01$ & $10.96 \pm 0.01$ & $10.92 \pm 0.01$ &  $10.97 \pm 0.01$ & $10.91\pm 0.01$  & $10.91 \pm 0.01$ & $10.92 \pm 0.01$  \\

4438 &  $10.95\pm 0.03$ & $10.88 \pm 0.04$ & $10.92 \pm 0.02$ &  $11.08 \pm 0.07$  & $10.92 \pm 0.02$ & $10.91 \pm 0.03$ & $10.94 \pm 0.03$  \\

4471 &  $10.87 \pm 0.01$ & $10.96 \pm 0.01$ & $10.90 \pm 0.01$ &  $10.93 \pm 0.01$ & $10.91 \pm 0.01$ & $10.93 \pm 0.01$ & $10.91 \pm 0.01$  \\

4922 & $10.90 \pm 0.01$ & $10.94 \pm 0.01$ & $10.92 \pm 0.01$ & $10.94 \pm 0.01$ & $10.92 \pm 0.01$ & $10.92 \pm 0.01$ & $10.92 \pm 0.01$  \\

5016 & $10.87 \pm 0.01$ & $10.78 \pm 0.02$ & $10.88 \pm 0.01$ & $10.84 \pm 0.02$ & $10.88 \pm 0.01$ & $10.88 \pm 0.01$ & $10.87 \pm 0.01$  \\

6678 & $10.90 \pm 0.02$ & $10.94 \pm 0.02$ & $10.91 \pm 0.01$ & $10.93 \pm 0.02$  & $10.92 \pm 0.02$ & $10.90\pm 0.01$ & $10.92 \pm 0.02$   \\

7281 & $10.90\pm 0.03$ & $10.95 \pm 0.03$ & $10.92\pm 0.03$ & $10.93 \pm 0.04$  & $10.92 \pm 0.03$ & $10.91 \pm 0.03$ & $10.93 \pm 0.03$  \\

\textbf{Average} & \boldmath$10.89 \pm 0.02$ &\boldmath$10.95\pm 0.03$&\boldmath$10.90 \pm 0.01$&\boldmath$10.95 \pm 0.03 $&\boldmath$10.90 \pm 0.02$&\boldmath$10.91 \pm 0.02$&\boldmath$10.91 \pm 0.02$ \\

\hline
\end{tabular}
\end{table*}

\begin{sidewaystable*}
\centering
\caption{O$^{2+}$ abundances based on RLs. The units are logarithmic with $n(\text{H})=12$.}
\label{tab:OII_abundances}
\begin{tabular}{cccccccccc}
\hline
 & & & \multicolumn{1}{c}{Cut 1} & \multicolumn{2}{c}{Cut 2} &  \multicolumn{2}{c}{Cut 3} & \multicolumn{1}{c}{Cut 4}&\\
Mult.& Transition& $\lambda_0$ &  Nebula & HH~529~II &  Nebula & HH~529~III &  Nebula &  Nebula & Combined Cuts \\
\hline
1& 3s$^{4}$P-3p$^{4}$D$^{0}$  &4638.86 & $8.547 \pm 0.038$  & $8.938 \pm 0.027$  & $8.546 \pm 0.019$ &  $8.742^{+0.113} _{-0.111}$  & $8.566 \pm 0.067$ & $8.519^{+0.031} _{-0.032}$ & $8.632 \pm 0.044$ \\
&&4641.81 & $8.433^{+0.017} _{-0.018}$ & $8.809^{+0.017} _{-0.018}$  & $8.516^{+0.014} _{-0.013}$ &  $8.824^{+0.027} _{-0.026}$  & $8.502^{+0.014} _{-0.013}$ & $8.509^{+0.017} _{-0.018}$ & $8.575^{+0.017} _{-0.018}$   \\
&&4649.13 & $8.473^{+0.024} _{-0.022}$ & $8.793 \pm 0.014 $ & $8.520 \pm 0.011$ & $8.782^{+0.026} _{-0.027}$ & $8.536 \pm 0.015$ & $8.534^{+0.018} _{-0.016}$ & $8.589^{+0.016} _{-0.015}$   \\
&&4650.84 & $8.469^{+0.032} _{-0.031}$ & $8.912 \pm 0.027 $ & $8.509 \pm 0.019$ & $9.016^{+0.044} _{-0.043}$ & $8.521 \pm 0.023$ & $8.558^{+0.029} _{-0.028}$ & $8.602 \pm 0.028$   \\
&&4661.63 & 8.534: &  $9.021^{+0.027} _{-0.026}$ & 8.550: & $8.974^{+0.032} _{-0.031}$ & 8.602: & 8.609:  & 8.672:  \\
&&4673.73 & $8.579^{+0.152} _{-0.149}$ & $9.041 \pm 0.120 $ & $8.478^{+0.070} _{-0.069}$ & 8.960: & $8.790^{+0.062} _{-0.060}$ & $8.749^{+0.088} _{-0.087}$ & $8.730^{+0.099} _{-0.095}$   \\
&&4676.23 & $8.400^{+0.052} _{-0.053}$ &  $8.748^{+0.047} _{-0.049}$ & $8.475 \pm 0.022$ & $8.926^{+0.068} _{-0.069}$ & $8.497 \pm 0.026$ & $8.507^{+0.031} _{-0.030}$ & $8.543^{+0.034} _{-0.035}$   \\
&&\textbf{Average} & \boldmath$8.465 \pm 0.043$ & \boldmath$8.830 \pm 0.073$ & \boldmath$8.515 \pm 0.018$ & \boldmath$8.843 \pm 0.085$ & \boldmath$8.517 \pm 0.033$&\boldmath$8.525 \pm 0.026$ &\boldmath$8.584 \pm 0.022$\\
2& 3s$^{4}$P-3p$^{4}$P$^{0}$  &4317.14 & $8.498^{+0.052} _{-0.054}$ & $9.099^{+0.030} _{-0.031}$ & $8.671 \pm 0.022$ &-& $8.644^{+0.047} _{-0.048}$ & $8.656 \pm 0.035$ & 8.723: \\
&&4345.56 & $8.582 \pm 0.057$ & $9.194^{+0.043} _{-0.044}$ & $8.714 \pm 0.026$ & $9.139^{+0.072} _{-0.073}$ & $8.659 \pm 0.031$ & $8.726 \pm 0.030$ & $8.829^{+0.043} _{-0.044}$  \\
&&4349.43 & $8.659 \pm 0.040$ & $9.015^{+0.022} _{-0.021}$ & $8.678 \pm 0.013$ & $8.996^{+0.051} _{-0.052}$  & $8.650^{+0.021} _{-0.022}$ & $8.715^{+0.026} _{-0.025}$ & $8.767 \pm 0.030$  \\
&&4366.89 & $8.631^{+0.041} _{-0.040}$ &  $9.237^{+0.026} _{-0.025}$ & $8.710 \pm 0.022$ &$9.210^{+0.056} _{-0.055}$& $8.672^{+0.030} _{-0.031}$ & $8.724^{+0.031} _{-0.030}$ & $8.835 \pm 0.030$  \\
&&\textbf{Average} & $8.595 \pm 0.062$  &  $9.085 \pm 0.091$ &$8.686 \pm 0.016$& $9.074 \pm 0.097$  & $8.656 \pm 0.009$&$8.706 \pm 0.027$&$8.802 \pm 0.033$\\
5& 3s$^{2}$P-3p$^{2}$D$^{0}$  &4414.90 &$8.807^{+0.053} _{-0.051}$& $8.939^{+0.046} _{-0.047}$ & $8.754 \pm 0.025$ &-& $8.772^{+0.026} _{-0.027}$  & $8.689^{+0.053} _{-0.052}$ & $8.753^{+0.036} _{-0.035}$   \\
&&4416.97&$8.622^{+0.069} _{-0.063}$ &- & $8.769^{+0.031} _{-0.030}$ & -& $8.734 \pm 0.039$ & $8.712^{+0.043} _{-0.044}$ & $8.680^{+0.036} _{-0.034}$  \\
&&\textbf{Average} & $8.710 \pm 0.093$ & $8.939^{+0.046} _{-0.047}$  &$8.760 \pm 0.007$&-&$8.759 \pm 0.018$&$8.702 \pm 0.011$&$8.712 \pm 0.037$\\
10& 3p$^{4}$D$^{0}$-3d$^{4}$F  &4069.62 & $8.459^{+0.064} _{-0.060}$& -& 8.246: &-& 8.352:  & $8.657^{+0.071} _{-0.069}$ & 8.299:  \\
&&4069.88 & $8.352^{+0.050} _{-0.049}$ & 8.431: &8.297: & 8.341: &8.386: & $8.412^{+0.096} _{-0.091}$ & 8.353: \\
&&4072.15 & $8.449 \pm 0.026$ & $8.608^{+0.035} _{-0.034}$ & $8.421^{+0.017} _{-0.018}$ &  $8.758 \pm 0.064$  & $8.423^{+0.021} _{-0.022}$ & $8.372^{+0.027} _{-0.026}$ & $8.467 \pm 0.026$  \\
&&4075.86 & 8.436: & 8.757: &-&-&-& $8.407^{+0.030} _{-0.029}$ & - \\
&&4078.84 & $8.299^{+0.166} _{-0.159}$ & $8.998^{+0.093} _{-0.092}$ & $8.503^{+0.040} _{-0.039}$ & -& 8.787: & -&- \\
&&4085.11 & $8.212 \pm 0.150$ & 9.385: & $8.509 \pm 0.035$ & 9.385:  & $8.408 \pm 0.065$ & $8.699^{+0.044} _{-0.043}$ & 8.810: \\
&&4092.93 & -&  8.550: & $8.598^{+0.049} _{-0.047}$ & 8.553: & $8.453^{+0.073} _{-0.071}$ & 8.627: & 9.013: \\
&&\textbf{Average} & $8.421 \pm 0.048$ & $8.623 \pm 0.093$  &$8.450 \pm 0.052$&  $8.758 \pm 0.064$ &$8.423 \pm 0.009$&$8.413 \pm 0.099$ & $8.467 \pm 0.026$ \\
15& 3s$^{2}$D-3p$^{2}$F$^{0}$  &4590.97 & $8.414^{+0.062} _{-0.060}$ & $8.595^{+0.062} _{-0.061}$  & $8.385^{+0.034} _{-0.033}$ &  $8.681^{+0.073} _{-0.072}$  & $8.387 \pm 0.031$ & $8.415 \pm 0.057$ & $8.433^{+0.045} _{-0.043}$  \\
19& 3p$^{4}$P$^{0}$-3d$^{4}$P &4121.46 & 8.791:& -& $8.865 \pm 0.030$ & -& 8.806: & 8.862: & 8.762: \\
&&4132.80 & $8.451^{+0.069} _{-0.071}$ & $9.029 \pm 0.056 $ & $8.565^{+0.027} _{-0.026}$ & $9.023^{+0.154} _{-0.148}$ & $8.512 \pm 0.035$  & $8.579 \pm 0.044$ & $8.651^{+0.053} _{-0.051}$ \\
&&4153.30 & $8.622^{+0.038} _{-0.037}$ & $9.059 \pm 0.031 $ & $8.605 \pm 0.019$ & $8.976^{+0.053} _{-0.051}$ & $8.603^{+0.023} _{-0.024}$ & $8.623 \pm 0.037$& $8.725 \pm 0.032$  \\
&&\textbf{Average} & $8.565 \pm 0.079$ & $9.051 \pm 0.013$ &$8.616 \pm 0.093$& $8.812 \pm 0.153$ &$8.568 \pm 0.044$&$8.603 \pm 0.022$ & $8.701 \pm 0.034$\\
20& 3p$^{4}$P$^{0}$-3d$^{4}$D &4104.99 & -& 8.761: & $8.363 ^{+0.126} _{-0.123}$& -&$8.458^{+0.102} _{-0.098}$ & $8.494 \pm 0.097$ & -\\
&&4110.79 & 8.811: & 9.584: & 8.805: & 9.167: & 8.766:& 9.250: & 9.030: \\
&&4119.22 & $8.563^{+0.049} _{-0.048}$ & $8.842^{+0.061} _{-0.059}$  & $8.626^{+0.023} _{-0.022}$ &-& $8.475^{+0.072} _{-0.074}$ & $8.722^{+0.036} _{-0.034}$ & $8.605^{+0.044} _{-0.043}$ \\
&&\textbf{Average}  & $8.563^{+0.049} _{-0.048}$& $8.842^{+0.061} _{-0.059}$ &$8.626 \pm 0.087$& 9.167: &$8.469 \pm 0.008$&$8.674 \pm 0.087$&$8.605^{+0.044} _{-0.043}$\\
36& 3p$^{2}$F$^{0}$-3d$^{2}$G &4185.44 & $8.095^{+0.062} _{-0.061}$  & $8.034 \pm 0.112 $ & $7.971^{+0.045} _{-0.043}$  &  8.425: & $8.079^{+0.065} _{-0.066}$ & $8.108^{+0.071} _{-0.070}$ & $8.074^{+0.092} _{-0.088}$   \\
&&4189.79 & $8.344^{+0.047} _{-0.044}$ & $8.638 \pm 0.048 $ & $8.307^{+0.031} _{-0.029}$ & $8.808 \pm 0.120 $ & $8.329^{+0.071} _{-0.070}$ & $8.380^{+0.055} _{-0.052}$ & $8.418^{+0.062} _{-0.063}$   \\
&&\textbf{Average} & $8.200 \pm 0.128$ &  $8.279 \pm 0.323$ & $8.107 \pm 0.172$ & $8.808 \pm 0.120 $  &$8.146 \pm 0.119$&$8.217 \pm 0.138$&$8.211 \pm 0.176$\\
3d-4f& 3d$^{4}$F-4fG$^{2}$$\left[3\right]^{0}$ &4087.15 & $8.588 \pm 0.092$ & $8.868^{+0.117} _{-0.119}$ & $8.644^{+0.065} _{-0.067}$ & $9.142 \pm 0.099$ & $8.431^{+0.057} _{-0.056}$ & $8.526^{+0.096} _{-0.095}$ & $8.651 \pm 0.082$  \\
&3d$^{4}$F-4fG$^{2}$$\left[5\right]^{0}$ &4089.29 & $8.451^{+0.058} _{-0.057}$ & 8.939: & $8.518^{+0.032} _{-0.030}$ & 8.752:& $8.422^{+0.028} _{-0.027}$ & $8.412^{+0.051} _{-0.048}$  & $8.555 \pm 0.036$ \\
&3d$^{4}$F-4fG$^{2}$$\left[3\right]^{0}$ &4095.64 & - & $8.824 \pm 0.100 $  & $8.499^{+0.054} _{-0.053}$ &  9.385:  & 8.545: &-&-  \\
&3d$^{4}$F-4fG$^{2}$$\left[4\right]^{0}$&4097.26 & $8.547^{+0.059} _{-0.056}$ & -& $8.551^{+0.029} _{-0.028}$ & -& $8.577 \pm 0.031$ & $8.592^{+0.039} _{-0.040}$ & $8.497^{+0.036} _{-0.035}$ \\
&3d$^{4}$D-4fF$^{2}$$\left[4\right]^{0}$ &4275.55 & $8.464^{+0.070} _{-0.068}$ &  $8.680^{+0.106} _{-0.110}$ & $8.520^{+0.046} _{-0.047}$ & - & $8.475^{+0.053} _{-0.052}$  & $8.554^{+0.063} _{-0.060}$ & $8.497^{+0.062} _{-0.061}$  \\

&&\textbf{Average} & $8.494 \pm 0.052$& $8.767 \pm 0.082$ &$8.534 \pm 0.032$& $9.142 \pm 0.099$ &$8.468 \pm 0.067$&$8.507 \pm 0.081$&$8.525 \pm 0.041$\\
&Mult. 1, 2, 10, 20 and 3d-4f transitions & Average&  $8.465 \pm 0.072$ & $8.838 \pm 0.143$ &  $8.539 \pm 0.084$ & $8.857 \pm 0.115$  & $8.512 \pm 0.075$ &$8.523 \pm 0.108$ &$8.541 \pm 0.137$ \\
\hline
\end{tabular}
\end{sidewaystable*}

\begin{table*}
\centering
\caption{O$^+$, C$^{2+}$ and Ne$^{2+}$ abundances based on RLs. The units are logarithmic with $n(\text{H})=12$.}
\label{tab:other_rls_abundances}
\begin{tabular}{cccccccccc}
\hline
 & & & \multicolumn{1}{c}{Cut 1} & \multicolumn{2}{c}{Cut 2} &  \multicolumn{2}{c}{Cut 3} & \multicolumn{1}{c}{Cut 4}&\\
Mult.& Transition& $\lambda_0$ &  Nebula & HH~529~II &  Nebula & HH~529~III &  Nebula &  Nebula & Combined Cuts \\
\hline
\noalign{\vskip3pt}
  \multicolumn{10}{c}{\bf{O$^{+}$} }\\
\noalign{\vskip3pt}
1&3s$^{5}$S$^{0}$-3p$^{5}$P & 7771.94&\multirow{ 3}{*}{$8.344 \pm 0.100$}&\multirow{ 3}{*}{<7.91}&\multirow{ 3}{*}{$8.250 \pm 0.064$}&\multirow{ 3}{*}{<7.95}&\multirow{ 3}{*}{$8.275 \pm 0.073$}&\multirow{ 3}{*}{$8.274^{+0.068} _{-0.069}$}&\multirow{ 3}{*}{ $8.187 \pm 0.073$}\\
&&7774.17&\\
&&7775.39&\\
\noalign{\vskip3pt}
  \multicolumn{10}{c}{\bf{C$^{2+}$} }\\
\noalign{\vskip3pt}  
6&3d$^2$D–4f$^2$F$^0$ & 4267.00&\multirow{ 3}{*}{$8.349^{+0.030} _{-0.031}$}&\multirow{ 3}{*}{$8.457 \pm 0.017 $}&\multirow{ 3}{*}{$8.347 \pm 0.017$}&\multirow{ 3}{*}{$8.557 \pm 0.026 $}&\multirow{ 3}{*}{$8.339 \pm 0.013$}&\multirow{ 3}{*}{$8.328^{+0.021} _{-0.022}$}&\multirow{ 3}{*}{$8.371 \pm 0.026$}\\
&&4267.18&\\
&&4267.26&\\

16.04 &4d$^2$D–6f$^2$F$^{0}$& 6151.27&\multirow{ 2}{*}{-}&\multirow{2}{*}{9.054:}&\multirow{ 2}{*}{8.376:}&\multirow{ 2}{*}{-}&\multirow{ 2}{*}{$8.441 \pm 0.120$}&\multirow{ 2}{*}{-}&\multirow{ 2}{*}{-}\\
&&6151.53&\\

17.02 &4f$^2$F$^{0}$–5g$^2$G& 9903.46&\multirow{ 2}{*}{$8.326^{+0.043} _{-0.045}$}&\multirow{2}{*}{$8.465^{+0.066} _{-0.063}$}&\multirow{ 2}{*}{$8.363 \pm 0.035$}&\multirow{ 2}{*}{$8.622^{+0.138} _{-0.136}$ }&\multirow{ 2}{*}{$8.301^{+0.057} _{-0.056}$}&\multirow{ 2}{*}{$8.353^{+0.043} _{-0.042}$}&\multirow{ 2}{*}{$8.377^{+0.057} _{-0.056}$}\\
&&9903.89&\\

17.04 &4f$^2$F$^{0}$–6g$^2$G& 6461.95&\multirow{ 2}{*}{$8.354 \pm 0.091$}&\multirow{2}{*}{8.704:}&\multirow{ 2}{*}{$8.298 \pm 0.065$}&\multirow{ 2}{*}{-}&\multirow{ 2}{*}{$8.318 \pm 0.060$}&\multirow{ 2}{*}{$8.351 \pm 0.071$}&\multirow{ 2}{*}{$8.353^{+0.118} _{-0.116}$}\\
&&6462.13&\\

17.06 &4f$^2$F$^{0}$–7g$^2$G& 5342.38&\multirow{ 2}{*}{8.619:}&\multirow{2}{*}{-}&\multirow{ 2}{*}{$8.449^{+0.059} _{-0.060}$}&\multirow{ 2}{*}{-}&\multirow{ 2}{*}{$8.502^{+0.064} _{-0.065}$}&\multirow{ 2}{*}{-}&\multirow{ 2}{*}{-}\\
&&5342.50&\\

&&\textbf{Adopted}&\boldmath${8.342\pm 0.030}$&\boldmath${8.458 \pm 0.021}$&\boldmath${ 8.351\pm 0.025}$&\boldmath${8.560 \pm 0.026}$&\boldmath${8.340 \pm 0.029}$&\boldmath${8.334\pm0.022}$&\boldmath${8.371\pm 0.026}$\\

\noalign{\vskip3pt}
 \multicolumn{10}{c}{\bf {Ne$^{2+}$} }\\
\noalign{\vskip3pt}

1&3s$^{4}$P-3p$^{4}$P$^{0}$&3694.21&-&$8.643^{+0.072} _{-0.073}$ & $8.095^{+0.063} _{-0.065}$ &  -&8.315:&-&-\\
&&3766.26 &-& $8.515^{+0.137} _{-0.135}$ & $8.034^{+0.089} _{-0.092}$ &  -&$8.036^{+0.153} _{-0.150}$&-&-  \\
&&\textbf{Adopted}&-&\boldmath${8.603 \pm 0.057}$&\boldmath${ 8.072\pm0.029}$&-&\boldmath${ 8.036\pm0.150}$&-&-\\

\hline
\end{tabular}
\end{table*}

\begin{table*}
\centering
\caption{Values of $t^2$ estimated for each component, based on the combination of $T_{\rm e}(\text{He\thinspace I})$, $T_{\rm e}(\text{[O\thinspace III]})$, $T_{\rm e}(\text{[S\thinspace III]})$ and $T_{\rm e}(\text{[N\thinspace I]I})$. }
\label{tab:t2_per_comp}
\begin{tabular}{cccccccc}
\hline
Cut &  Component & $t^2(\text{O}^{2+})$& $t^2(\text{S}^{2+})$& $t^2(\text{N}^{+})$\\
\hline
1&Nebular&$0.004 \pm 0.012$&$0.040 \pm 0.026$&$0.053 \pm 0.018$\\
2&HH~529~II&$0.025 \pm 0.013$&$0.062 \pm 0.026$&$0.095 \pm 0.024$\\
2&Nebular&$0.008 \pm 0.012$&$0.039 \pm 0.025$&$0.058 \pm 0.018$\\
3&HH~529~III&$0.030 \pm 0.017$ & $0.072 \pm 0.034$ & $0.120 \pm 0.038$ \\
3&Nebular&$0.010 \pm 0.013$&$0.036 \pm 0.026$&$0.064 \pm 0.019$\\
4&Nebular&$0.022 \pm 0.014$&$0.062 \pm 0.027$&$0.079 \pm 0.020$\\
\hline
\end{tabular}
\end{table*}

\begin{table*}
\centering
\caption{Average velocities and FWHM for the observed ions in each component. The number of lines used in the average are shown in parentheses.}
\label{tab:kin_tab}
\begin{tabular}{cccccccc}
\hline
 & & \multicolumn{2}{c}{Nebular Cut 2} & \multicolumn{2}{c}{HH~529~II} &  \multicolumn{2}{c}{HH~529~III}\\

& I.P. & $\langle V \rangle$ & $\langle \text{FWHM} \rangle$ & $\langle V \rangle $ & $\langle \text{FWHM} \rangle $ & $\langle V \rangle $ & $\langle \text{FWHM} \rangle $\\

Ion& (eV) & $( \text{Km s}^{-1})$ & $( \text{Km s}^{-1})$ & $( \text{Km s}^{-1})$ & $( \text{Km s}^{-1})$& $( \text{Km s}^{-1})$ & $( \text{Km s}^{-1})$\\
\hline

\mbox{[O}\thinspace \mbox{I]} & 0.00 & $28.80 \pm 0.03 \left( 2 \right)$ & $11.80 \pm 0.02 \left(2\right) $&-&-&-&-\\ 

\mbox{[C}\thinspace \mbox{I]} & 0.00 & $28.70 \pm 0.03 \left( 3 \right)$& $9.49 \pm 0.10 \left(3\right) $&-&-&-&-\\ 

\mbox{[N}\thinspace \mbox{I]} & 0.00 & $30.25 \pm 0.64 \left( 3 \right)$ & $9.51 \pm 0.60 \left(3\right) $&-&-&-&-\\  

\mbox{[Cr}\thinspace \mbox{II]} & 6.77 & $28.28 \pm 2.66  \left( 5 \right)$& $9.71 \pm 1.19 \left(5\right) $& $-26.50 \pm 1.00 \left(1\right) $& $25.60 \pm 1.00 \left(1\right) $& $-27.20 \pm 1.00 \left(1\right) $& $39.40 \pm 1.00 \left(1\right) $\\

\mbox{[Ni}\thinspace \mbox{II]} & 7.64 & $30.20 \pm 0.08  \left( 10 \right)$& $11.75 \pm 0.63 \left(10\right) $& $-30.20 \pm 6.09 \left(3\right) $& $15.81 \pm 2.83 \left(3\right) $&-&-\\

\mbox{[Fe}\thinspace \mbox{II]} & 7.90 & $27.33 \pm 1.36  \left( 34 \right)$& $11.84 \pm 2.29 \left(34\right) $&  $-28.76 \pm 1.21 \left(2\right) $& $13.66 \pm 0.97 \left(2\right) $&-&-\\ 

\mbox{[S}\thinspace \mbox{II]} & 10.36 & $23.10 \pm 0.86  \left( 4 \right)$& $21.52 \pm 0.90 \left(4\right) $& $-27.80 \pm 0.03 \left(4\right) $& $16.29 \pm 0.45 \left(4\right) $& $-21.94 \pm 1.46 \left(6\right) $& $22.92 \pm 1.00 \left(6\right) $\\

\mbox{[Cl}\thinspace \mbox{II]} & 12.97 & $24.80 \pm 1.00  \left( 1 \right)$& $20.40 \pm 1.00 \left(1\right) $& $-23.80 \pm 1.00 \left(1\right) $&$18.90 \pm 1.00 \left(1\right) $&-&-\\ 

\mbox{H}\thinspace \mbox{I} & 13.60 & $16.39 \pm 0.78  \left( 51 \right)$& $24.95 \pm 0.16 \left(51\right)$ & $-29.08 \pm 0.36 \left(51\right) $& $27.20 \pm 0.01 \left(51\right)$ & $-23.90 \pm 0.89 \left(45\right) $& $33.12 \pm 0.44 \left(45\right)$ \\

\mbox{[O}\thinspace \mbox{II]} & 13.62 & $17.85 \pm 1.65  \left( 2 \right)$& $18.69 \pm 0.69 \left(2\right) $& $-28.40 \pm 1.20 \left(2\right) $& $20.04 \pm 1.10 \left(2\right) $& $-21.60 \pm 0.03 \left(2\right) $& $25.27 \pm 0.53 \left(2\right) $\\   

\mbox{O}\thinspace \mbox{I} & 13.62 & $18.73 \pm 0.50  \left( 3 \right)$& $23.29 \pm 0.86 \left(3\right)$&-&-&-&-\\

\mbox{[N}\thinspace \mbox{II]} & 14.53 & $20.67 \pm 0.95  \left( 4 \right)$& $19.45 \pm 0.05 \left(4\right) $ & $-28.70 \pm 0.80 \left(4\right) $& $18.13 \pm 0.45 \left(4\right) $& $-19.75 \pm 1.05 \left(3\right) $& $25.98 \pm 0.19 \left(3\right) $\\

\mbox{[Fe}\thinspace \mbox{III]} & 16.19 & $13.83 \pm 1.09  \left( 21 \right)$ & $12.10 \pm 0.32 \left(21\right) $& $-31.90 \pm 0.02 \left(17\right) $& $13.56 \pm 0.34 \left(17\right) $& $-26.43 \pm 1.07 \left(13\right) $& $25.11 \pm 2.47 \left(13\right) $\\

\mbox{[Cr}\thinspace \mbox{III]} & 16.49 & $12.40 \pm 2.02  \left( 4 \right)$& $16.98 \pm 4.86 \left(4\right) $ & $-34.54 \pm 1.56 \left(3\right) $& $20.24 \pm 1.56 \left(3\right) $& $-23.60 \pm 1.00 \left(1\right) $& $10.80 \pm 1.00 \left(1\right) $\\ 

\mbox{[Ni}\thinspace \mbox{III]} & 18.17 & $20.28 \pm 0.15\left(2\right) $& $15.36 \pm 0.87 \left(2\right) $& $-24.54 \pm 0.24 \left(2\right) $& $12.80 \pm 1.57 \left(2\right) $& $-19.20 \pm 1.00 \left(1\right) $& $25.70 \pm 1.00 \left(1\right) $\\

\mbox{[S}\thinspace \mbox{III]} & 23.34 & $15.00 \pm 1.00  \left( 1 \right)$& $12.70 \pm 1.00 \left(1\right) $& $-30.60 \pm 1.00 \left(1\right) $& $15.10 \pm 1.00 \left(1\right) $& $-25.90 \pm 1.00 \left(1\right) $& $25.10 \pm 1.00 \left(1\right) $\\ 

\mbox{[Cl}\thinspace \mbox{III]} & 23.81 & $14.20 \pm 0.30  \left( 4 \right)$& $12.44 \pm 0.29 \left(4\right) $ & $-30.98 \pm 1.91 \left(3\right) $ & $16.16 \pm 1.06 \left(3\right) $& $-27.24 \pm 0.08 \left(2\right) $& $25.81 \pm 1.08 \left(2\right) $\\  

\mbox{C}\thinspace \mbox{II} & 24.38 & $18.63 \pm 1.95  \left( 6 \right)$& $13.71 \pm 0.69 \left(6\right)$& $-26.00 \pm 2.18 \left(5\right) $& $21.61 \pm 1.82 \left(5\right)$& $-23.39 \pm 1.93 \left(2\right) $& $46.05 \pm 4.30 \left(2\right)$ \\

\mbox{He}\thinspace \mbox{I} & 24.59 & $15.87 \pm 0.53  \left( 75 \right)$& $15.81 \pm 0.33 \left(75\right)$ & $-29.08 \pm 0.90 \left(66\right) $& $21.52 \pm 1.23 \left(66\right)$ & $-24.42 \pm 0.60 \left(42\right) $& $26.51 \pm 0.63 \left(42\right)$ \\

\mbox{[Ar}\thinspace \mbox{III]} & 27.63 & $15.40 \pm 0.04  \left( 2 \right)$& $11.10 \pm 0.04 \left(2\right) $& $-30.40 \pm 0.04 \left(2\right) $& $16.10 \pm 0.04 \left(2\right) $& $-25.00 \pm 1.00 \left(1\right) $& $25.10 \pm 1.00 \left(1\right) $\\ 

\mbox{[Fe}\thinspace \mbox{IV]} & 30.65 & $21.20 \pm 1.00  \left( 1 \right)$& $12.30 \pm 1.00 \left(1\right) $& $-23.70 \pm 1.00 \left(1\right) $& $19.40 \pm 1.00 \left(1\right) $&-&-\\  

\mbox{[Cr}\thinspace \mbox{IV]} & 30.96 & $18.40 \pm 1.00  \left( 1 \right)$& $12.10 \pm 1.00 \left(1\right) $& $-25.70 \pm 1.00 \left(1\right) $ & $21.90 \pm 1.00 \left(1\right) $& $-30.20 \pm 1.00 \left(1\right) $& $39.40 \pm 1.00 \left(1\right) $\\ 

\mbox{[O}\thinspace \mbox{III]} & 35.12 & $16.13 \pm 1.37  \left( 3 \right)$ & $11.75 \pm 0.45 \left(3\right) $& $-29.07 \pm 1.03 \left(3\right) $& $17.60 \pm 0.01 \left(3\right) $& $-25.03 \pm 0.42 \left(3\right) $& $25.02 \pm 0.70 \left(3\right) $\\ 

\mbox{O}\thinspace \mbox{II} & 35.12 & $15.80 \pm 0.03  \left( 6 \right)$& $12.64 \pm 0.44 \left(6\right)$ & $-29.38 \pm 0.12 \left(6\right) $& $18.89 \pm 0.43 \left(6\right)$ & $-25.23 \pm 0.68 \left(6\right) $& $31.78 \pm 3.46 \left(6\right)$ \\

\mbox{[Cl}\thinspace \mbox{IV]} & 39.61 & $17.80 \pm 1.00  \left( 1 \right)$ & $10.50 \pm 1.00 \left(1\right) $& $-27.60 \pm 1.00 \left(1\right) $& $21.90 \pm 1.00 \left(1\right) $ & $-20.50 \pm 1.00 \left(1\right) $& $33.10 \pm 1.00 \left(1\right) $\\ 

\mbox{[Ar}\thinspace \mbox{IV]} & 40.74 & $15.35 \pm 1.55  \left( 2 \right)$& $11.69 \pm 0.69 \left(2\right) $& $-29.45 \pm 2.65 \left(2\right) $& $21.20 \pm 2.20 \left(2\right) $& $-26.83 \pm 0.07 \left(2\right) $& $27.88 \pm 2.92 \left(2\right) $\\  

\mbox{[Ne}\thinspace \mbox{III]} & 40.96 & $15.80 \pm 0.10  \left( 2 \right)$ & $12.35 \pm 0.05 \left(2\right) $ & $-28.95 \pm 0.05 \left(2\right) $& $16.22 \pm 0.04 \left(2\right) $& $-25.90 \pm 0.03 \left(2\right) $&  $25.50 \pm 0.03 \left(2\right) $\\  

\mbox{Ne}\thinspace \mbox{II} & 40.96 & $14.92 \pm 0.33 \left( 2 \right)$& $14.44 \pm 1.49 \left(2\right)$& $-26.89 \pm 0.37 \left(2\right) $& $22.28 \pm 0.14 \left(2\right)$ &-&-\\

\mbox{N}\thinspace \mbox{I} & 14.53 & $29.82 \pm 0.25  \left( 16 \right)$& $9.47 \pm 0.20 \left(16\right)$ &$-29.80 \pm 1.00 \left(1\right) $& $36.00 \pm 1.00 \left(1\right)$ & $-25.40 \pm 1.00 \left(1\right) $& $29.40 \pm 1.00 \left(1\right)$ \\

\mbox{Si}\thinspace \mbox{II} & 16.35 & $19.25 \pm 2.45  \left( 10 \right)$& $19.42 \pm 1.35 \left(10\right)$& $-29.18 \pm 1.06 \left(8\right) $& $17.35 \pm 1.78 \left(8\right)$ & $-26.90 \pm 1.16 \left(7\right) $& $23.89 \pm 2.25 \left(7\right)$ \\

\mbox{Si}\thinspace \mbox{III} &33.49& $12.91 \pm 1.00 \left(1\right)$&$13.23 \pm 2.79 \left(1\right)$& $-31.80 \pm 1.00 \left(1\right) $& $10.80 \pm 1.00 \left(1\right)$&-&- \\

\mbox{Ne}\thinspace \mbox{I} & 21.57 & $16.66 \pm 2.38  \left( 6 \right)$& $15.66 \pm 0.65 \left(6\right)$& $-31.78 \pm 5.50 \left(3\right) $& $23.81 \pm 2.20 \left(3\right)$&-&- \\

\mbox{S}\thinspace \mbox{II} & 23.34 & $16.50 \pm 0.69  \left( 5 \right)$& $16.69 \pm 4.21 \left(5\right)$& $-26.88 \pm 0.44 \left(3\right) $& $25.49 \pm 3.34 \left(3\right)$ & $-34.90 \pm 1.00 \left(1\right) $& $33.10 \pm 1.00 \left(1\right)$ \\

\mbox{N}\thinspace \mbox{II} & 29.60 & $15.07 \pm 2.57  \left( 25 \right)$& $13.65 \pm 1.94 \left(25\right)$ & $-29.93 \pm 1.10 \left(10\right) $& $18.33 \pm 2.29 \left(10\right)$ & $-25.31 \pm 3.86 \left(7\right) $& $24.49 \pm 7.94 \left(7\right)$ \\

\mbox{S}\thinspace \mbox{III} & 34.79 & $16.06 \pm 0.80 \left( 4 \right)$& $13.83 \pm 2.61 \left(4\right)$ & $-29.04 \pm 1.87 \left(4\right) $& $25.21 \pm 3.48 \left(4\right)$ &  $-38.86 \pm 4.34 \left(3\right) $& $15.85 \pm 4.12 \left(3\right)$ \\

\mbox{N}\thinspace \mbox{III} & 47.45 & $11.52 \pm 0.81 \left( 2 \right)$& $11.94 \pm 0.16 \left(2\right)$ & $-36.50 \pm 1.00 \left(1\right) $& $22.70 \pm 1.00 \left(1\right)$ &-&-\\ 

\hline
\end{tabular}
\end{table*}


\bsp	
\label{lastpage}
\end{document}